\documentclass[3p,12pt]{elsarticle}

\usepackage{indentfirst}
\usepackage{color,graphicx}
\usepackage[caption=false]{subfig}
\usepackage{caption,booktabs,placeins}
\usepackage{threeparttable}
\usepackage{array}
\usepackage{multirow}
\usepackage{amsmath}
\allowdisplaybreaks[4]	% allow page break in the middle of a multi-line equation
\usepackage{amssymb}
\usepackage{bm}
\usepackage{hyperref}
\usepackage{lineno}
\biboptions{numbers,sort&compress}	% [1,2,3]--> [1-3]

\journal{Journal of Computational Physics}

\newcommand{\dXi}{\Delta\Xi}
\newcommand{\DD}{\mathcal D}
\newcommand{\RR}{\mathcal R}
\newcommand{\CC}{\mathcal C}
\newcommand{\GG}{\mathcal G}
\newcommand{\ih}{\hat{\imath}}
\newcommand{\kh}{\hat{k}}
\newcommand{\bh}{\hat{b}}
\newcommand{\ibh}{\hat{\imath}\hat{b}}
\newcommand{\hh}{\hat{h}}
\newcommand{\vv}{\vec v}
\newcommand{\vn}{\vec n}
\newcommand{\bpsi}{\bm\psi}
\newcommand{\bU}{\bm U}
\newcommand{\bW}{\bm W}
\newcommand{\bR}{\bm R}
\newcommand{\bH}{\bm H}
\newcommand{\bF}{\bm F}

\newcommand{\bK}{\mathsf K}
\newcommand{\bw}{\bm w}
\newcommand{\bwh}{\hat{\bm w}}
\newcommand{\bWh}{\hat{\bm W}}
\newcommand{\bCh}{\hat{\bm C}}
\newcommand{\NS}{\mathrm{NS}}
\newcommand{\BM}{\mathrm{BM}}
\newcommand{\rM}{\mathrm{rM}}
\newcommand{\Heavi}{\Theta}
\newcommand{\dd}{\mathrm d}
\newcommand{\vx}{\vec x}
\newcommand{\vxi}{\vec\xi}
\newcommand{\balpha}{\bm\alpha}
\newcommand{\PP}{\mathcal P}

\begin{document}

\begin{frontmatter}

%----------------------------------------------------------
% 1. Title
%----------------------------------------------------------

\title{Discrete adjoint method for the general synthetic iterative scheme for optimization of rarefied and continuum gas flows}

  \author[hkust1]{ Jianan Zeng}

\author[sustech]{Ruifeng Yuan~\corref{cor1}}
\cortext[cor1]{ Corresponding author: yuanrf@sustech.edu.cn (Ruifeng Yuan). }

  \author[hkust1,hkust2]{ Wei Su}

 \affiliation[hkust1]{organization={Division of Emerging Interdisciplinary Areas, Academy of Interdisciplinary Studies, The Hong Kong University of Science and Technology},
            city={Hong Kong SAR},
            % postcode={518055}, 
 %           state={},
            country={China}}

 \affiliation[sustech]{organization={Department of Mechanics and Aerospace Engineering, Southern University of Science and Technology},
 %           addressline={}, 
            city={Shenzhen},
            postcode={518055}, 
 %           state={},
            country={China}}

  \affiliation[hkust2]{organization={Department of Mathematics, The Hong Kong University of Science and Technology},
    city={Hong Kong SAR},
    % postcode={518055}, 
%           state={},
    country={China}}

\begin{abstract}
A discrete adjoint method is developed for the general synthetic iterative scheme (GSIS) to enable efficient gradient-based optimization of gas flows across rarefied and continuum regimes. To retain the macroscopic acceleration of the primal solver, the primal GSIS is expressed as a fixed-point system of kinetic and macroscopic synthetic equations, and both sets of residuals are incorporated into a discrete Lagrangian. Imposing stationarity of this Lagrangian then yields a coupled mesoscopic--macroscopic adjoint system that directly inherits the primal acceleration structure, without a separate Chapman--Enskog or asymptotic analysis. The resulting system is solved by alternating kinetic and macroscopic updates analogous to those in the primal GSIS, with the inner subproblems treated using established discrete adjoint techniques for the discrete velocity method and Navier--Stokes equations. Tests on the NACA0012 airfoil show that adjoint sensitivities closely agree with central finite-difference estimates, while the adjoint solver achieves convergence efficiency comparable to that of the primal GSIS, with both solvers converging within tens of iterations in rarefied and near-continuum flows. The method is further validated by multi-objective shape optimization and boundary-flow-parameter optimization. For a three-dimensional re-entry capsule, shape optimization reduces peak wall heat flux by 14.17\%, 10.68\%, and 6.54\% while achieving respective lift-to-drag gains of 10\%, 20\%, and 25\%. For a fixed-geometry showerhead chamber, optimization of inlet-flow allocation reduces the coefficient of variation of wafer pressure from 1.49\% to 0.06\% at fixed total gas supply, while maintaining the mean pressure near its initial value. Beyond these applications, the formulation provides a systematic and extensible framework for constructing multiscale adjoint solvers directly from the fixed-point representations of their primal counterparts.
\end{abstract}


\end{frontmatter}

%\linenumbers

%----------------------------------------------------------
% 3. Main text
%----------------------------------------------------------

\section{Introduction}\label{sec:intro}

Rarefied gas flows are central to a broad range of engineering systems, including atmospheric re-entry vehicles~\cite{reed2010investigation} and very-low-Earth-orbit satellites~\cite{walsh2021drag}, as well as vacuum pumps~\cite{sharipov2005numerical}, extreme ultraviolet lithography equipment~\cite{bakshi2009euv}, and fusion exhaust systems~\cite{tantos2020deterministic}. Many such systems also span multiple flow regimes: a re-entry vehicle can experience free-molecular, transition, slip, and continuum flows along its trajectory~\cite{tsien1946superaerodynamics,reed2010investigation}, while different regions of a vacuum device can involve markedly different Knudsen numbers~\cite{sharipov2010numerical}.  In these applications, geometric design strongly affects aerodynamic forces, heat loading, and pressure distributions. Because the Navier--Stokes (NS) equations lose validity as rarefaction increases, a kinetic description~\cite{Cercignanibook1988,Lei2022} is required for consistent modeling across different flow regimes. Efficient optimization methods applicable from continuum to rarefied flows are therefore of considerable practical importance.

For this purpose, surrogate-based optimization~\cite{bhosekar2018advances} is often the most straightforward approach to implement, as it is nonintrusive and can directly use an existing rarefied-flow solver~\cite{walsh2021drag,Liu2025Aerodynamic,li2025surrogate,niu2026cooptimization}. However, the number of high-fidelity samples required for constructing a reliable surrogate grows rapidly with the dimension of the design space, and each sample is expensive for kinetic simulations~\cite{Mieussens2000DISCRETE,bird1994molecular}. Restricting the number of design variables or the admissible shape space alleviates this sampling burden, but sacrifices geometric detail and design freedom. This dilemma motivates adjoint-based gradient methods~\cite{nadarajah2000comparison,peter2010numerical}, for which the cost of sensitivity evaluation is largely independent of the number of design variables.

Recent years have seen a growing number of studies focused on adjoint methods for rarefied gas flows.
Caflisch et al.~\cite{caflisch2021adjoint} developed an adjoint direct simulation Monte Carlo (DSMC) formulation, which was later extended to general collision models~\cite{yang2023adjoint} and to spatially inhomogeneous problems with general boundary conditions~\cite{caflisch2026adjointDSMC}. Guan et al.~\cite{guan2023topology} extended the analysis and developed topology optimization using adjoint information-preserving DSMC. Within deterministic frameworks, Sato et al.~\cite{sato2019topology} developed a continuous-adjoint formulation for topology optimization and employed the discrete velocity method (DVM) for its numerical solution.

Adjoint methods have also been extended across continuum and rarefied regimes. Yuan and Wu developed multiscale adjoint methods for density-based topology optimization~\cite{yuan2024design} and body-fitted shape optimization~\cite{yuan2025adjoint},  with both the primal and adjoint kinetic equations solved using the discrete unified gas-kinetic scheme (DUGKS)~\cite{guo2013discrete} with macroscopic prediction acceleration. Following this framework, Yanbing Zhang et al.~\cite{zhang2026adjointGSIS} subsequently incorporated the general synthetic iterative scheme (GSIS)~\cite{su2020can} into adjoint shape optimization, and then Li and Wu~\cite{li2027adjointOscillatory} extended the method to oscillatory rarefied gas flows. Using a discretize-then-differentiate strategy, Yue Zhang et al.~\cite{zhang2026adjointUGKS} recently developed an asymptotic-preserving discrete adjoint unified gas-kinetic scheme (UGKS)~\cite{Xu2010A} for sensitivity analysis, offering a potential route to gradient-based optimization across flow regimes.

Despite these advances, constructing an efficient multiscale adjoint solver remains a substantial challenge. For primal steady-state computations, macroscopic-prediction-based UGKS and DUGKS~\cite{Zhu2016Implicit,yuan2021multi} use Euler- or NS-type prediction equations to accelerate kinetic iteration. GSIS~\cite{su2020can} likewise couples mesoscopic kinetic equations with macroscopic synthetic equations that combine continuum-level components with higher-order kinetic corrections. These macroscopic steps are essential for rapid convergence, particularly in near-continuum flows, and incorporating them into the adjoint formulation is central to developing an efficient method.

Existing multiscale adjoint methods do not yet inherit this acceleration structure automatically.
The continuous-adjoint route~\cite{yuan2024design,yuan2025adjoint,zhang2026adjointGSIS,li2027adjointOscillatory} derives and then discretizes the adjoint kinetic equation, requiring construction of a dedicated multiscale scheme that involves asymptotic analysis, derivation of macroscopic acceleration equations through a Chapman--Enskog expansion, and compatible flux and boundary treatments. The discrete adjoint UGKS of Yue Zhang et al.~\cite{zhang2026adjointUGKS} avoids deriving a continuous adjoint equation, but still requires applying ``macroscopic-moment projection'' and ``microscopic lifting'' to the adjoint system, followed by asymptotic analysis and the explicit construction of macroscopic acceleration equations.
Thus, neither route automatically inherits the primal acceleration structure; instead, both require a separate and intricate theoretical analysis of the adjoint system. This analysis is particularly challenging because adjoint variables and their kinetic moments often lack clear physical interpretations. As a result, developing, understanding, and applying multiscale adjoint methods require considerable expertise and effort, particularly when extending them to more complex physical models, such as gas mixtures~\cite{zeng2024mixtures} and coupled gas--radiation transport~\cite{zeng2026radiation}.

The present work addresses this difficulty through a discrete adjoint formulation of GSIS. We incorporate macroscopic acceleration into the iterative fixed-point representation of the primal GSIS and include both kinetic and macroscopic residuals in the Lagrangian. Discrete adjoint analysis then yields a coupled kinetic--macroscopic adjoint system that directly inherits the macroscopic acceleration equations. The adjoint system can be solved by alternating mesoscopic and macroscopic updates, which closely mirrors the primal GSIS. Moreover, the inner problems at both levels have the same forms as the discrete adjoints of conventional DVM and NS discretizations, respectively, and can therefore be solved using established techniques. Once the primal GSIS is specified, systematic differentiation and transposition of the discrete residuals yield the complete adjoint scheme, without a separate Chapman--Enskog or asymptotic analysis. The accuracy and efficiency of the method are assessed through sensitivity and convergence tests for flows around an airfoil, multi-objective shape optimization of a three-dimensional re-entry capsule, and inlet-flow allocation in a showerhead chamber.

The remainder of this paper is organized as follows. Section~\ref{sec:gsis} reviews the primal GSIS, Section~\ref{sec:method} presents the discrete adjoint formulation and iterative solution procedure, and Section~\ref{sec:sens} describes sensitivity evaluation. Section~\ref{sec:result} reports the numerical verification and optimization results, and Section~\ref{sec:conc} summarizes the main findings.

\section{Method of GSIS}\label{sec:gsis}
The discrete adjoint formulation developed here starts from the coupled mesoscopic--macroscopic system solved by the primal GSIS. We therefore first introduce the kinetic model and its discretization, and then describe the macroscopic acceleration and boundary treatment in a form suitable for subsequent differentiation.

\subsection{Gas-kinetic model}\label{sec:kinetic-model}
In gas-kinetic theory, the gas state is described by the molecular velocity distribution function $f(t,\vx,\vv)$, where $t$, $\vx$, and $\vv$ denote time, position, and molecular velocity, respectively. The macroscopic conservative variables are obtained by taking moments in molecular velocity space~\cite{Cercignanibook1988,Lei2022}:
\begin{equation}
\bW=\int_{\mathbb R^3}\bpsi f\,\dd\Xi,
\qquad
\bW=(\rho,\rho\vec u,\rho E)^{\mathsf T},
\qquad
\bpsi=(1,\vv,|\vv|^2/2)^{\mathsf T},
\label{eq:kinetic-moments}
\end{equation}
where $\dd\Xi=\dd v_1\dd v_2\dd v_3$, and $\rho$, $\vec u$, and $E$ are the density, macroscopic velocity, and specific total energy. For a monatomic gas,
\begin{equation}
E=\frac{|\vec u|^2}{2}+\frac{RT}{\gamma-1},
\qquad \gamma=\frac{5}{3},
\qquad p=\rho RT,
\label{eq:gas-state}
\end{equation}
with $T$ the temperature, $p$ the pressure, and $R$ the specific gas constant.

For the sake of conciseness, we first use the Bhatnagar--Gross--Krook (BGK) model~\cite{bhatnagar1954model},
\begin{equation}
\frac{\partial f}{\partial t}+\vv\cdot\nabla f
=\frac{g-f}{\tau},
\label{eq:bgk-model}
\end{equation}
whose equilibrium state is the Maxwellian distribution
\begin{equation}
g=g_{\mathrm M}(\rho,\vec u,T)
=\frac{\rho}{(2\pi RT)^{3/2}}
\exp\left(-\frac{|\vv-\vec u|^2}{2RT}\right).
\label{eq:maxwellian}
\end{equation}
The relaxation time and   viscosity power law are
\begin{equation}
\tau=\frac{\mu}{p},
\qquad
\mu(T)=\mu_{\mathrm{ref}}\left(\frac{T}{T_{\mathrm{ref}}}\right)^\omega,
\label{eq:relaxation-viscosity}
\end{equation} 
where $\mu_{\mathrm{ref}}$ is the viscosity at the reference temperature $T_{\mathrm{ref}}$.  Maxwell molecules ($\omega=1$) are used throughout this work.  Taking the conserved moments of Eq.~\eqref{eq:bgk-model} gives
\begin{equation}
\frac{\partial\bW}{\partial t}+\nabla\cdot\vec{\bF}=0,
\qquad
\vec{\bF}=\int\vv\bpsi f\,\dd\Xi,
\label{eq:moment-transport}
\end{equation}
because the collision term conserves mass, momentum, and energy. This moment equation provides the starting point for the macroscopic synthetic equation introduced below. The extension of the formulation to the Shakhov model~\cite{shakhov1968generalization} is given in \ref{app:shakhov-extension}.

The degree of gas rarefaction is characterized by the Knudsen number, for which we adopt the definition
\begin{equation}
\mathrm{Kn}=\frac{\tau}{l_{\mathrm{ref}}}\sqrt{\frac{\pi RT}{2}},
\label{eq:knudsen}
\end{equation}
with $l_{\mathrm{ref}}$ the flow characteristic length. As order-of-magnitude guidelines, wall slip and temperature jump become relevant for $\mathrm{Kn}\gtrsim0.001$, while the Navier--Stokes (NS) equations themselves become inadequate when $\mathrm{Kn}\gtrsim0.1$~\cite{tsien1946superaerodynamics}. A kinetic description is therefore required for a consistent treatment of the continuum--rarefied regimes considered in this work.

\subsection{Mesoscopic discretization and CIS iteration}\label{sec:mesoscopic-cis}
Applying the finite volume method in physical space and the discrete velocity method (DVM) in molecular velocity space to the BGK equation~\eqref{eq:bgk-model} gives the following discrete equation for an interior cell:
\begin{equation}
\DD_{i,k}[f]=\sum_{j\in N(i)}A_{ij}\vv_k\cdot\vn_{ij}f_{ij,k}
=V_i\frac{g_k(\bW_i)-f_{i,k}}{\tau_i(\bW_i)}.
\label{eq:bgk-discrete}
\end{equation}
The time-derivative term is omitted because only steady problems are considered in this work. Here $f_{i,k}$ is the cell-averaged distribution in physical cell $i$ at velocity node $\vv_k$, and $V_i$ is the cell volume. The set $N(i)$ contains the cells sharing an interior face with cell $i$; $A_{ij}$ and $\vn_{ij}$ are the area and unit normal of face $ij$, with the normal pointing outward from cell $i$. The operator $\DD_{i,k}[f]$ denotes the interior transport contribution, while the right-hand side represents molecular relaxation.

The conservative state entering the collision term is calculated directly from the distribution:
\begin{equation}
\bW_i=\sum_k\bpsi_k f_{i,k}\dXi_k.
\label{eq:Wdef}
\end{equation}
The velocity nodes $\vv_k$ and quadrature weights $\dXi_k$ are fixed and do not depend on the flow state or design variables. The equilibrium value $g_k(\bW_i)$ is the Maxwellian~\eqref{eq:maxwellian} evaluated at $\vv_k$, and $\tau_i(\bW_i)$ follows Eq.~\eqref{eq:relaxation-viscosity}; both are determined by the direct moments $\bW_i$.

The face distribution in Eq.~\eqref{eq:bgk-discrete} is reconstructed according to the upwind direction:
\begin{equation}
f_{ij,k}=\begin{cases}
\CC_{ij}[f_k],&\vv_k\cdot\vn_{ij}>0,\\
\CC_{ji}[f_k],&\vv_k\cdot\vn_{ij}<0.
\end{cases}
\label{eq:fij}
\end{equation}
Here $\CC_{ij}$ and $\CC_{ji}$ reconstruct the distribution from the respective sides of the face. They act separately for each velocity index, without coupling different velocity nodes. Faces with $\vv_k\cdot\vn_{ij}=0$ contribute no transport flux. The reconstruction operators, including their boundary counterparts, are described in \ref{app:reconstruction}.

Note that the above discretization is written for the general three-dimensional velocity formulation. For two-dimensional cases, reduced distribution functions are used in the numerical implementation  (see Ref.~\cite{yuan2025adjoint}).

The conventional iterative scheme (CIS) solves Eq.~\eqref{eq:bgk-discrete} for $f$ with $\bW_i$, and hence $g_k(\bW_i)$ and $\tau_i(\bW_i)$, fixed. It then updates $\bW_i$ from the new distribution using Eq.~\eqref{eq:Wdef}. These two operations are repeated until convergence. The macroscopic acceleration of this iteration is introduced next.

\subsection{Macroscopic acceleration and primal GSIS}\label{sec:primal-gsis}
In near-continuum flows, the relaxation time $\tau$ is small and $f$ is close to its local equilibrium. The alternating updates of $f$ and its moments in CIS then advance the macroscopic flow only on a time scale comparable to $\tau$, resulting in slow convergence. GSIS overcomes this difficulty by introducing macroscopic synthetic equations: their NS part provides rapid evolution on macroscopic scales, while kinetic corrections retain the nonequilibrium information required across flow regimes~\cite{su2020can}.

To express the coupling, we introduce the predicted macroscopic state
\begin{equation}
\bU_i=\bW_i+\bw_i,
\label{eq:Udef}
\end{equation}
where $\bw_i$ is the correction supplied by the macroscopic solver relative to the direct kinetic moments $\bW_i$. For an interior cell, the predicted state $\bU_i$ is governed by the macroscopic synthetic equation
\begin{equation}
\sum_{j\in N(i)}A_{ij}\left(\bF_{\NS,ij}[\bU]+\bH_{ij}\right)
=\bm 0 .
\label{eq:synthetic-macro-discrete}
\end{equation}
Here $\bF_{\NS,ij}[\bU]$ is the numerical NS flux evaluated from the predicted macroscopic field $\bU$, and $\bH_{ij}$ is a higher-order kinetic flux correction obtained from the distribution $f$. This equation expresses the steady macroscopic flux balance, with the NS contribution evaluated from $\bU$ and the departure from the NS constitutive relations supplied by the kinetic solution.   

The interface NS flux $\bF_{\NS,ij}$ and details of the associated reconstruction are discussed in \ref{app:reconstruction}.
 The kinetic correction at the face is reconstructed from the cell-center quantities $\vec{\bH}_i$ as
\begin{equation}
\bH_{ij}=\frac{1}{2}\vn_{ij}\cdot\left(\CC_{ij}[\vec{\bH}]+\CC_{ji}[\vec{\bH}]\right),
\label{eq:Hij}
\end{equation}
where each cell-center correction is the difference between the kinetic flux moment and the NS flux evaluated from the direct moments $\bW$:
\begin{equation}
\vec{\bH}_i=\sum_k\vv_k\bpsi_k f_{i,k}\dXi_k-\vec{\bF}_{\NS,i}[\bW],
\label{eq:Hcell}
\end{equation}
with
  \begin{equation}
\vec{\bF}_{\NS,i}[\bW]
=\vec{\bF}_{\NS}\big(\bW_i,\GG_i[\bW]\big),
\qquad \GG_i[\bW]=\big((\nabla\vec u)_i,(\nabla T)_i\big)^\top .
\label{eq:FNS-cell}
\end{equation}

The cell-center NS flux uses the same reference constitutive law as the interface flux. The operator  $\GG_i$ converts the conservative states in the stencil to primitive variables and then reconstructs the velocity and temperature gradients as described in \ref{app:reconstruction}. The cell-center NS flux therefore depends on the states in the full reconstruction stencil. The face reconstruction of  $\vec{\bH}$ uses weights fixed by the mesh, so the kinetic field determines $\bH_{ij}$ in Eq.~\eqref{eq:synthetic-macro-discrete}  while $\bU$   supplies  the macroscopic prediction.

To express the coupled iteration compactly, define the two residual operators

\begin{equation}
\RR_{\NS,i}[\bU]=\sum_{j\in N(i)}A_{ij}\bF_{\NS,ij}[\bU],
\qquad
\RR_{H,i}[\bH]=\sum_{j\in N(i)}A_{ij}\bH_{ij}.
\label{eq:RNS}
\end{equation}

The predicted state $\bU_i$ then replaces the direct moments $\bW_i$ in the collision target and relaxation time of Eq.~\eqref{eq:bgk-discrete}. Together with Eq.~\eqref{eq:synthetic-macro-discrete}, this gives the interior GSIS updates
\begin{align}
\DD_{i,k}[f]+\frac{V_i}{\tau_i}f_{i,k}
&=\frac{V_i}{\tau_i}g_k(\bU_i),\label{eq:gsis1}\\
\RR_{\NS,i}[\bU]&=-\RR_{H,i}[\bH],\label{eq:gsis2}
\end{align}
where $\tau_i=\tau_i(\bU_i)$. The solution to the above two equations can be briefly described as: {\bf freeze the right-hand side and update (solve) the left-hand side}, alternating between the two equations. In the mesoscopic substep, the predicted macroscopic state, and hence $g_k$ and $\tau_i$, are fixed while Eq.~\eqref{eq:gsis1} updates $f$. The resulting distribution supplies the direct moments and the higher-order correction through Eqs.~\eqref{eq:Wdef} and~\eqref{eq:Hcell}. In the macroscopic substep, this higher-order correction is fixed while Eq.~\eqref{eq:gsis2} updates $\bU$, providing the prediction for the next mesoscopic solve. The two substeps alternate until convergence.

Omitting the correction $\bw$ and the macroscopic acceleration, GSIS recovers the CIS update described above. The inner solvers for the kinetic and NS subproblems are specified in \ref{app:primal-inner-solvers}. The boundary contributions needed to complete these interior updates are introduced next.

\subsection{Boundary treatment in the GSIS iteration}\label{sec:primal-boundary}
To complete the primal system, boundary fluxes are added separately from the interior operator $\DD_{i,k}$. Let $B(i)$ denote the boundary faces of cell $i$, with face $ib$ having area $A_{ib}$ and outward unit normal $\vn_{ib}$ relative to the computational domain. Thus $\vv_k\cdot\vn_{ib}>0$ denotes molecules leaving the domain, or incident on a solid wall, whereas $\vv_k\cdot\vn_{ib}<0$ denotes molecules entering the domain, or reflected from the wall. The same inflow/outflow convention is used at open boundaries, where the incoming distribution does not represent physical wall reflection. Define $M(ib,\pm)=\{k:\vv_k\cdot\vn_{ib}\gtrless0\}$ and $\bm f_{ib}^{+}=\{f_{ib,k}:k\in M(ib,+)\}$. The boundary distribution is
\begin{equation}
f_{ib,k}=\begin{cases}
\CC_{ib}[f_k],&\vv_k\cdot\vn_{ib}>0,\\
g_{\mathrm r,ib,k}(\bm f_{ib}^{+}),&\vv_k\cdot\vn_{ib}<0,
\end{cases}
\label{eq:fib}
\end{equation}
where $\CC_{ib}$ reconstructs the outgoing distribution from the interior field, and $g_{\mathrm r,ib,k}$ denotes the incoming distribution (or the reflection kernel) that distinguishes different boundary conditions.

For diffuse reflection, the incoming distribution is constructed from the wall Maxwellian with the normalization
\begin{equation}
\bar g_{\mathrm{rw},ib,k}
=-\frac{g_{\mathrm M,k}(1,\vec u_{\mathrm w,ib},T_{\mathrm w,ib})}{\displaystyle\sum_{\hat k\in M(ib,-)}\vv_{\hat k}\cdot\vn_{ib}
g_{\mathrm M,\hat k}(1,\vec u_{\mathrm w,ib},T_{\mathrm w,ib})\dXi_{\hat k}},\label{eq:g-rw}
\end{equation} 
which enforces zero net mass flux, as
\begin{equation}
g_{\mathrm r,ib,k}(\bm f_{ib}^{+})
=\bar g_{\mathrm{rw},ib,k}\sum_{\hat k\in M(ib,+)}
\vv_{\hat k}\cdot\vn_{ib}f_{ib,\hat k}\dXi_{\hat k}.\label{eq:g-diffuse}
\end{equation}
Here $\vec u_{\mathrm w,ib}$ and $T_{\mathrm w,ib}$ are the prescribed wall velocity and temperature. In the present work, the wall velocity is zero, i.e.~$\vec u_{\mathrm w,ib}=\vec0$. For a Dirichlet boundary, the incoming distribution is prescribed independently of the outgoing distribution:
\begin{equation}
g_{\mathrm r,ib,k}(\bm f_{ib}^{+})=g_{\mathrm{rd},ib,k},
\qquad k\in M(ib,-),
\label{eq:g-dirichlet}
\end{equation}
where $g_{\mathrm{rd},ib,k}$ is typically a Maxwellian evaluated from prescribed exterior conditions.

At a showerhead inlet (Section~\ref{sec:cvd-allocation}), a prescribed net mass influx $j_{ib}>0$ is added to the diffuse normalization:
\begin{equation}
g_{\mathrm r,ib,k}=\bar g_{\mathrm{rw},ib,k}
\left(\sum_{\hat k\in M(ib,+)}\vv_{\hat k}\cdot\vn_{ib}f_{ib,\hat k}\dXi_{\hat k}+j_{ib}\right),
\qquad k\in M(ib,-).
\label{eq:diffuse-net-inlet}
\end{equation}
The resulting net boundary mass flux is $-j_{ib}$ under the outward-normal convention. For $j_{ib}=0$, this condition reduces to the diffuse-reflection condition in Eq.~\eqref{eq:g-diffuse}.

The boundary contribution to the discrete mesoscopic equation is
\begin{equation}
r_{\mathrm B,i,k}=\sum_{b\in B(i)}A_{ib}\vv_k\cdot\vn_{ib}f_{ib,k}
=r_{\mathrm B,i,k}^{\mathrm{inc}}+r_{\mathrm B,i,k}^{\mathrm{ref}},
\label{eq:rB}
\end{equation}
with the outgoing and incoming parts written as
\begin{equation}
\begin{aligned}
r_{\mathrm B,i,k}^{\mathrm{inc}}
&=\sum_{b\in B(i)}A_{ib}\vv_k\cdot\vn_{ib}\Heavi(\vv_k\cdot\vn_{ib})\CC_{ib}[f_k],\\
r_{\mathrm B,i,k}^{\mathrm{ref}}
&=\sum_{b\in B(i)}A_{ib}\vv_k\cdot\vn_{ib}\Heavi(-\vv_k\cdot\vn_{ib})g_{\mathrm r,ib,k}(\bm f_{ib}^{+}),
\end{aligned}
\label{eq:rB-split}
\end{equation}
where $\Heavi$ is the Heaviside function. With the outgoing part updated on the left and the incoming part frozen on the right, the full mesoscopic update becomes
\begin{equation}
\DD_{i,k}[f]+\frac{V_i}{\tau_i}f_{i,k}+r_{\mathrm B,i,k}^{\mathrm{inc}}
=\frac{V_i}{\tau_i}g_k(\bU_i)-r_{\mathrm B,i,k}^{\mathrm{ref}}.
\label{eq:gsis1-b}
\end{equation}
Since the incoming normal velocity is negative, moving its flux to the right-hand side gives the positive incoming contribution.

For the macroscopic update, the boundary flux is treated in a similar way as the interior flux~\cite{zhang2026efficient,zeng2026gsisdg}: it separates into the Maxwellian part evaluated from the macroscopic state and a boundary higher-order correction supplied by the kinetic solution. The Maxwellian part is assembled from the boundary Maxwellian flux
\begin{equation}
\bF_{\BM}(\bW,\vn)
=\int_{\vv\cdot\vn>0}\vv\cdot\vn\bpsi g_{\mathrm M}(\bW)\,\dd\Xi
+\int_{\vv\cdot\vn<0}\vv\cdot\vn\bpsi g_{\rM}(\bW,\vn)\,\dd\Xi,
\label{eq:FBM}
\end{equation}
whose outgoing and incoming parts follow the same diffuse-reflection, prescribed-net-inflow, or Dirichlet rule as above, with $g_{\rM}$  the incoming boundary distribution obtained when the outgoing distribution is Maxwellian. The boundary contribution is the Maxwellian part plus the interface higher-order correction,
\begin{align}
\bR_{\mathrm B,i}
&=\RR_{\BM,i}[\bU]+\Big[\sum_k\bpsi_k r_{\mathrm B,i,k}\dXi_k-\RR_{\BM,i}[\bW]\Big],\label{eq:RB}\\
\RR_{\BM,i}[\bW]
&=\sum_{b\in B(i)}A_{ib}\bF_{\BM}\big(\CC_{ib}[\bW],\vn_{ib}\big).\label{eq:RBM}
\end{align} 
The bracket is the departure of the kinetic boundary flux from its Maxwellian part, evaluated from the latest kinetic distribution and frozen during the macroscopic update.
The definition of $\RR_{\BM,i}[\bU]$ is identical to Eq.~\eqref{eq:RBM}, with $\bW$ replaced by $\bU$. The full macroscopic update is therefore
\begin{equation}
\RR_{\NS,i}[\bU]+\RR_{\BM,i}[\bU]
=-\RR_{H,i}[\bH]+\RR_{\BM,i}[\bW]-\sum_k\bpsi_k r_{\mathrm B,i,k}\dXi_k.
\label{eq:gsis2-b}
\end{equation}
Only the predicted macroscopic state on the left is updated in the substep; the kinetic correction, direct moments, and kinetic boundary flux on the right are frozen. Equations~\eqref{eq:gsis1-b} and~\eqref{eq:gsis2-b}, together with the boundary relation~\eqref{eq:fib}, provide the complete primal system from which the discrete adjoint will be constructed.

\section{Discrete adjoint method for GSIS}\label{sec:method}
Rather than deriving a continuous adjoint equation and constructing its acceleration separately, we differentiate the coupled discrete system introduced in Section~\ref{sec:gsis}. The key step is to retain both the kinetic and macroscopic equations in the fixed-point constraints. Their stationarity conditions then determine the two levels of the adjoint system, including the macroscopic acceleration.

\subsection{Residual formulation and discrete Lagrangian} \label{sec:discrete-lagrangian}
Recombining the updated and frozen terms in Eqs.~\eqref{eq:gsis1-b} and~\eqref{eq:gsis2-b} gives the full residuals
\begin{align}
r_{i,k}&=\DD_{i,k}[f]-V_i\frac{g_k(\bU_i)-f_{i,k}}{\tau_i}+r_{\mathrm B,i,k},\label{eq:rmeso-b}\\
\bR_i&=\RR_{\NS,i}[\bU]+\RR_{H,i}[\bH]+\bR_{\mathrm B,i}.\label{eq:Rmacro-b}
\end{align}
The boundary relation~\eqref{eq:fib} is equivalently expressed as an auxiliary constraint,
\begin{equation}
r_{ib,k}=\begin{cases}
f_{ib,k}-\CC_{ib}[f_k],&\vv_k\cdot\vn_{ib}>0,\\
f_{ib,k}-g_{\mathrm r,ib,k}(\bm f_{ib}^{+}),&\vv_k\cdot\vn_{ib}<0.
\end{cases}
\label{eq:rib}
\end{equation}
At convergence, the frozen and updated quantities agree, and the GSIS solution satisfies
\begin{equation}
r_{i,k}=0,\qquad \bR_i=\bm0,\qquad r_{ib,k}=0.
\label{eq:gsis-fixed-point}
\end{equation}
These fixed-point equations, rather than the sequence of individual inner iterations, are the constraints used in the adjoint analysis, which serve as the direct starting point of the present discrete adjoint construction.

For differentiation, $f_{i,k}$, $\bw_i$, and $f_{ib,k}$ are treated as independent variables, whereas $\bW_i$ and $\bU_i$ follow Eqs.~\eqref{eq:Wdef} and~\eqref{eq:Udef}. Introducing $f_{ib,k}$ independently only makes the boundary relation explicit; the constraints~\eqref{eq:rib} ensure that no additional physical degrees of freedom are introduced. Let $J$ denote an optimization objective, whose specific form will be introduced in Section~\ref{sec:boundary-adjoint}. With mesoscopic, macroscopic, and boundary multipliers $\phi_{i,k}$, $\bwh_i$, and $\varphi_{ib,k}$, respectively, the discrete Lagrangian is
\begin{equation}
L=J+\sum_i\sum_k\phi_{i,k}r_{i,k}\dXi_k
+\sum_i\bwh_i\cdot\bR_i
+\sum_{ib}\sum_k\varphi_{ib,k}r_{ib,k}A_{ib}\dXi_k.
\label{eq:lagrangian-full}
\end{equation}
The factors $A_{ib}\dXi_k$ specify the scaling of the boundary multipliers. The adjoint equations follow from stationarity with respect to the three sets of independent state variables:
\begin{equation}
\frac{\partial L}{\partial f_{i,k}}=0,\qquad
\frac{\partial L}{\partial\bw_i}=\bm0,\qquad
\frac{\partial L}{\partial f_{ib,k}}=0.
\label{eq:stationarity}
\end{equation}
The three conditions in Eq.~\eqref{eq:stationarity} yield the mesoscopic, macroscopic, and boundary adjoint equations, respectively. On a flow satisfying Eq.~\eqref{eq:gsis-fixed-point}, $L=J$; imposing these conditions then reduces the design derivative to explicit derivatives of the Lagrangian.

\subsection{Interior discrete adjoint equations for GSIS}\label{sec:interior-adjoint}
We first set aside the contribution of $J$ and the explicit boundary terms to isolate the interior adjoint operators. The objective and boundary coupling will be addressed in Section~\ref{sec:boundary-adjoint}. 
Considering the stationarity Eq.~\eqref{eq:stationarity}, differentiation of the residuals in the discrete Lagrangian requires the chain rule.
For example, the dependence of the predicted macroscopic state on the distribution is
\begin{equation}
\frac{\partial\bU_i}{\partial f_{\ih,\kh}}
=\frac{\partial\bW_i}{\partial f_{\ih,\kh}}
=\frac{\partial}{\partial f_{\ih,\kh}}\left(\sum_k\bpsi_k f_{i,k}\dXi_k\right)
=\delta_{i\ih}\bpsi_{\kh}\dXi_{\kh},
\label{eq:dUdf}
\end{equation}
with the independent correction $\bw$ held fixed. Here and below, hatted indices identify the variable with respect to which differentiation is performed, and $\delta$ denotes the Kronecker delta. The detailed chain-rule calculations are given in \ref{app:adjoint-derivations}.

To write the coupled equations compactly, define
\begin{equation}
\bWh_i=\sum_k\phi_{i,k}\frac{\partial g_k(\bU_i)}{\partial\bU_i}\dXi_k,
\qquad
\bCh_i=\left(\sum_k\phi_{i,k}\frac{g_k(\bU_i)-f_{i,k}}{\tau_i^2}\dXi_k\right)
\frac{\partial\tau_i}{\partial\bU_i}.
\label{eq:CWhat}
\end{equation}
These weighted velocity moments account for the derivatives of the collision target and relaxation time. In particular, $\bWh_i$ is not the macroscopic adjoint variable $\bwh_i$. The nonlocal derivative of the cell-center NS flux is collected in the matrix
\begin{equation}
\bK_{i\ih}=\sum_j
\frac{\partial\RR_{H,i}[\bH]}{\partial\vec{\bH}_j}
:\frac{\partial\vec{\bF}_{\NS,j}[\bW]}{\partial\bW_{\ih}}.
\label{eq:Kdef}
\end{equation}
Its rows and columns correspond to the components of the macroscopic residual and conservative state. The colon denotes contraction over the appropriate spatial and conservative-variable indices of the flux or gradient tensor, leaving all other indices free.
Note that the transposed sums will extend over the full dependency stencil, not only direct face neighbors.

With the above definitions, expanding the stationarity conditions~\eqref{eq:stationarity} gives the interior mesoscopic adjoint equation,
  \begin{equation}
\begin{aligned}
\sum_i\phi_{i,\kh}\frac{\partial\DD_{i,\kh}[f]}{\partial f_{\ih,\kh}}
+\frac{V_{\ih}}{\tau_{\ih}}\phi_{\ih,\kh}=&\frac{V_{\ih}}{\tau_{\ih}}\bWh_{\ih}\cdot\bpsi_{\kh}
-V_{\ih}\bCh_{\ih}\cdot\bpsi_{\kh}
-\sum_i\bwh_i\cdot\frac{\partial\RR_{\NS,i}[\bU]}{\partial\bU_{\ih}}\cdot\bpsi_{\kh}\\
&-\sum_i\bwh_i\cdot\left[
\frac{\partial\RR_{H,i}[\bH]}{\partial\vec{\bH}_{\ih}}:(\vv_{\kh}\bpsi_{\kh})
-\bK_{i\ih}\cdot\bpsi_{\kh}\right],
\end{aligned}
\label{eq:adj-meso}
\end{equation} 
and the interior macroscopic adjoint equation,
\begin{equation}
\sum_i\bwh_i\cdot\frac{\partial\RR_{\NS,i}[\bU]}{\partial\bU_{\ih}}
=\frac{V_{\ih}}{\tau_{\ih}}\bWh_{\ih}-V_{\ih}\bCh_{\ih}.
\label{eq:adj-macro}
\end{equation}
In these adjoint equations we simply arrange the meso-macro-uncoupled terms on the left-hand sides and all interlevel coupling terms on the right-hand sides.
Note that the higher-order correction depends on the kinetic moments but not on the independent $\bw$, so it produces no corresponding term in Eq.~\eqref{eq:adj-macro}.

The left-hand sides of Eqs.~\eqref{eq:adj-meso} and~\eqref{eq:adj-macro} are precisely the discrete adjoint counterparts of the left-hand sides of the primal GSIS updates~\eqref{eq:gsis1} and~\eqref{eq:gsis2}, respectively. They therefore admit {\bf the same alternating solution strategy}: update the adjoint variable on the left while freezing the coupling terms on the right during each substep. The two left-hand-side problems are the discrete adjoint of the kinetic solve used in CIS and the discrete adjoint of the NS equations, respectively, both of which can be treated by standard iterative techniques. Their inner solvers are described in \ref{app:adjoint-inner-solvers}. Thus the macroscopic acceleration equation is already contained in the adjoint system derived from the stationarity conditions~\eqref{eq:stationarity}, no additional construction through asymptotic analysis is required.

\subsection{Boundary adjoint equations and objective coupling}\label{sec:boundary-adjoint}
We now specify the objective and show how its contribution enters the adjoint system. As a representative example, consider a linear wall-flux objective,
\begin{equation}
J=\sum_{ib}\sum_k m_{J,ib,k}(\vv_k\cdot\vn_{ib})f_{ib,k}A_{ib}\dXi_k.
\label{eq:Jdef}
\end{equation}
The face dependence of the weight is written explicitly: $m_{J,ib,k}$ is zero outside the wall surface on which the objective is evaluated. On that surface, representative choices are $m_{J,ib,k}=v_{x,k}$ for the force in the $x$ direction and $m_{J,ib,k}=|\vv_k|^2/2$ for heat transfer.
With the domain-outward normal used here, these quantities are the force exerted by the gas on the wall and the net energy transferred from the gas to the stationary wall.

The objective~\eqref{eq:Jdef} depends on the state only through the independent boundary distributions and therefore supplies no direct objective source to the interior equations in Section~\ref{sec:interior-adjoint}. Instead, the boundary constraints transfer the objective information to the interior adjoint through the reconstruction and reflection relations. From Eqs.~\eqref{eq:g-diffuse} and~\eqref{eq:g-dirichlet}, the reflection-kernel derivative, for $k\in M(ib,-)$ and $\hat k\in M(ib,+)$, is
\begin{equation}
\frac{\partial g_{\mathrm r,ib,k}}{\partial f_{ib,\hat k}}
=\begin{cases}
\bar g_{\mathrm{rw},ib,k}\vv_{\hat k}\cdot\vn_{ib}\dXi_{\hat k},&\text{diffuse reflection},\\
0,&\text{Dirichlet boundary}.
\end{cases}
\label{eq:reflection-derivative}
\end{equation}
For the prescribed-net-inflow boundary in Eq.~\eqref{eq:diffuse-net-inlet}, $j_{ib}$ is held fixed when differentiating with respect to the outgoing distribution. The resulting derivative is therefore given by the diffuse-reflection branch of Eq.~\eqref{eq:reflection-derivative}.
Meanwhile, the objective derivative is
\begin{equation}
\frac{\partial J}{\partial f_{ib,k}}
=m_{J,ib,k}\vv_k\cdot\vn_{ib}A_{ib}\dXi_k.
\label{eq:dJ}
\end{equation}
Substituting these derivatives into Eq.~\eqref{eq:stationarity} gives the incoming and outgoing boundary adjoints:
\begin{align}
\varphi_{ib,k}
&=-\vv_k\cdot\vn_{ib}\big(\phi_{i,k}+\bwh_i\cdot\bpsi_k+m_{J,ib,k}\big),
&&k\in M(ib,-),\label{eq:phi-final-neg}\\
\varphi_{ib,k}
&=-\vv_k\cdot\vn_{ib}\big(\phi_{i,k}+\bwh_i\cdot\bpsi_k+m_{J,ib,k}\big)
+\frac{1}{\dXi_k}\hat B_{ib,k},
&&k\in M(ib,+),\label{eq:phi-final-pos}
\end{align}
where
\begin{equation}
\hat B_{ib,k}
=\sum_{\hat k\in M(ib,-)}\varphi_{ib,\hat k}\dXi_{\hat k}
\frac{\partial g_{\mathrm r,ib,\hat k}}{\partial f_{ib,k}}.
\label{eq:Bhat}
\end{equation}
The term $\hat B$ describes redistribution through the reflection kernel and vanishes for a Dirichlet boundary. The weighted combination of the mesoscopic and macroscopic adjoint variables in Eqs.~\eqref{eq:phi-final-neg} and~\eqref{eq:phi-final-pos} results from the appearance of the kinetic boundary flux in both primal residuals.

All objectives considered in this work depend on the flow through boundary distributions. For more complex differentiable objectives constructed from multiple boundary moments, their boundary derivatives can be obtained using the chain rule. An objective with explicit interior-state dependence would additionally contribute its known state derivatives to the corresponding interior adjoint equations. These extensions are straightforward and are therefore not discussed further here.

To obtain the complete mesoscopic equation from Eq.~\eqref{eq:stationarity}, substitute Eq.~\eqref{eq:phi-final-pos} into the boundary reconstruction contribution to $\partial L/\partial f_{\ih,\kh}$. The term directly proportional to $\phi$ is placed on the left, and the other boundary terms on the right. Including the derivatives of the macroscopic boundary flux gives
\begin{equation}
\begin{aligned}
&\sum_i\phi_{i,\kh}\frac{\partial\DD_{i,\kh}[f]}{\partial f_{\ih,\kh}}
+\frac{V_{\ih}}{\tau_{\ih}}\phi_{\ih,\kh}+\sum_{ib}A_{ib}\Heavi(\vv_{\kh}\cdot\vn_{ib})\vv_{\kh}\cdot\vn_{ib}
\phi_{i,\kh}\frac{\partial\CC_{ib}[f_{\kh}]}{\partial f_{\ih,\kh}}\\
&=\frac{V_{\ih}}{\tau_{\ih}}\bWh_{\ih}\cdot\bpsi_{\kh}
-V_{\ih}\bCh_{\ih}\cdot\bpsi_{\kh}
-\sum_i\bwh_i\cdot\left[
\frac{\partial\RR_{\NS,i}[\bU]}{\partial\bU_{\ih}}
+\frac{\partial\RR_{\BM,i}[\bU]}{\partial\bU_{\ih}}
-\frac{\partial\RR_{\BM,i}[\bW]}{\partial\bW_{\ih}}\right]\cdot\bpsi_{\kh}\\
&\quad-\sum_i\bwh_i\cdot\left[
\frac{\partial\RR_{H,i}[\bH]}{\partial\vec{\bH}_{\ih}}:(\vv_{\kh}\bpsi_{\kh})
-\bK_{i\ih}\cdot\bpsi_{\kh}\right]\\
&\quad+\sum_{ib}A_{ib}\Heavi(\vv_{\kh}\cdot\vn_{ib})
\frac{\partial\CC_{ib}[f_{\kh}]}{\partial f_{\ih,\kh}}
\left[\frac{\hat B_{ib,\kh}}{\dXi_{\kh}}
-\vv_{\kh}\cdot\vn_{ib}\big(\bwh_i\cdot\bpsi_{\kh}+m_{J,ib,\kh}\big)\right].
\end{aligned}
\label{eq:adj-meso-full}
\end{equation} 
Similarly, retaining the boundary terms in Eq.~\eqref{eq:stationarity} gives the complete macroscopic equation,
\begin{equation}
\sum_i\bwh_i\cdot\left[
\frac{\partial\RR_{\NS,i}[\bU]}{\partial\bU_{\ih}}
+\frac{\partial\RR_{\BM,i}[\bU]}{\partial\bU_{\ih}}\right]
=\frac{V_{\ih}}{\tau_{\ih}}\bWh_{\ih}-V_{\ih}\bCh_{\ih}.
\label{eq:adj-macro-b}
\end{equation}
Equations~\eqref{eq:phi-final-neg}--\eqref{eq:Bhat}, \eqref{eq:adj-meso-full}, and~\eqref{eq:adj-macro-b} constitute the full discrete adjoint system. Boundary contributions act over the entire boundary reconstruction stencil, which can include neighboring interior cells. Equations~\eqref{eq:Hcell} and~\eqref{eq:FNS-cell} use only cell states for $\GG_i[\bW]$ and $\vec{\bH}_i$, so their nonlocal state dependence only enters the interior stationarity conditions. The derivation and the cancellation of the velocity weights are detailed in \ref{app:boundary-stationarity}.

\subsection{Coupled iterative solution of the adjoint GSIS}\label{sec:adjoint-iteration}
The adjoint is solved on the converged primal flow. All discrete Jacobians, including $\bK$, are evaluated at that flow and remain fixed during the adjoint iterations; their construction nevertheless retains the full state dependence described above. Mirroring the primal mesoscopic--macroscopic alternation, the coupled adjoint solve proceeds as follows:
\begin{enumerate}
\item Update the mesoscopic adjoint $\phi$ using Eq.~\eqref{eq:adj-meso-full}. Before this inner solve, compute $\bWh$ and $\bCh$ from the latest available $\phi$, and assemble the remaining right-hand side using the latest $\bwh$, $\hat B$, and the known objective derivatives. Hold this right-hand side fixed, while updating the boundary term directly proportional to $\phi$ together with the other left-hand-side terms.
\item Recompute $\bWh$ and $\bCh$ from the newly updated $\phi$, freeze them, and solve Eq.~\eqref{eq:adj-macro-b} for $\bwh$.
\item Using the latest $\phi$ and $\bwh$, first update the incoming boundary adjoint by Eq.~\eqref{eq:phi-final-neg}, then compute $\hat B$ by Eq.~\eqref{eq:Bhat}, and finally update the outgoing boundary adjoint by Eq.~\eqref{eq:phi-final-pos}. This prepares the redistribution source for the next mesoscopic substep without changing the implicit treatment of the direct $\phi$ contribution.
\end{enumerate}
 These steps are repeated until the coupled adjoint residuals satisfy the convergence criterion. Freezing the velocity-integrated coupling quantities avoids solving a fully coupled physical--velocity-space system within a single mesoscopic substep. The inner algorithms for the transposed kinetic and NS problems are given in \ref{app:adjoint-inner-solvers}. The Shakhov extension modifies the source-term evaluation but preserves this alternating structure, as shown in \ref{app:shakhov-extension}.

\section{Design sensitivity}\label{sec:sens}
The preceding section determines the adjoint variables needed to eliminate the implicit response of the converged flow to a design change. We now use these variables to evaluate the design gradient, with particular attention to the dependence introduced by deformation of a body-fitted mesh.

\subsection{Sensitivity evaluation and mesh adjoint}\label{sec:sensitivity-evaluation}
Let $\balpha$ denote the design-variable vector. When the primal constraints~\eqref{eq:gsis-fixed-point} and adjoint stationarity conditions~\eqref{eq:stationarity} are satisfied,
\begin{equation}
\frac{\dd J}{\dd\balpha}=\frac{\partial L}{\partial\balpha}.
\label{eq:sens-general}
\end{equation}
The right-hand side includes the explicit dependence of both $J$ and all discrete residuals on $\balpha$, with the independent flow variables and adjoint multipliers held fixed. Gradients of flow-dependent design constraints are obtained in the same way, using the corresponding scalar functional in place of $J$.

For inlet-flow design on a fixed mesh (i.e.~Section~\ref{sec:cvd-allocation}), the prescribed net mass influx $j_{ib}=j_{ib}(\balpha)$ introduces explicit design dependence through Eq.~\eqref{eq:diffuse-net-inlet}. With the inlet temperature and velocity fixed,
\[
\frac{\partial g_{\mathrm r,ib,k}}{\partial\balpha}
=\frac{\partial g_{\mathrm r,ib,k}}{\partial j_{ib}}
\frac{\partial j_{ib}}{\partial\balpha}
=\bar g_{\mathrm{rw},ib,k}
\frac{\partial j_{ib}}{\partial\balpha},
\qquad k\in M(ib,-).
\]
This derivative enters Eq.~\eqref{eq:sens-general} through the boundary constraints~\eqref{eq:rib}. Any explicit dependence of the macroscopic boundary-flux terms on the same controls is differentiated consistently in the full residual~\eqref{eq:RB}. No mesh-adjoint solve is required for this fixed-mesh design.

For shape design in which the parameters $\balpha$ act through mesh coordinates, let $\vx_s$ denote the nodes whose coordinates are prescribed by the geometry parameterization or held fixed, and let $\vx_h$ denote the remaining nodes, whose coordinates are determined by mesh deformation driven by changes in $\vx_s$. The design sensitivity follows from the dependence of $L$ on the mesh coordinates (i.e.~$\partial L/\partial\vx$) and that of the mesh coordinates on $\balpha$. The flow-state response has already been eliminated by the flow adjoint and must not be included again in $\partial L/\partial\vx$. These partial derivatives contain the geometry dependence of the objective, face areas, face normals, cell volumes, reconstruction weights, and boundary reflection kernels. In particular, fixing the discrete state does not fix the reconstructed gradients as the mesh moves: the geometry dependence of $\GG_i$, and hence of the cell-center NS flux and higher-order correction, must also be differentiated. The velocity nodes and quadrature weights remain fixed. In the present implementation, the required local Jacobians and geometry derivatives are evaluated by forward-mode automatic differentiation using CoDiPack~\cite{sagebaum2019codipack} and propagated to the mesh coordinates through the chain rule.

The treatment of the mesh response depends on how $\vx_h$ is determined. If both $\vx_s$ and $\vx_h$ are explicit functions of $\balpha$, the chain rule gives
\begin{equation}
\frac{\dd J}{\dd\balpha}
=\sum_s\frac{\partial L}{\partial\vx_s}\cdot\frac{\partial\vx_s}{\partial\balpha}
+\sum_h\frac{\partial L}{\partial\vx_h}\cdot\frac{\partial\vx_h}{\partial\balpha}.
\label{eq:sens-explicit}
\end{equation}
This explicit treatment is used for the airfoil in Section~\ref{sec:naca-verification}, where the mesh displacements are obtained by radial-basis-function (RBF) interpolation.

When the remaining mesh coordinates of $\vx_h$ are instead defined implicitly, they satisfy the discrete mesh-deformation constraints
\begin{equation}
\PP_h[\vx]=\vec0.
\label{eq:mesh-constraint}
\end{equation}
The mesh adjoint eliminates the response of the implicitly determined nodes. Introduce the mesh-adjoint variables $\vxi_h$ and the augmented Lagrangian
\begin{equation}
L_\PP=L+\sum_h\vxi_h\cdot\PP_h[\vx].
\label{eq:LP}
\end{equation}
Stationarity with respect to each implicitly determined node gives
\begin{equation}
\frac{\partial L}{\partial\vx_{\hh}}
+\sum_h\vxi_h\cdot\frac{\partial\PP_h[\vx]}{\partial\vx_{\hh}}=\vec0.
\label{eq:mesh-adjoint}
\end{equation}
After solving this transposed mesh problem, the design gradient is
\begin{equation}
\frac{\dd J}{\dd\balpha}
=\frac{\partial L_\PP}{\partial\balpha}
=\sum_s\left[\frac{\partial L}{\partial\vx_s}
+\sum_h\vxi_h\cdot\frac{\partial\PP_h[\vx]}{\partial\vx_s}\right]
\cdot\frac{\partial\vx_s}{\partial\balpha}.
\label{eq:sens-implicit}
\end{equation}
The mesh adjoint transfers the sensitivity associated with the implicitly determined nodes to the prescribed coordinates, whose design derivatives appear in Eq.~\eqref{eq:sens-implicit}. This formulation is used for the re-entry capsule in Section~\ref{sec:orion-optimization}, where the mesh deformation is governed by discrete mesh equations.

\subsection{Overall gradient-evaluation and optimization procedure}\label{sec:adjoint-workflow}
Combining the primal solve, the flow adjoint, and the mesh sensitivity gives the following procedure for each design evaluation:
\begin{enumerate}
\item Solve the primal GSIS described in Section~\ref{sec:primal-gsis} to obtain the converged $(f_{i,k},\bw_i,f_{ib,k})$ and evaluate the objective and constraints.
\item Solve for $(\phi,\bwh,\varphi)$ by the coupled iterations in Section~\ref{sec:adjoint-iteration}.
\item Compute the design sensitivity $\dd J/\dd \balpha $ by the method introduced in Section~\ref{sec:sensitivity-evaluation}. This involves   assembly  of the explicit derivatives of the Lagrangian, including the objective and all residual contributions.
\item Repeat the corresponding adjoint solve and derivative assembly for any other scalar objective or constraint requiring a gradient. Purely geometric constraints do not require a flow-adjoint solve.
\item Supply the objective and constraint values and their gradients to the optimizer, update the design, and proceed to the next design evaluation until the optimization stopping criterion is met.
\end{enumerate}
This sequence separates the primal and adjoint inner solves, their mesoscopic--macroscopic outer iterations, and the outermost design updates. It provides the basis for the sensitivity verification and optimization studies in Section~\ref{sec:result}. All optimization examples use sequential quadratic programming (SQP)~\cite{kraft1988software} as implemented in the NLopt library~\cite{johnson2007nLopt}, with the implementation settings given in Section~\ref{sec:orion-optimization}. Case-specific constraints, variable bounds, and stopping criteria are given in Section~\ref{sec:result}.

\section{Numerical Results}\label{sec:result}

In this section, three numerical examples are presented to assess the accuracy and efficiency of the proposed discrete adjoint method and demonstrate its application to shape and flow-parameter optimization. First, the flow around a NACA0012 airfoil is used to verify the discrete shape sensitivities and compare the convergence of the primal and adjoint solvers. Next, multi-objective shape optimization of a three-dimensional re-entry capsule is performed to reduce the peak wall heat flux and increase the lift-to-drag ratio. Finally, the inlet flow rates distribution in a fixed-geometry showerhead chamber is optimized to improve pressure uniformity over the target wafer region.

Before presenting the individual examples, we briefly describe the initialization procedure and convergence criteria. The initial state of each primal solve is prepared using a preprocessing NS solver (NS-pre, see \ref{app:ns-pre}), followed by ten CIS steps to establish the starting distribution from the resulting macroscopic field. For the chamber case in Section~\ref{sec:cvd-allocation}, 1{,}000 NS-pre steps are used. The primal convergence is measured by the relative change of the primitive variables between successive iterates,
\begin{equation}
\mathcal E_{\mathrm p}=\max_{Q}\frac{\big[\sum_i V_i(Q_i^{n}-Q_i^{n-1})^2\big]^{1/2}}{\big[\sum_i V_i(Q_i^{n-1})^2\big]^{1/2}},
\qquad Q\in\{\rho,\vec u,T\},
\label{eq:primal-convergence}
\end{equation}
where $V_i$ denotes the cell area in two dimensions and the cell volume in three dimensions. The adjoint convergence is measured by the maximum relative change of the macroscopic adjoint and the two collision-weighted moments. For CIS, which involves no macroscopic adjoint, only the latter two are monitored. The primal and adjoint outer tolerances are $10^{-8}$ and $10^{-6}$, with inner tolerances of $10^{-9}$ and $10^{-7}$, respectively. At the outermost design-optimization level, the optimization terminates when the relative change in the objective between successive major iterations falls below a prescribed tolerance or the prescribed iteration limit is reached. These specific values are given in each example.

\newcommand{\nacapanelwidth}{.40\textwidth}
\subsection{Sensitivity verification and convergence test}
\label{sec:naca-verification}

We first use the test case of flow over NACA0012 airfoil to verify the discrete shape sensitivities and assess the convergence of the primal and adjoint GSIS iterations at two Knudsen numbers. The sensitivities are checked against finite differences, and the convergence histories are compared with those of CIS.

\begin{figure}[!htbp]
\centering
\subfloat[Physical mesh near the airfoil]{\includegraphics[width=.40\textwidth]{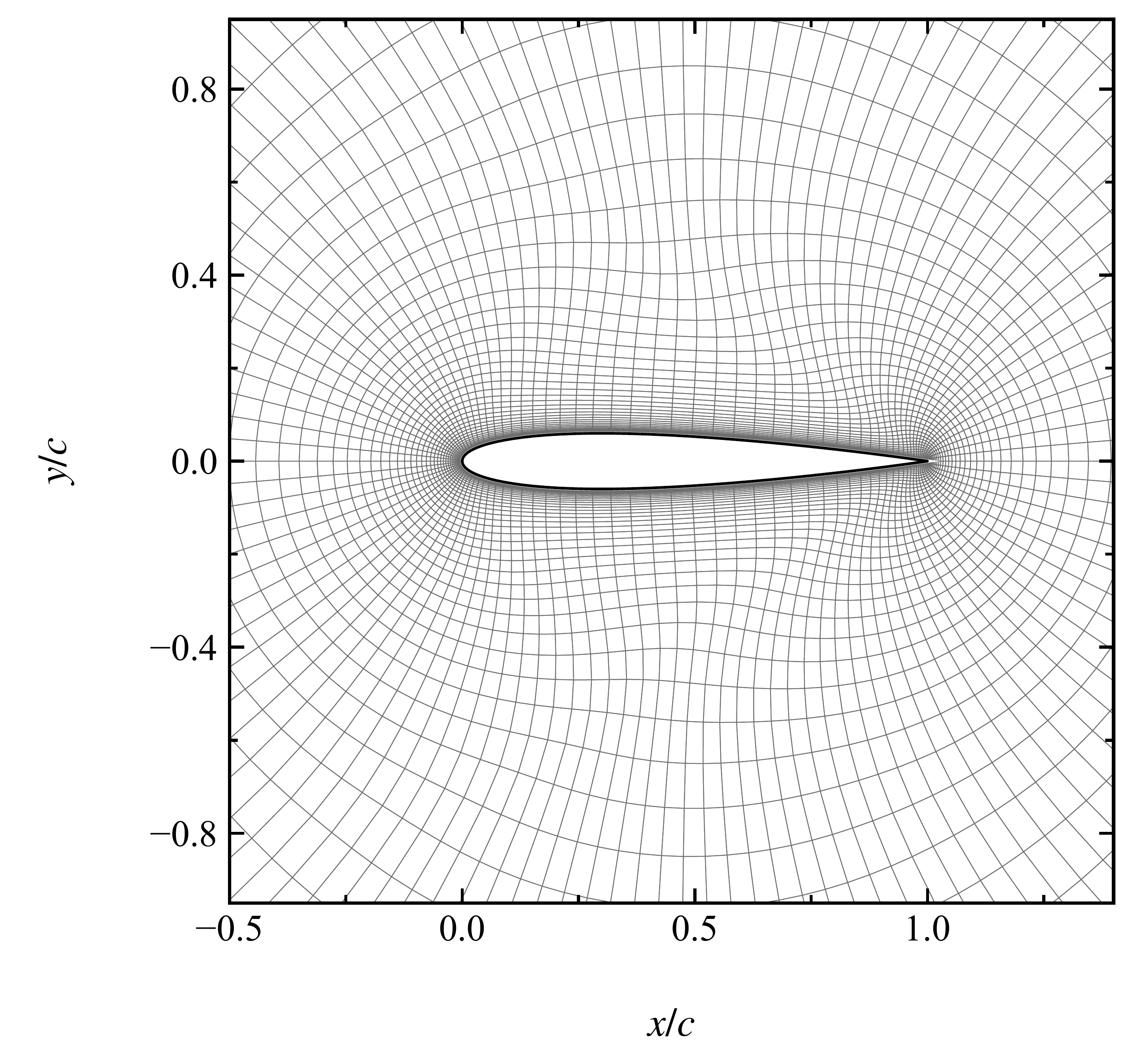}}\hspace{.02\textwidth}%
\subfloat[Quadtree velocity mesh]{\includegraphics[width=.40\textwidth]{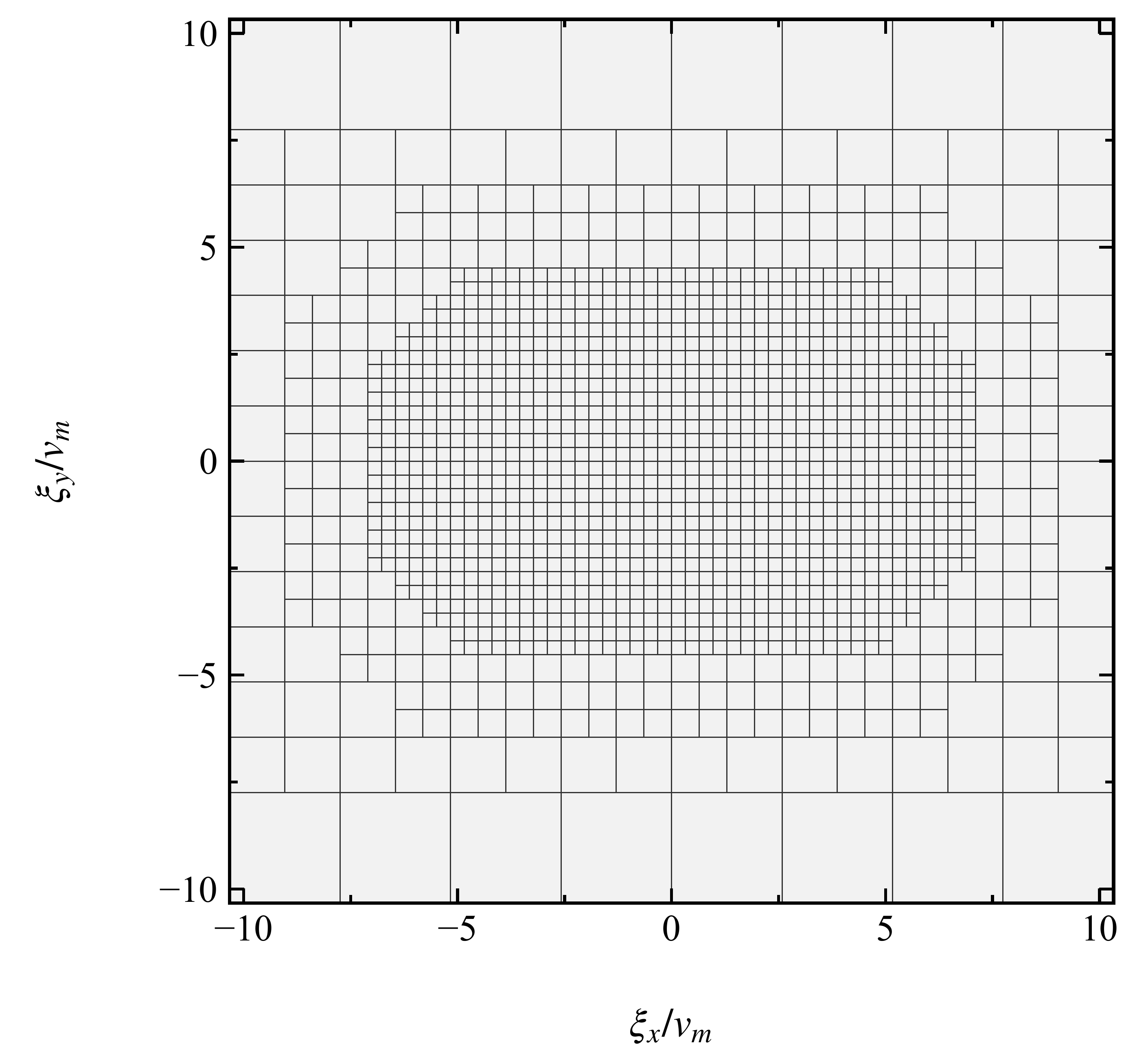}}
\caption{Physical and velocity meshes for NACA0012.}
\label{fig:naca-meshes}
\end{figure}

The NACA0012 test uses $\mathrm{Ma}_\infty=2$ at zero angle of attack, with Knudsen numbers of $0.01$ and $0.5$ based on the chord $c$. The stationary wall is diffusely reflecting at $T_w=T_\infty$, with free-stream conditions at the outer boundary. Both cases use 5,880 physical cells and 1,444 quadtree velocity cells (Fig.~\ref{fig:naca-meshes}), with a velocity domain of approximately $[-10v_m,10v_m]^2$, where $v_m$ is a characteristic velocity defined as $v_m=\sqrt{RT}$ and is evaluated at $T_\infty$. Each surface $s\in\{u,l\}$ is described by a sixth-order class function/shape function transformation (CST) parameterization~\cite{yuan2025adjoint}, with the trailing edge and chord length fixed:
\begin{equation}
y_s(x)=x^{N_1}(1-x)^{N_2}\sum_{j=0}^{6}A_j^{s}\,S_j(x),
\qquad
S_j(x)=\binom{6}{j}x^{j}(1-x)^{6-j},
\label{eq:cst}
\end{equation}
giving sixteen design variables: seven coefficients $A_j^s$ per surface, $j=0,\ldots,6$, and two shared class exponents $N_1,N_2$.

During the shape sensitivity test, the airfoil surface displacements prescribed by the CST parameterization are interpolated to the interior mesh nodes using a Wendland $C^0$ radial basis function~\cite{yuan2025adjoint} with a support radius equal to the chord, while the far-field boundary nodes remain fixed.

\begin{figure}[!htbp]
\centering
\subfloat[Upper-surface sensitivities]{\includegraphics[width=\nacapanelwidth,height=.26\textheight,keepaspectratio]{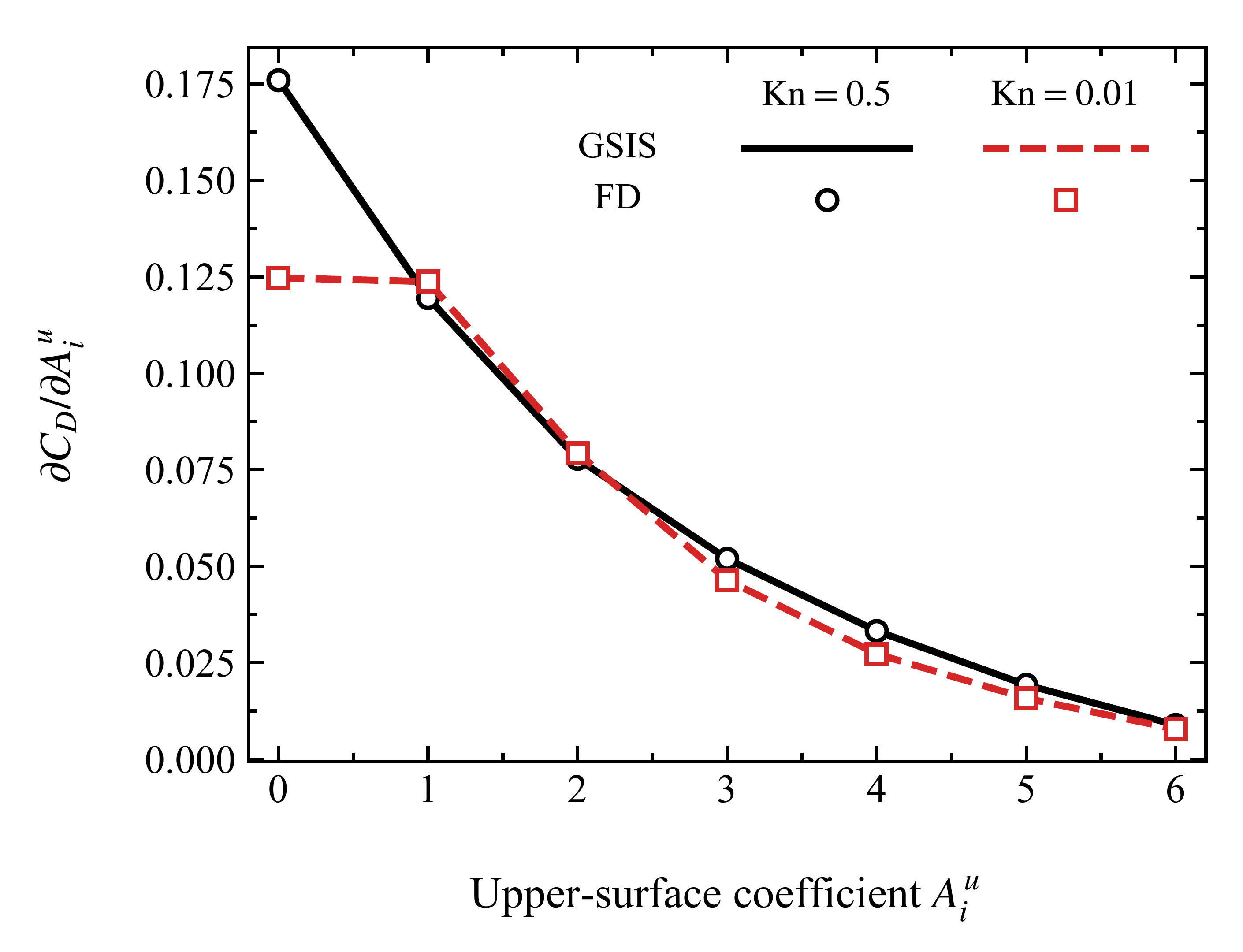}}\hspace{.02\textwidth}%
\subfloat[Step-size comparisons for $A_0^u$]{\includegraphics[width=\nacapanelwidth,height=.26\textheight,keepaspectratio]{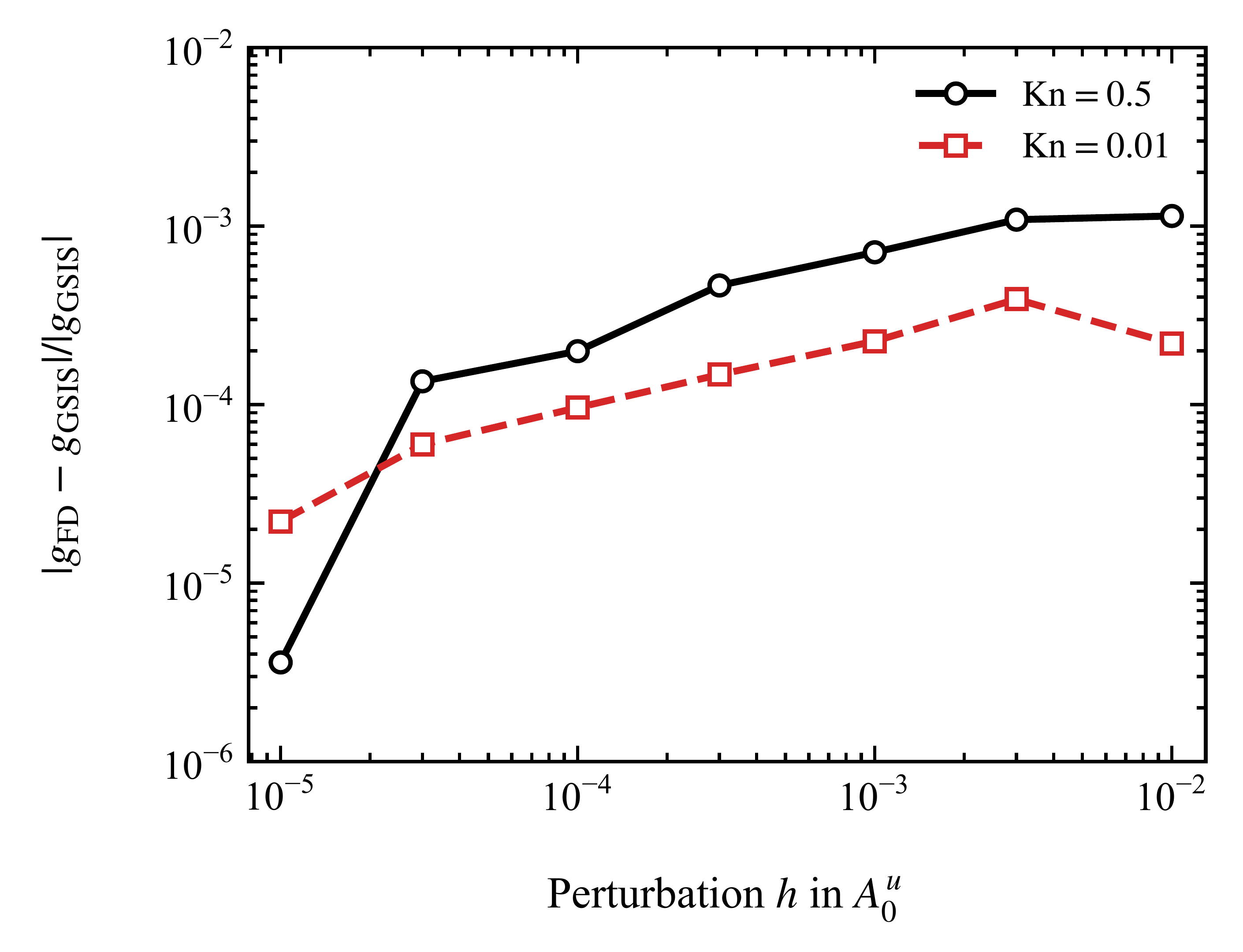}}
\caption{Sensitivity verification: (a) upper-surface CST sensitivities from the adjoint method (lines) and central finite differences (open symbols); (b) relative difference for $A_0^u$ versus perturbation size. Black and red denote $\mathrm{Kn}=0.5$ and $0.01$, respectively.}
\label{fig:naca-sensitivity}
\end{figure}

The drag sensitivities are verified against central finite differences at the same spatial and velocity space resolutions. At difference step $h=10^{-4}$, Fig.~\ref{fig:naca-sensitivity}(a) compares the gradients for the seven upper-surface CST coefficients. At zero angle of attack the configuration is symmetric about the chord, so the lower-surface sensitivities share the same accuracy. The relative $L_2$ differences over these coefficients, normalized by the $L_2$ norm of the adjoint gradient, are $0.03\%$ at $\mathrm{Kn}=0.01$ and $0.02\%$ at $\mathrm{Kn}=0.5$. Figure~\ref{fig:naca-sensitivity}(b) then tests the dependence on the perturbation size for $A_0^u$, the coefficient with the largest sensitivity in Fig.~\ref{fig:naca-sensitivity}(a), over $10^{-5}\le h\le10^{-2}$. At the smallest tested step, the relative differences are $2.2\times10^{-5}$ and $3.6\times10^{-6}$, respectively. Note that, to resolve the small perturbations, the perturbed primal flows use tightened mesoscopic and macroscopic tolerances of $10^{-13}$ and $10^{-14}$. Note that these tolerances are already close to machine precision in double-precision arithmetic, combined with the small finite-difference step of $10^{-5}$, the observed relative sensitivity differences of order $10^{-5}$--$10^{-6}$ provide a stringent verification of the discrete adjoint sensitivities.

\begin{figure}[!htbp]
\centering
\subfloat[Primal convergence]{\includegraphics[width=.40\textwidth,height=.26\textheight,keepaspectratio]{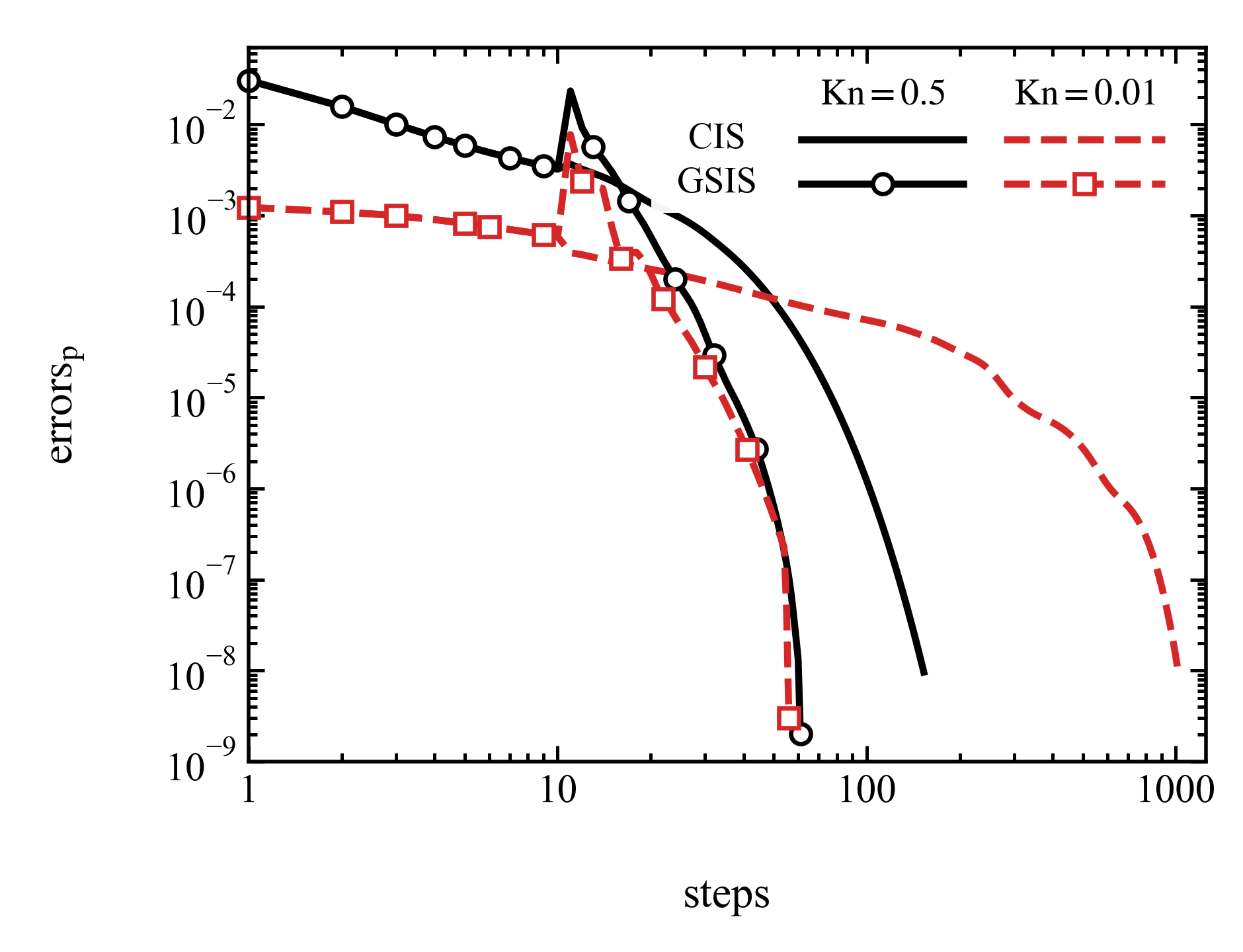}}\hspace{.02\textwidth}%
\subfloat[Adjoint convergence]{\includegraphics[width=.40\textwidth,height=.26\textheight,keepaspectratio]{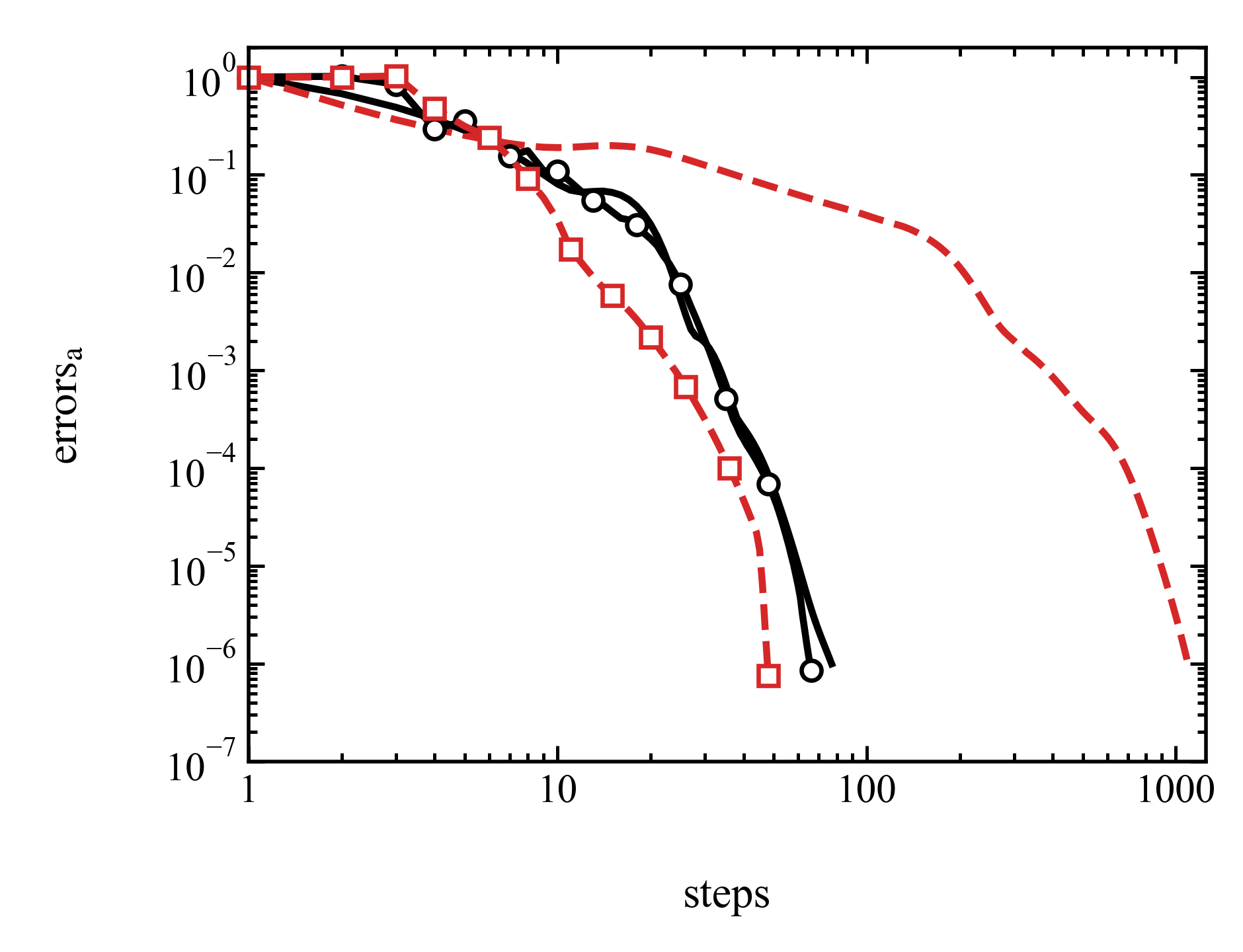}}
\caption{Convergence of (a) the primal and (b) the adjoint iterations. Red dashed and black solid curves correspond to $\mathrm{Kn}=0.01$ and $0.5$, respectively. Symbols indicate GSIS.}
\label{fig:naca-iteration}
\end{figure}

Figure~\ref{fig:naca-iteration} compares the iteration counts of CIS and GSIS, starting from the same initial primal field and zero adjoint fields. At $\mathrm{Kn}=0.01$, GSIS reduces the primal and adjoint counts from 1,017 and 1,096 to 46 and 48, by factors of about $22$ and $23$, respectively. At $\mathrm{Kn}=0.5$, the corresponding counts fall from 153 and 77 to 51 and 66, by factors of about $3.0$ and $1.2$. The primal GSIS counts exclude the ten initial CIS steps.

The above results verify the accuracy of the GSIS discrete adjoint sensitivities and demonstrate that the adjoint solver retains the rapid convergence of the primal GSIS. Both the primal and adjoint solvers converge within tens of iterations in the rarefied and near-continuum gas flows. In the near-continuum case, the adjoint GSIS requires over an order of magnitude fewer iterations than CIS, demonstrating the effectiveness of the macroscopic acceleration in the adjoint solution.

\FloatBarrier

\subsection{Multi-object optimization of a re-entry capsule}\label{sec:orion-optimization}

The second example considers the multi-objective shape optimization of a three-dimensional re-entry capsule in hypersonic flow, aiming to reduce the peak wall heat flux and increase the lift-to-drag ratio under geometric constraints. These two objectives address thermal protection and flight control, respectively: reducing the peak wall heat flux alleviates local thermal loading, whereas increasing the lift-to-drag ratio is motivated by the need for trajectory adjustment and landing-site targeting~\cite{graves1972apollo,rea2010orion}. In this context, lift provides maneuvering capability alongside the deceleration provided by drag.

\begin{figure}[!htbp]
\centering
\subfloat[Geometry and near-body mesh, $z=0$]{\includegraphics[width=.44\textwidth]{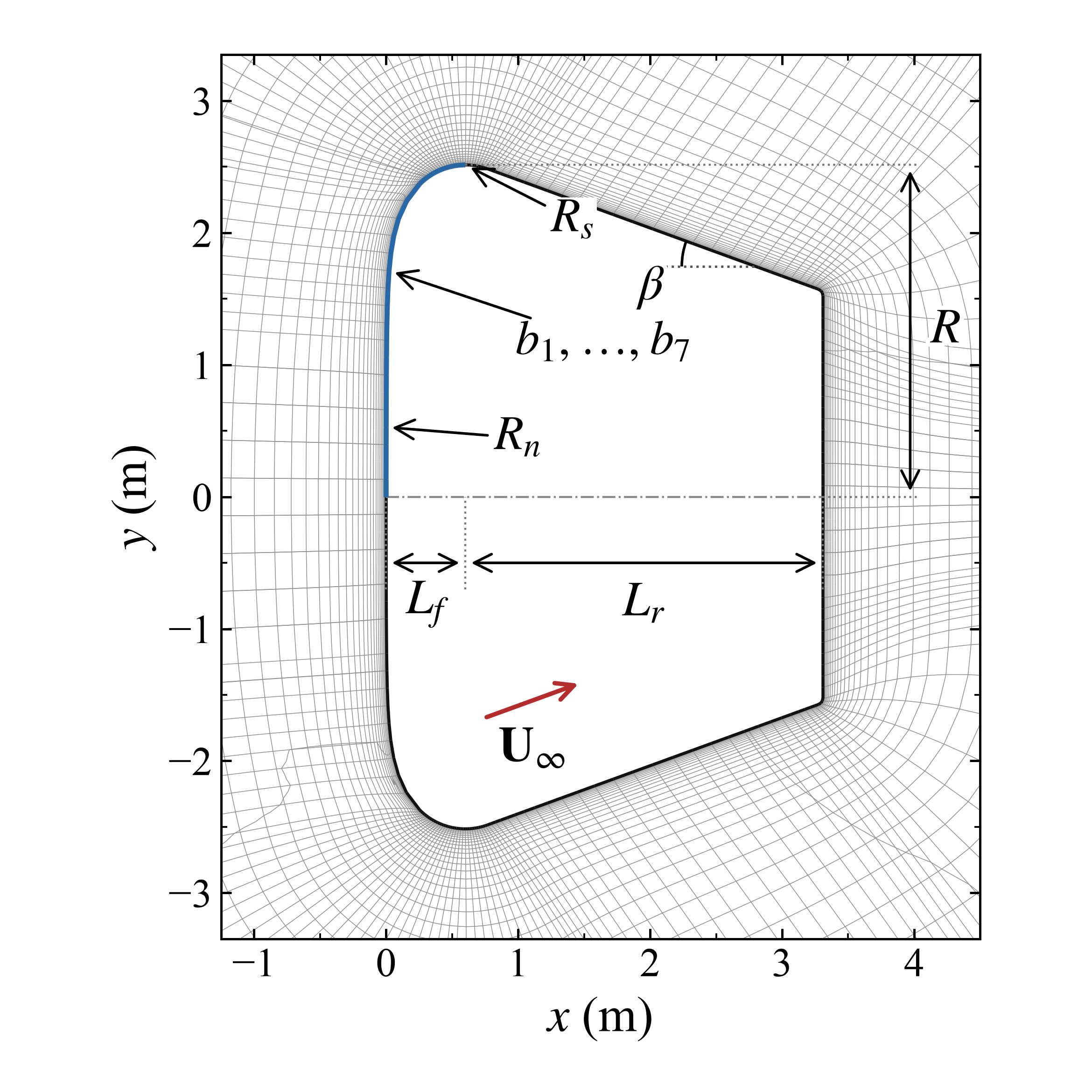}}\hspace{.02\textwidth}%
\subfloat[Exploded surface mesh]{\includegraphics[width=.44\textwidth]{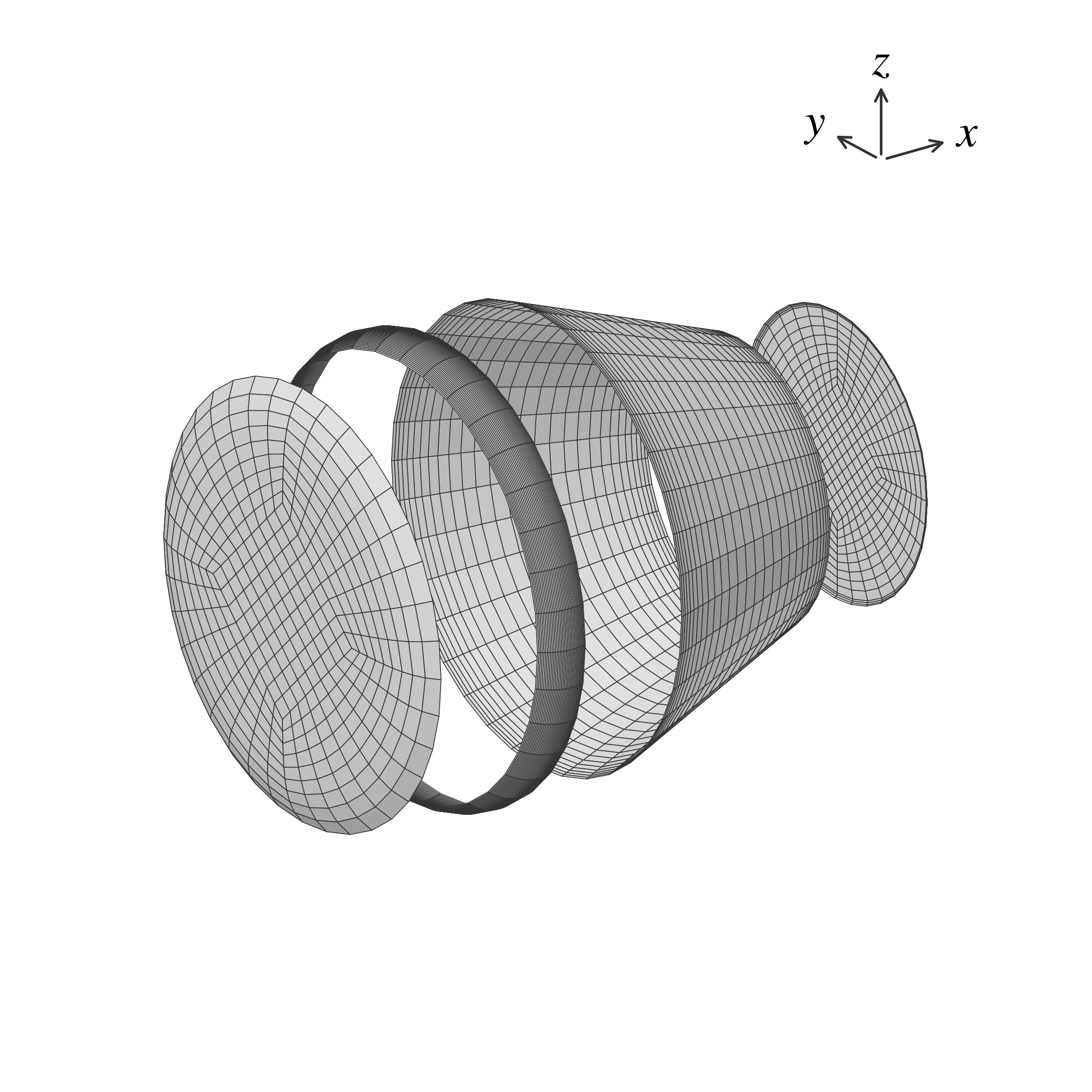}}\\[.2em]
% Center the replacement mesh in a square panel without changing its aspect ratio.
\subfloat[Full physical-domain section]{\parbox[b][.44\textwidth][c]{.44\textwidth}{%
\centering\includegraphics[width=\linewidth]{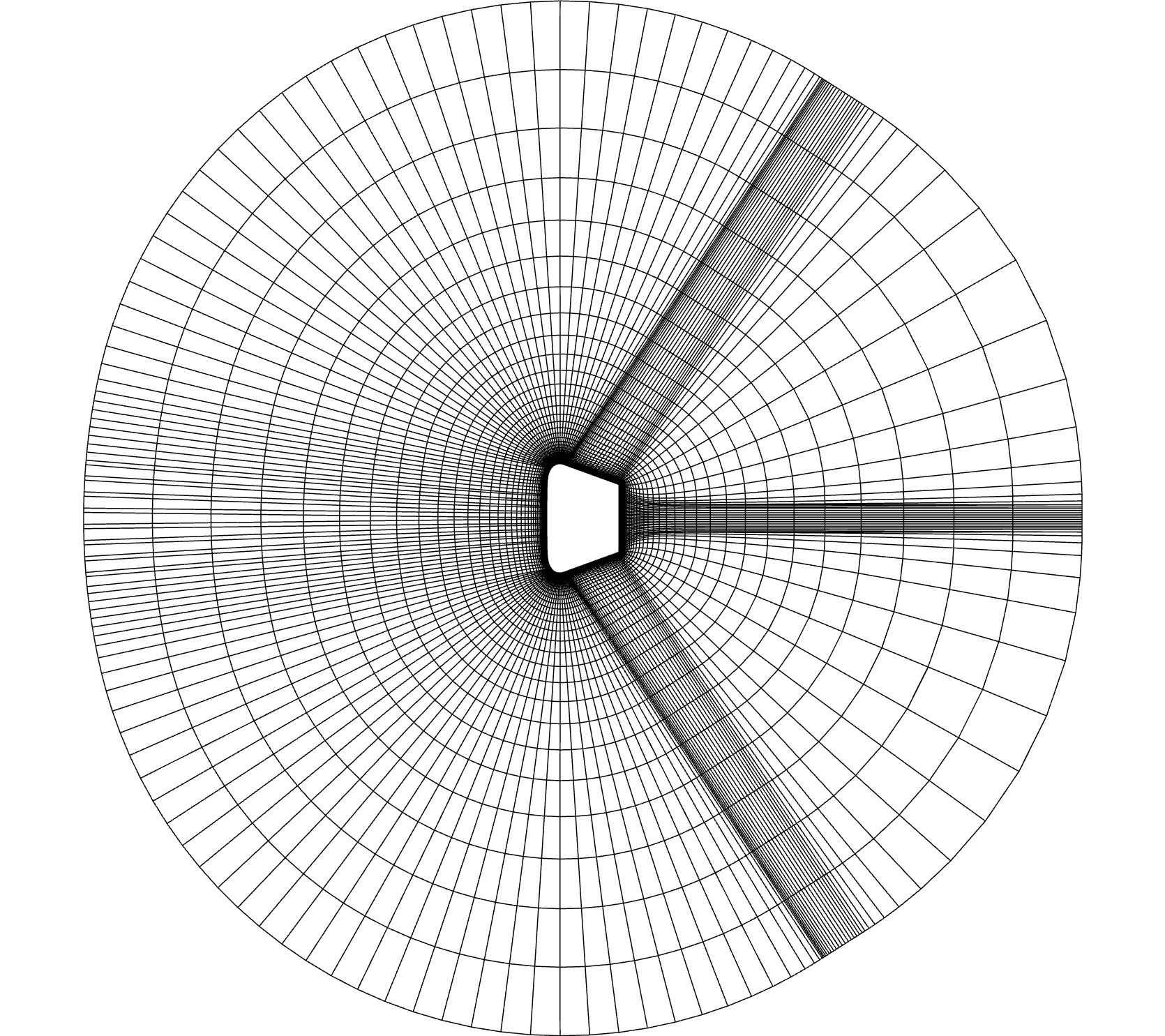}%
}}\hspace{.02\textwidth}%
\subfloat[Octree velocity-mesh section]{\includegraphics[width=.44\textwidth]{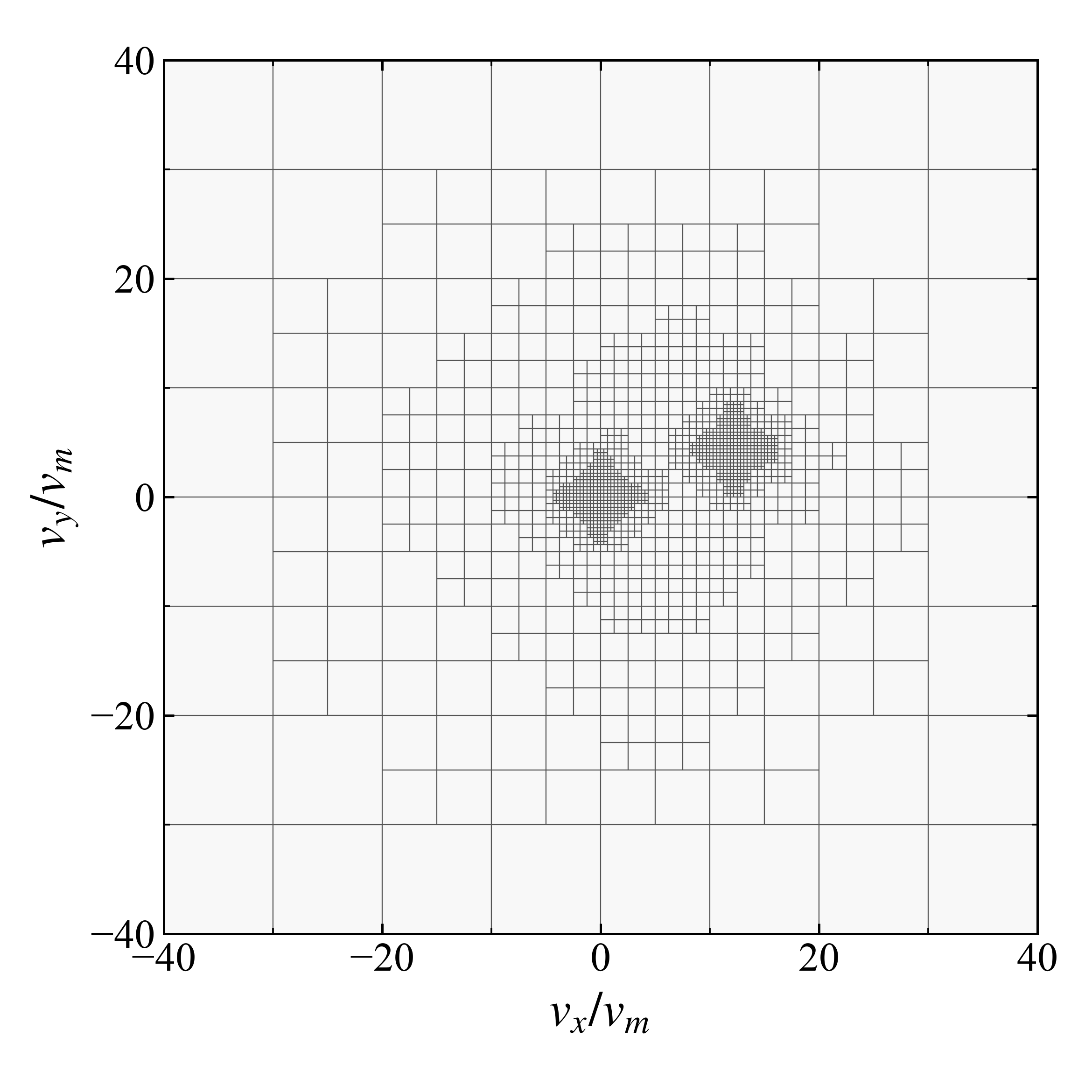}}
\caption{Initial re-entry capsule geometry and meshes: (a) parameters and near-body mesh; (b) surface mesh with the heat shield, shoulder, cone, and rear cap separated axially; (c) full physical-domain section at $z=0$; (d) velocity-mesh section at $v_z=0$.}
\label{fig:orion-setup}
\end{figure}

The capsule in Fig.~\ref{fig:orion-setup} is evaluated at $\mathrm{Ma}_\infty=10$ and an angle of attack of $20^\circ$. Free-stream conditions are imposed at the outer boundary, and the capsule surface is diffusely reflecting at $T_w=T_\infty$. The initial maximum diameter is $D_0\simeq5.03$~m, and the Knudsen number based on this diameter is $\mathrm{Kn}_{D_0}=5\times10^{-3}$. Note that although this global Knudsen number is small, it is based on the maximum capsule diameter and does not characterize local rarefaction throughout the flow field. The strong bow shock introduces steep gradients over length scales much smaller than $D_0$, while the low density in the leeward wake increases the local molecular mean free path. Since wall heat flux is sensitive to local nonequilibrium effects, accurately capturing these effects is particularly important for the present peak-heat-flux optimization. The coexistence of near-continuum regions and locally nonequilibrium flow therefore motivates a multiscale kinetic method that retains both kinetic accuracy and computational efficiency across these regimes.

The geometry of the capsule is parameterized by thirteen design variables, comprising the maximum radius $R$, forebody length $L_f$, nose radius $R_n$, shoulder radius $R_s$, rear length $L_r$, cone angle $\beta$, and seven forebody curvature coefficients $b_j$. Their bounds and optimized values are listed in Table~\ref{tab:orion-optimization}, with $R_0=D_0/2$ the initial radius. The forebody profile is generated from a positive curvature distribution over the local radius coordinate $u=r/Y$,
\begin{equation}
\frac{\dd^2x}{\dd r^2}(u)=\sum_{i=0}^{128}c_i\binom{128}{i}u^i(1-u)^{128-i},\qquad c_i>0,
\label{eq:orion-nose-curvature}
\end{equation}
where $r$ and $x$ are the radial and axial coordinates. The endpoint coefficients $c_0=1/R_n$ and $c_{128}=1/(R_s\sin^3\alpha)$ with $\alpha=45^\circ$ fix the nose-tip curvature and the curvature at the shoulder join $(x,r)=(L_f-R_s\sin\alpha,\,R-R_s+R_s\cos\alpha)$ with the circular shoulder of radius $R_s$. The interior coefficients are modulated by the design variables as $c_i\propto\exp\big[\sum_{j=1}^{7}b_jP_{j+1}(2u_i-1)\big]$ at $u_i=i/128$, with $P_j$ the Legendre polynomial of degree $j$. The forebody profile is obtained by integrating the curvature distribution twice. Two integral constraints ensure that the resulting profile joins the circular shoulder at the prescribed position with a matching tangent.

The discretization uses 127,674 hexahedral cells and 20,000 velocity points. For mesh deformation in physical space, let $\vec d_\ell=\vx_\ell-\vx_\ell^0$ denote the displacement of node $\ell$ from the reference mesh. The prescribed set comprises the wall nodes and first off-wall layer, which follow the geometry map, and the fixed far-field nodes. The remaining displacements satisfy the weighted graph-harmonic equations
\begin{equation}
\PP_h[\vx]=\sum_{j\in N_v(h)}a_{hj}(\vec d_h-\vec d_j)=\vec0,
\qquad a_{hj}=\left(\frac{l_{\mathrm{ref}}}{|\vx_h^0-\vx_j^0|}\right)^2,
\label{eq:mesh-harmonic-implementation}
\end{equation}
where $N_v(h)$ contains the vertices connected to $h$ by an edge and the weights are fixed on the reference mesh. The corresponding design sensitivities are evaluated using the mesh-adjoint formulation in Section~\ref{sec:sensitivity-evaluation}.

The optimization minimizes a smooth approximation of the peak wall heat flux subject to a prescribed gain in lift-to-drag ratio and bounds on the capsule volume. The thermal and aerodynamic objectives are
\begin{equation}
J_q=\frac{1}{s}\log\sum_{f=1}^{N_w}
\exp\!\left(s\frac{q_{w,f}}{q_{\max,0}}\right),
\qquad J_{L/D}=-\frac{L}{D},
\label{eq:orion-metrics}
\end{equation}
where $N_w=3114$ and $q_{w,f}$ is the heat flux on wall face $f$. The smoothing parameter $s=\log N_w/0.01$ ensures that $J_q$ overestimates the normalized peak $q_{\max}/q_{\max,0}$ by at most $1\%$. The signed lift convention gives $J_{L/D}>0$, and $E=J_{L/D}$ is used in the normalized plots. To recast the multi-objective problem as a single-objective constrained optimization problem, we adopt the $\epsilon$-constraint method, retaining the thermal objective and imposing a prescribed gain in lift-to-drag ratio as a constraint. The resulting formulation is
\begin{equation}
\min_{\boldsymbol a}\frac{J_q(\boldsymbol a)}{J_{q,0}},
\qquad \frac{J_{L/D}(\boldsymbol a)}{J_{L/D,0}}\ge1+\epsilon,
\qquad 1\le\frac{V(\boldsymbol a)}{V_0}\le1.05.
\label{eq:orion-problem}
\end{equation}
Here, the subscript 0 denotes the fixed initial geometry, and $\epsilon$ specifies the minimum relative gain in lift-to-drag ratio with respect to this reference. Three target values, $\epsilon=0.1$, $0.2$, and $0.25$, are considered to examine the trade-off between aerodynamic performance and peak wall heating. The corresponding optimized designs are denoted by $A$, $B$, and $C$, respectively.

It is worth noting that, these three designs represent three sampled points on the approximate Pareto front. To sample this front efficiently over the range of interest, $\epsilon$ can be varied progressively, with the optimized design and converged primal and adjoint states at one target value serving as the initial guess for the optimization at the next target value.

In the present example, this continuation strategy is applied in two directions from the intermediate target $\epsilon=0.2$. The optimization at this target starts from the initial geometry and yields design $B$, which then provides the geometry and converged primal and adjoint states for two continuation runs. The target is relaxed to $\epsilon=0.1$ in one run and tightened to $\epsilon=0.25$ in the other, yielding designs $A$ and $C$, respectively.

For each fixed value of $\epsilon$, the constrained optimization problem is solved using the SQP implementation in NLopt (SLSQP), with BFGS Hessian approximations and an $L_1$ merit-function line search. Each fixed-$\epsilon$ optimization stage is limited to 20 major SQP iterations, with tolerances of $10^{-4}$ for the relative objective change, $10^{-5}$ for the absolute change in normalized design variables, and $10^{-9}$ for constraint satisfaction. The primal and adjoint residual tolerances are both $10^{-3}$ throughout the optimization.

\begin{figure}[htbp]
\centering
\subfloat[Thermal-objective history]{\includegraphics[width=.46\textwidth]{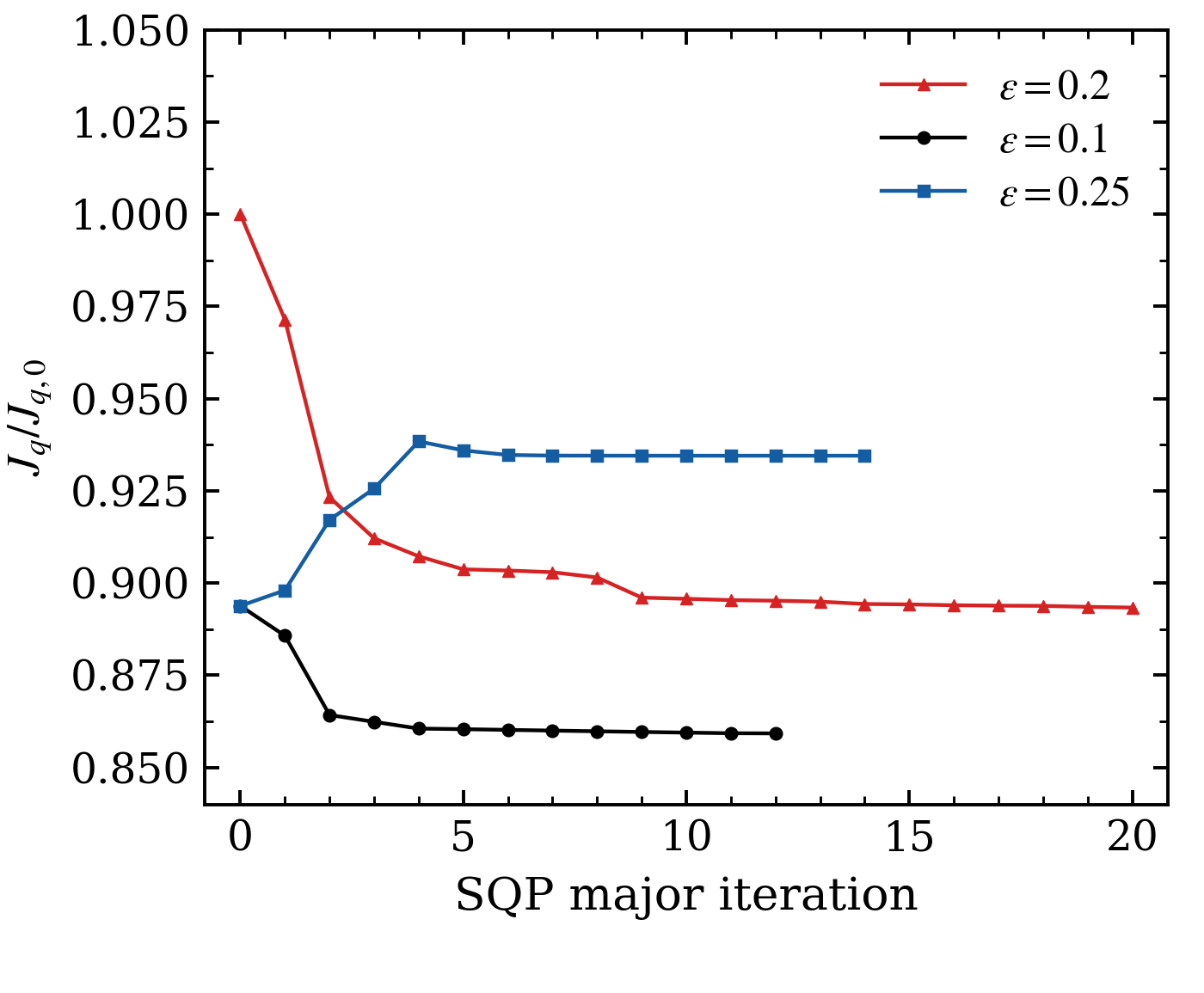}}\hspace{.02\textwidth}%
\subfloat[Objective-space trajectories]{\includegraphics[width=.46\textwidth]{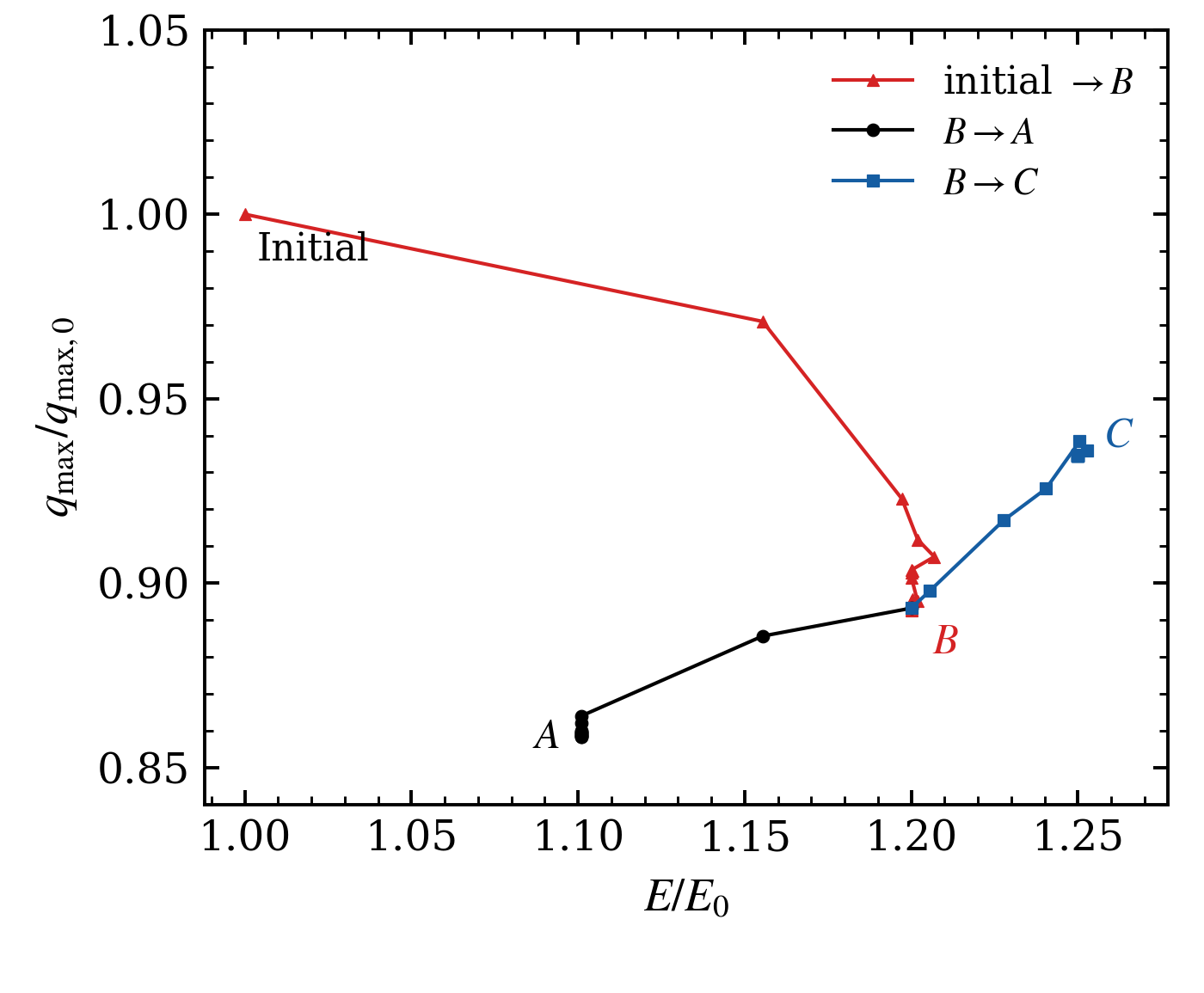}}
\caption{Capsule SQP histories. (a) Normalized thermal objective $J_q/J_{q,0}$ versus major iteration. Iteration 0 is the starting state of each stage. (b) Normalized peak heat flux $q_{\max}/q_{\max,0}$ versus lift-to-drag ratio $E/E_0$ for the accepted major iterates. Labels $A$, $B$, and $C$ denote the optimized designs for the 10\%, 20\%, and 25\% targets, respectively.}
\label{fig:orion-history}
\end{figure}

The optimized designs $A$, $B$, and $C$ reduce the peak wall heat flux by 14.17\%, 10.68\%, and 6.54\%, respectively, reflecting the trade-off between aerodynamic gain and thermal load. Figure~\ref{fig:orion-history} traces the initial optimization toward $B$ and the two continuations from $B$ toward $A$ and $C$. In the objective space, the two continuation trajectories nearly join into a single continuous trade-off curve, indicating that the intermediate iterates remain close to the approximate Pareto front. Continuation in $\epsilon$ therefore samples additional trade-off designs efficiently, without restarting the optimization from the initial geometry. The iteration counts and computing times of the three stages are listed in Table~\ref{tab:orion-optimization}. The initial optimization at $\epsilon=0.2$ uses 20 major iterations, and the two continuations from $B$ toward $A$ and $C$ use 12 and 14, taking 123, 58, and 124~min in total on 512 cores of the CPU model ``\emph{AMD EPYC 9755}'', respectively. Notably, the wall time is not proportional to the iteration count: the $C$ stage takes fewer major iterations than the $B$ stage (14 versus 20) but almost the same wall time. The reason is that the line search within a single major SQP iteration may probe several design points: with the tightest lift-to-drag requirement, the 14 major iterations of the $C$ stage trigger 57 flow evaluations in total, i.e., 43 extra line-search trials.

\begin{table}[htbp]
\centering
\caption{Capsule design variables and optimization results. The upper block lists the bounds, initial values, and optimized values of the thirteen design variables; lengths are in meters and $\beta$ in degrees, with $R_0=D_0/2$ the initial radius. The lower block lists the normalized objective and constraint values and the costs on 512 cores, where $\Delta q_{\max}$ is the peak heat-flux reduction from the initial geometry, $N_{\mathrm{SQP}}$ counts major iterations, and $N_{\mathrm{phys}}$ counts new flow evaluations, including failed trials. The three designs correspond to targets $\epsilon=0.1$, $0.2$, and $0.25$.}
\label{tab:orion-optimization}
{\small
\begin{tabular*}{\textwidth}{l@{\extracolsep{\fill}}rrrrrr}
\hline
 & \multicolumn{2}{c}{Bounds} & & & & \\
\cline{2-3}
Variable & Lower & Upper & Initial & $A$ & $B$ & $C$ \\
\hline
$R$ (m) & $0.85R_0$ & 3.00 & 2.51 & 3.00 & 3.00 & 3.00 \\
$L_f$ (m) & 0.55 & 1.30 & 0.60 & 0.85 & 0.58 & 0.55 \\
$R_n$ (m) & 7.50 & 14.00 & 10.01 & 10.05 & 10.04 & 7.60 \\
$R_s$ (m) & 0.15 & 0.65 & 0.49 & 0.18 & 0.15 & 0.15 \\
$L_r$ (m) & 1.50 & 3.30 & 2.71 & 3.02 & 3.02 & 3.05 \\
$\beta$ (deg) & 15.00 & 45.00 & 20.00 & 42.17 & 40.04 & 38.52 \\
\hline
$E/E_0$ & \multicolumn{2}{c}{$\ge1+\epsilon$} & 1.00 & 1.10 & 1.20 & 1.25 \\
$J_q/J_{q,0}$ & & & 1.00 & 0.86 & 0.89 & 0.93 \\
$\Delta q_{\max}$ (\%) & & & 0.00 & 14.17 & 10.68 & 6.54 \\
$V/V_0$ & 1.00 & 1.05 & 1.00 & 1.03 & 1.00 & 1.00 \\
$N_{\mathrm{SQP}}$ & & & -- & 12 & 20 & 14 \\
$N_{\mathrm{phys}}$ & & & -- & 12 & 20 & 57 \\
Wall time (min) & & & -- & 58 & 123 & 124 \\
\hline
\end{tabular*}

\smallskip
\begin{minipage}{\textwidth}
\footnotesize Optimized $(b_1,\ldots,b_7)$: (-0.50,-0.50,-0.50,-0.10,0.27,0.40,0.34) for $A$, (-0.50,-0.50,-0.50,-0.12,0.25,0.39,0.34) for $B$, and (0.46,0.47,0.47,0.48,0.50,0.50,0.50) for $C$; all coefficients are bounded in $[-0.5,0.5]$ and start from zero.
\end{minipage}}
\end{table}

\begin{figure}[p]
\centering
\captionsetup[subfloat]{captionskip=2pt,nearskip=0pt,farskip=0pt}
\subfloat[$A$, surface]{\includegraphics[width=.31\textwidth]{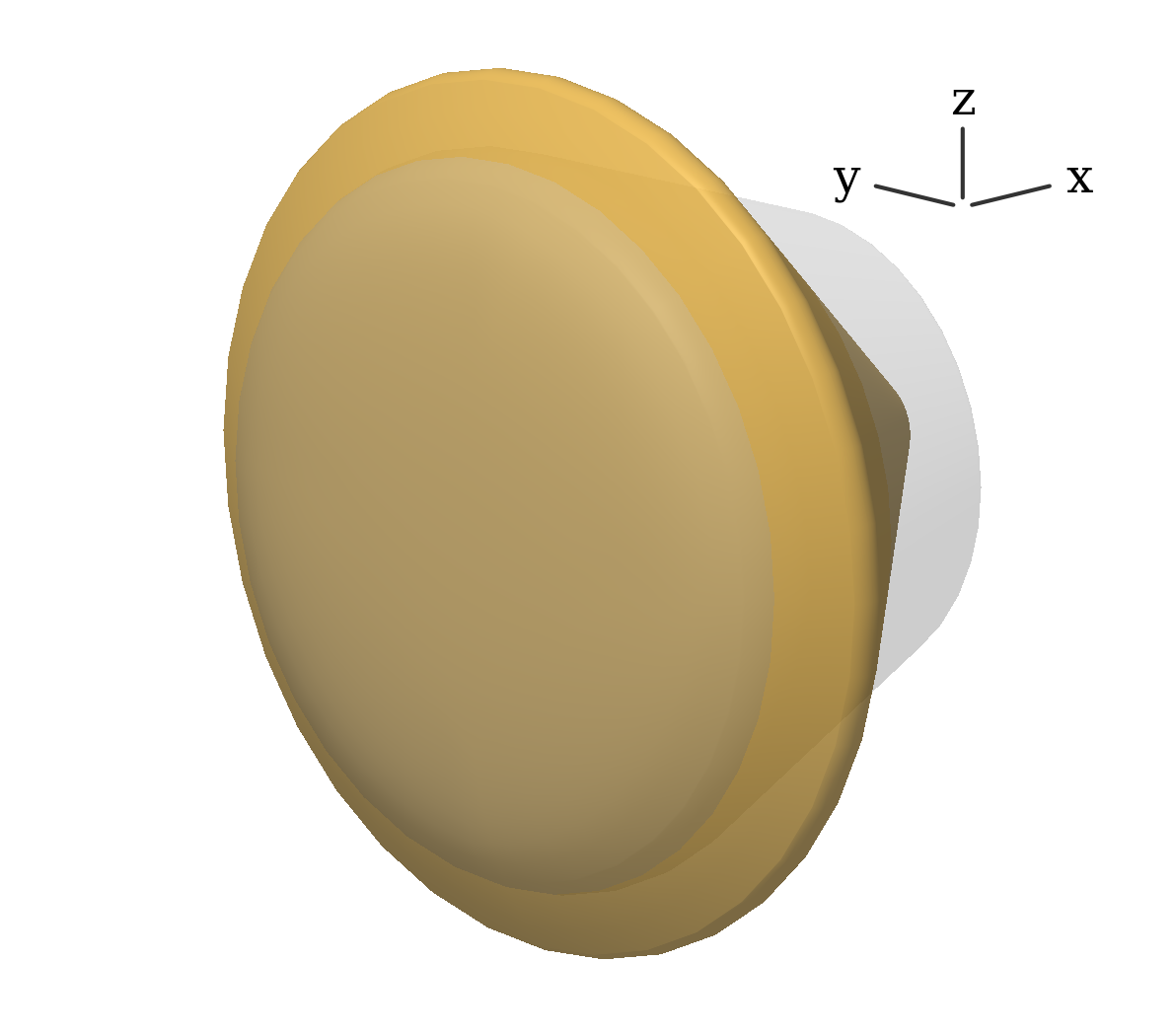}}\hspace{.02\textwidth}%
\subfloat[$B$, surface]{\includegraphics[width=.31\textwidth]{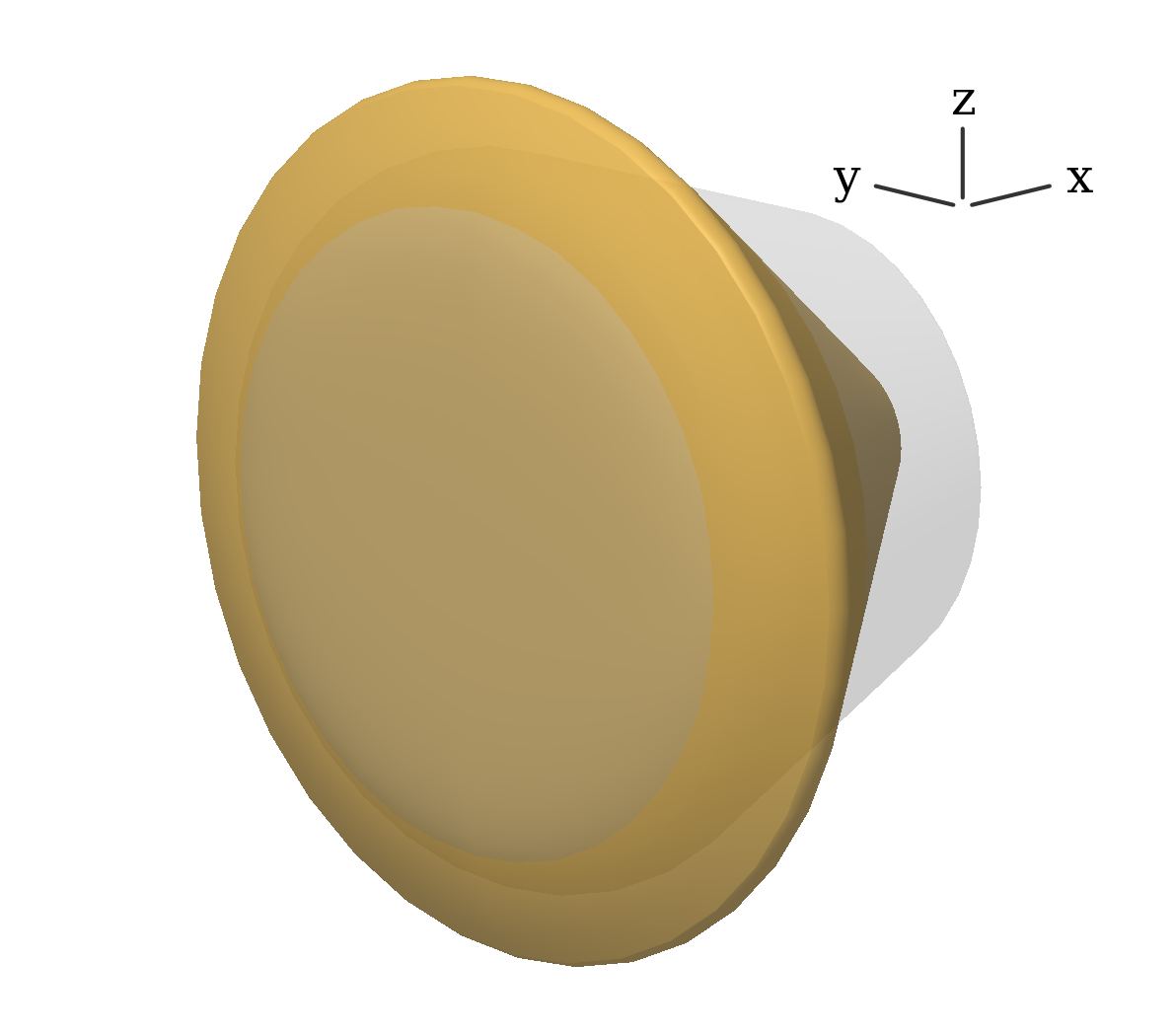}}\hspace{.02\textwidth}%
\subfloat[$C$, surface]{\includegraphics[width=.31\textwidth]{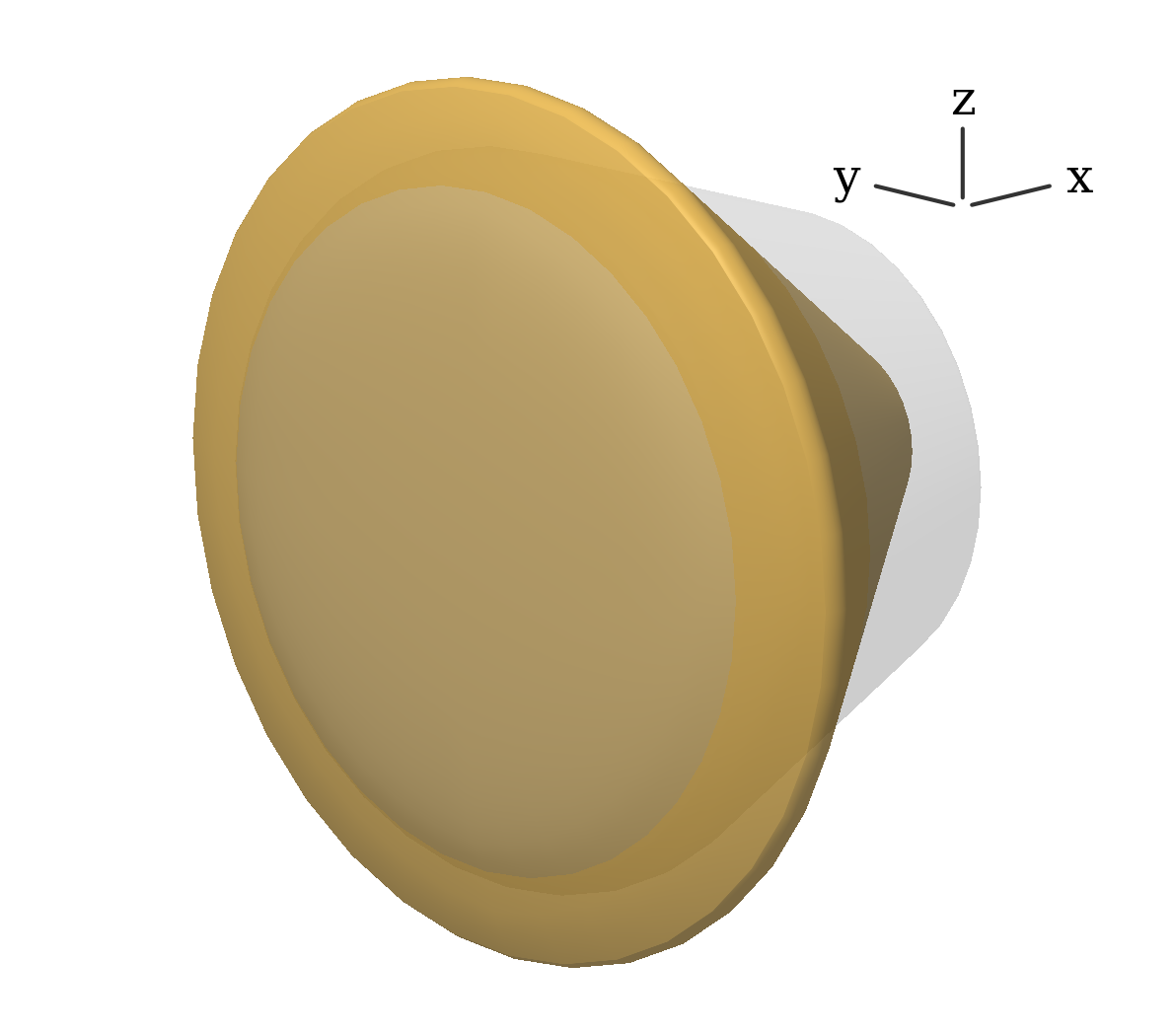}}\\[.45em]
\subfloat[$A$, half-section]{\includegraphics[width=.31\textwidth]{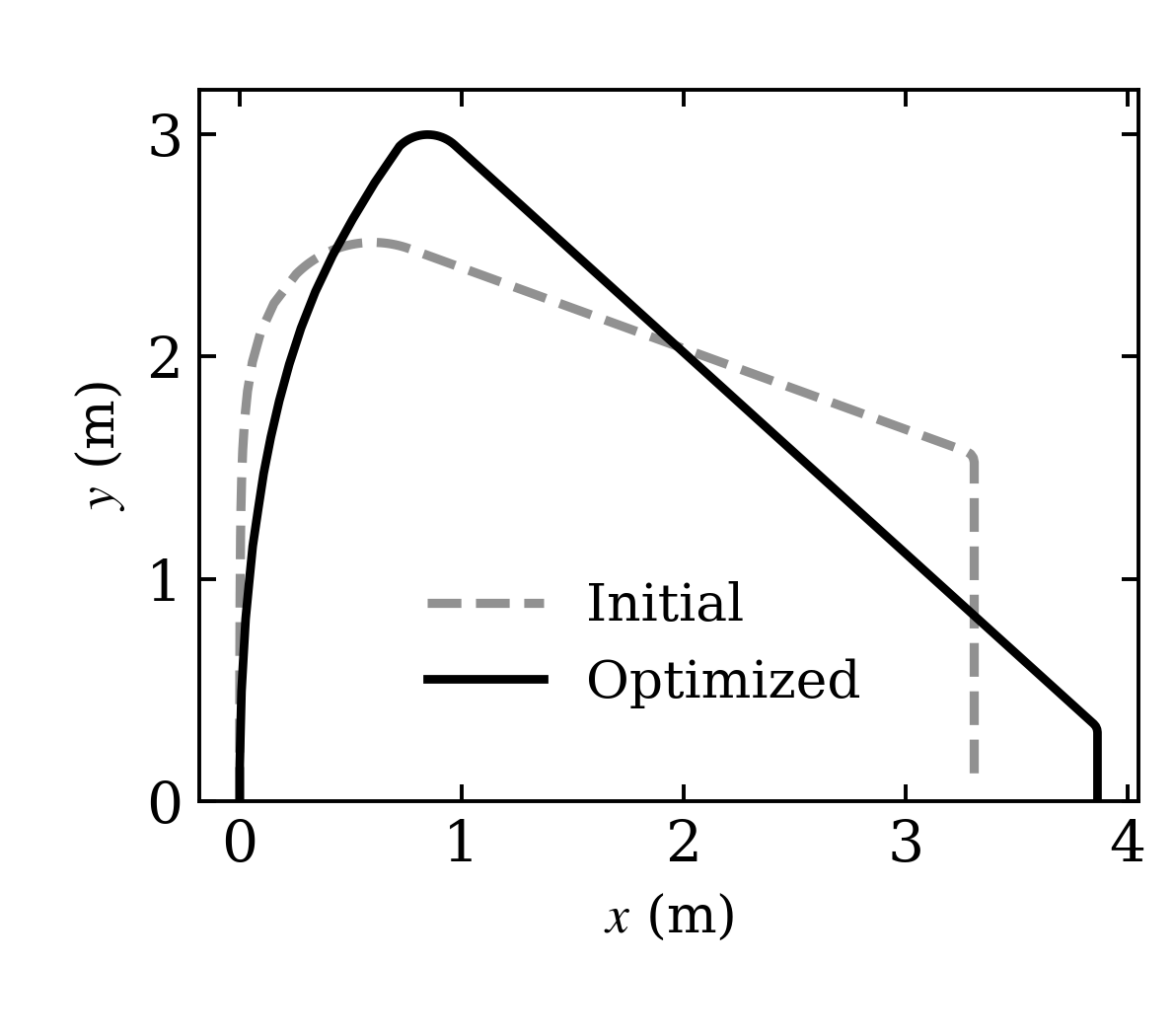}}\hspace{.02\textwidth}%
\subfloat[$B$, half-section]{\includegraphics[width=.31\textwidth]{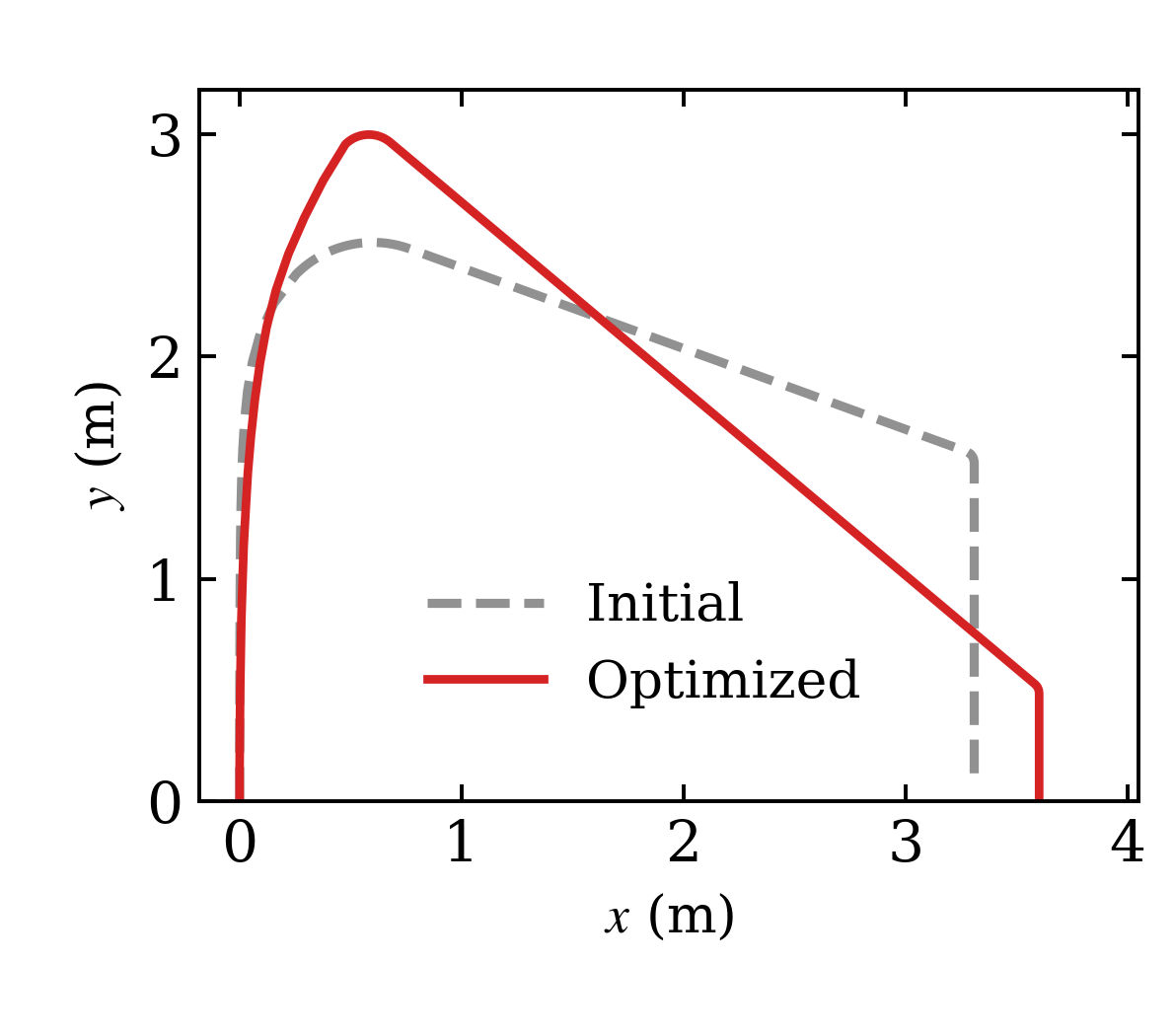}}\hspace{.02\textwidth}%
\subfloat[$C$, half-section]{\includegraphics[width=.31\textwidth]{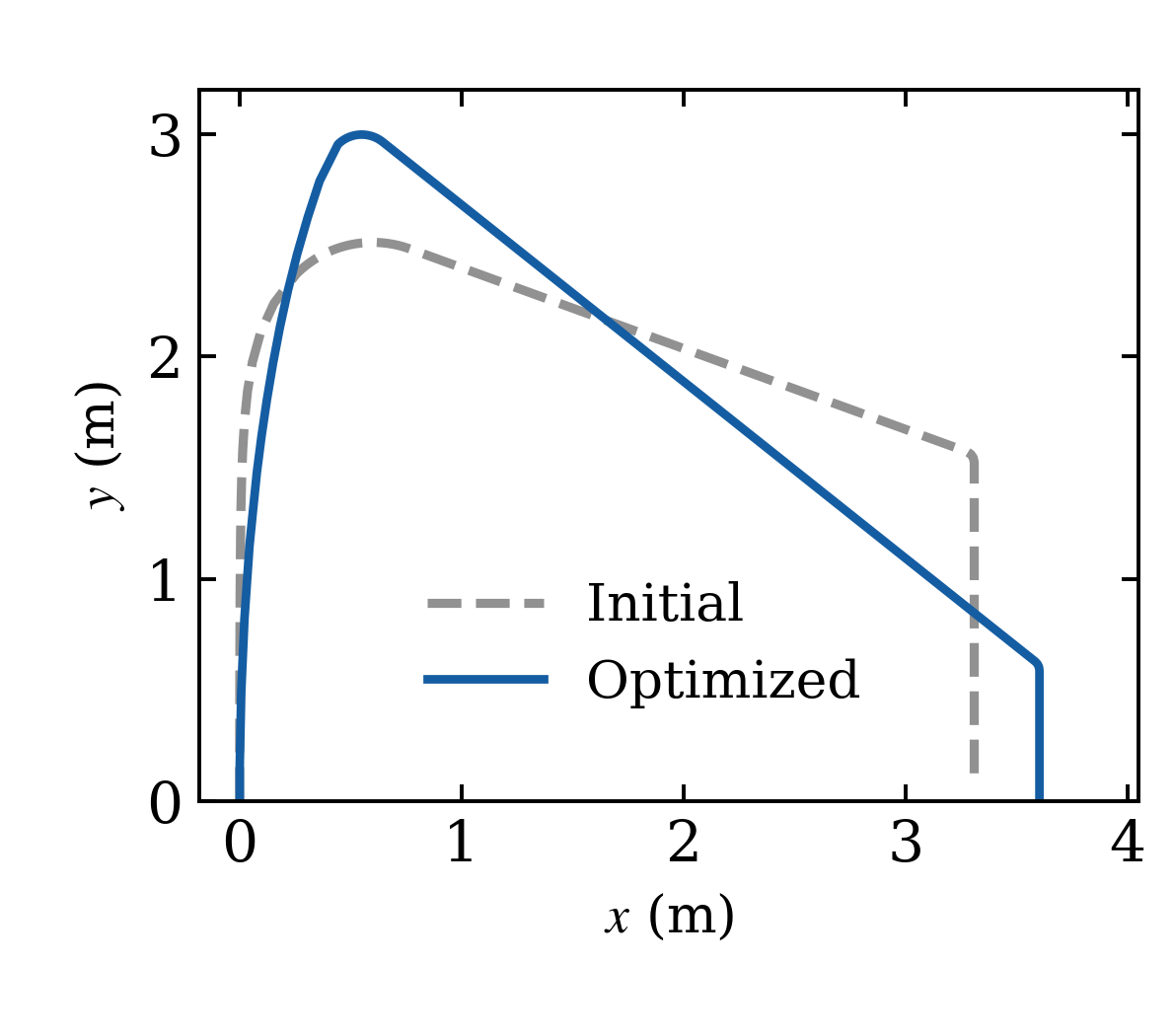}}\\[.45em]
\subfloat[$A$, flow field]{\makebox[.31\textwidth][c]{\includegraphics[width=.27\textwidth]{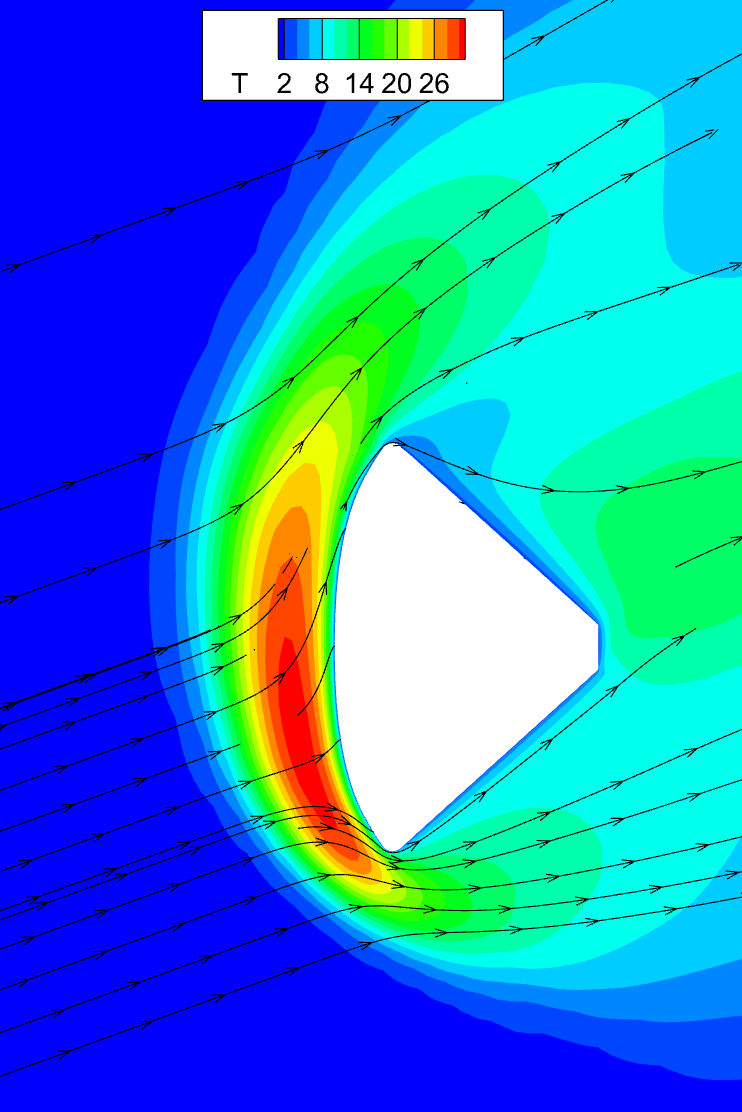}}}\hspace{.02\textwidth}%
\subfloat[$B$, flow field]{\makebox[.31\textwidth][c]{\includegraphics[width=.27\textwidth]{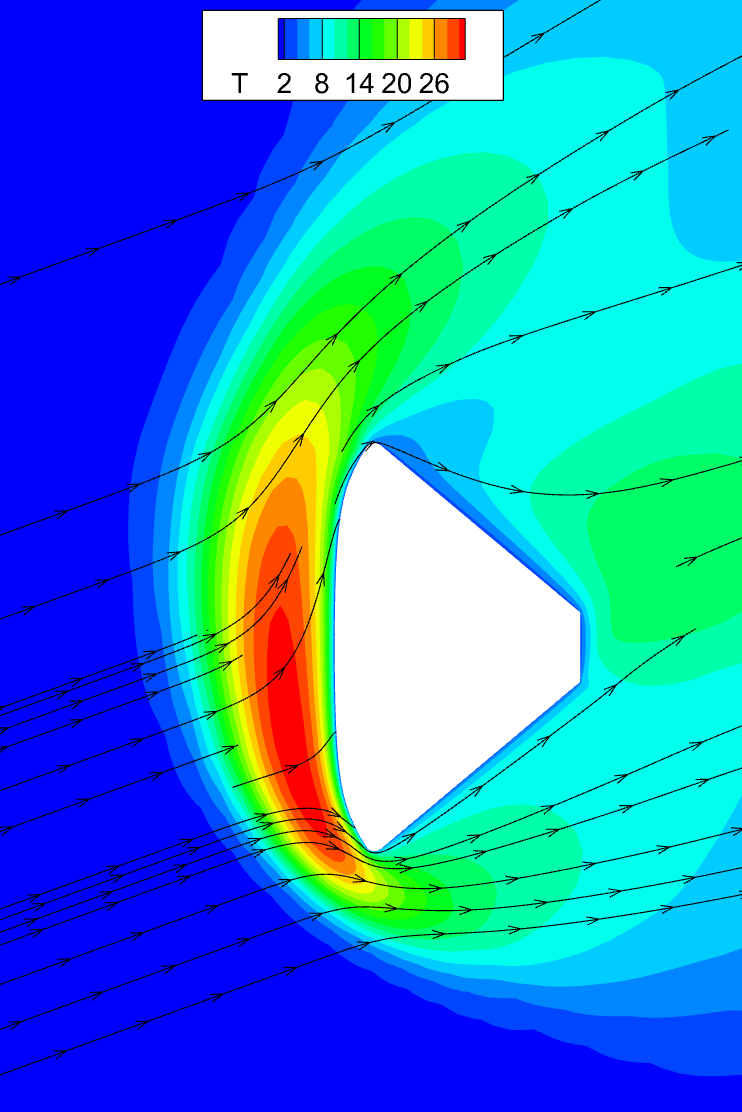}}}\hspace{.02\textwidth}%
\subfloat[$C$, flow field]{\makebox[.31\textwidth][c]{\includegraphics[width=.27\textwidth]{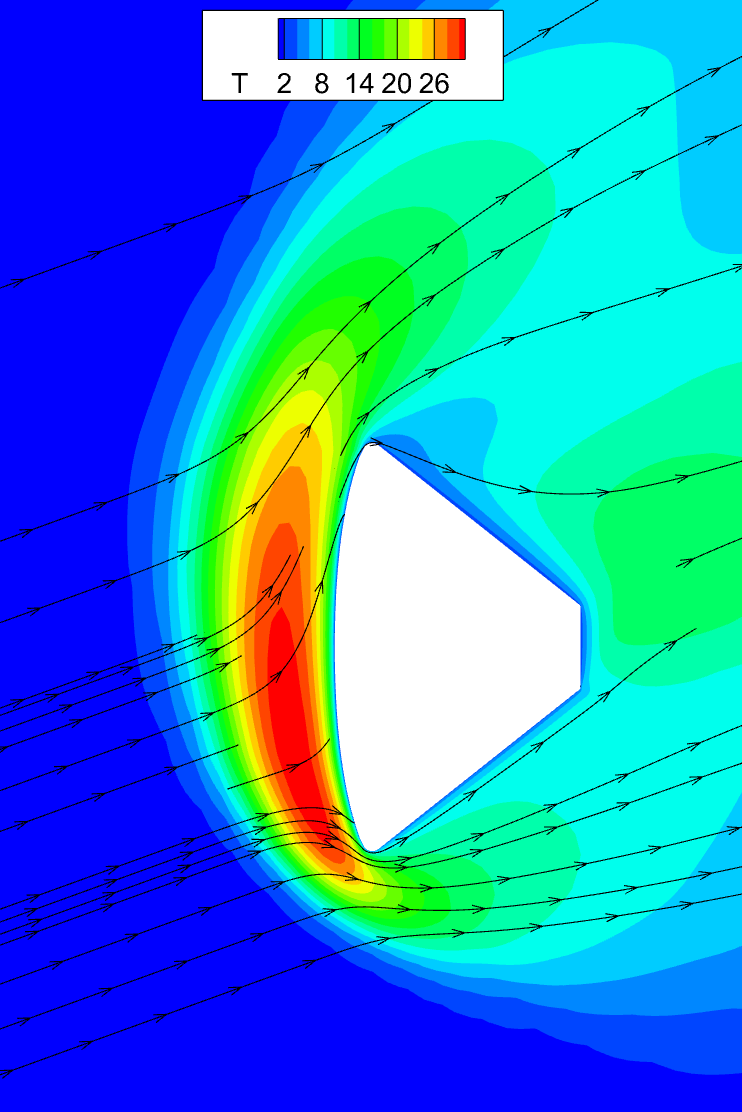}}}
\caption{Geometry and near-body flow for designs $A$, $B$, and $C$ (left to right). (a--c) Initial (gray) and optimized (gold) surfaces at a common physical scale. (d--f) Half-sections at $z=0$, $y\ge0$, with the initial section in dashed gray and the optimized sections in black, red, and blue. (g--i) Symmetry-plane temperature contours and streamlines on common spatial and $T/T_\infty$ scales. Black arrows indicate the streamline direction.}
\label{fig:orion-shapes}
\end{figure}

\begin{figure}[!htbp]
\centering
\subfloat[Front-view wall heat-flux comparisons]{\includegraphics[width=.9\textwidth]{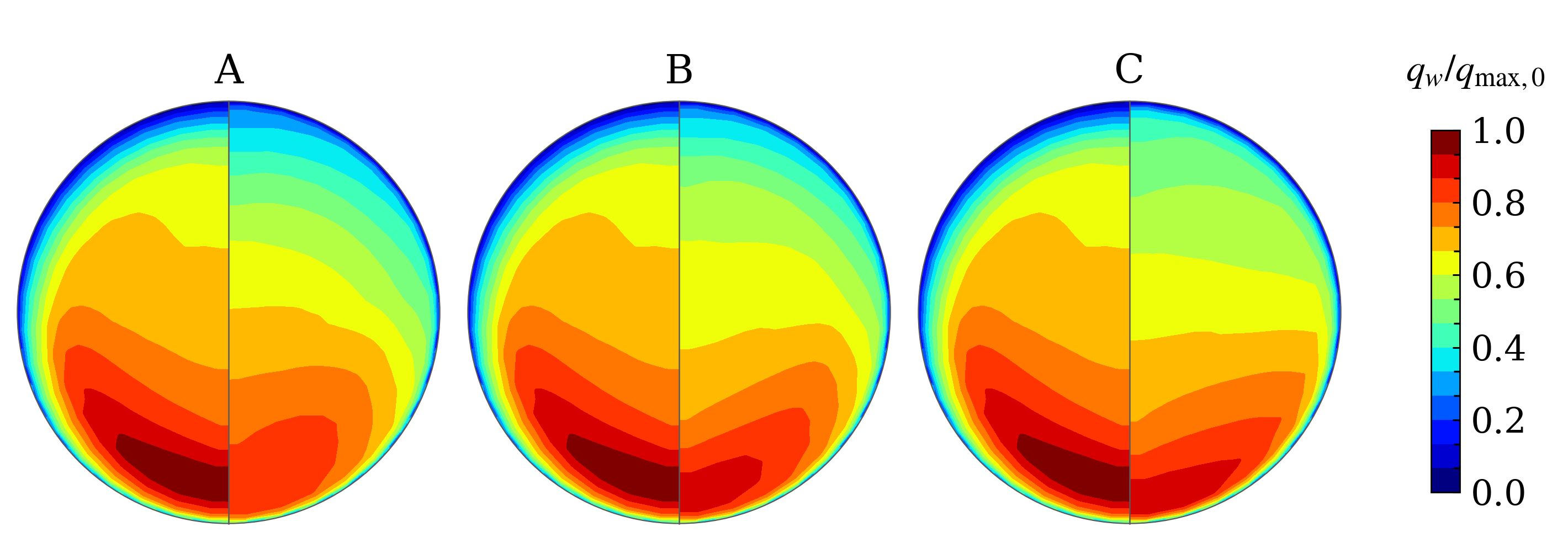}}\\
\subfloat[Centerline distribution]{\includegraphics[height=.25\textheight]{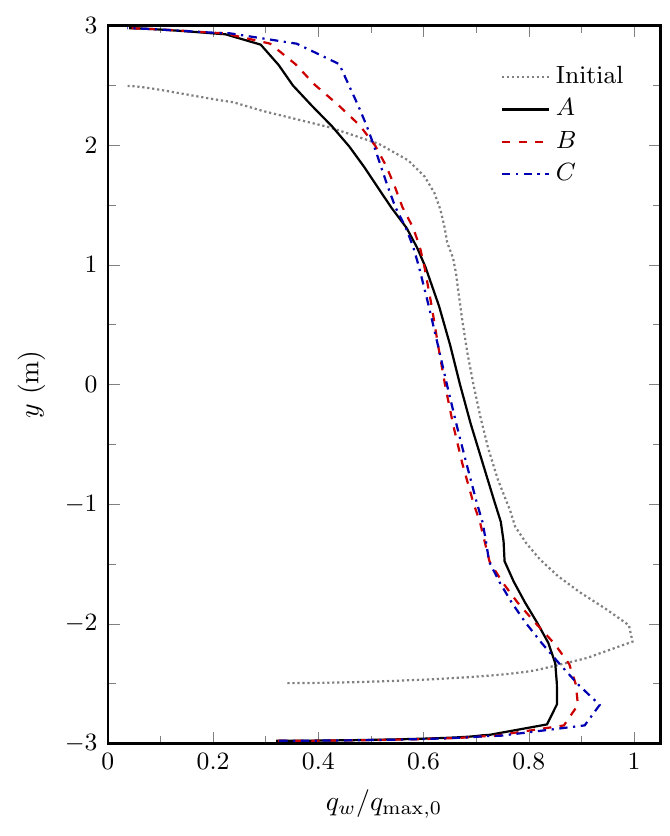}}\hspace{.03\textwidth}
\caption{Wall heat flux normalized by the initial peak $q_{\max,0}$. (a) Front-view comparisons for designs $A$, $B$, and $C$. Each panel shows the initial geometry on the left and the optimized geometry on the right. The interpolated contours use the respective maximum radii for the plotted coordinates and a common color scale. (b) Distribution along the vertical centerline of the symmetry plane ($z=0$); $y<0$ denotes the windward side. }
\label{fig:orion-heat}
\end{figure}

Geometrically (Fig.~\ref{fig:orion-shapes}), all three optimized designs reach the upper radius bound of $3$~m and have steeper rear cones than the initial capsule, and design $A$, which achieves the largest peak heat-flux reduction, has the longest forebody. The streamlines in Fig.~\ref{fig:orion-shapes}(g--i) show that, from $C$ to $A$, as the forebody becomes progressively longer, the stagnation point on the heat shield moves toward its middle. Figure~\ref{fig:orion-heat}(b) plots the wall heat flux along the vertical centerline of the symmetry plane ($z=0$): over the entire windward centerline, the optimized heat fluxes lie below the initial one, with the peak reduction ordered $A>B>C$; moreover, the initial profile exhibits a narrow, steep peak immediately adjacent to the windward shoulder, whereas the optimized profiles are broader and flatter in the peak region. Consistently, the front-view contours in Fig.~\ref{fig:orion-heat}(a) show that the peak-heating region moves away from the sharp windward shoulder corner toward the middle of the heat shield. These observations suggest the following mechanism: as the stagnation point moves toward the middle of the heat shield, away from the sharp edge of small curvature radius, the local flow scale near the stagnation region increases, and the resulting weaker velocity and temperature gradients reduce the peak wall heat flux.

\begingroup
\subsection{Chamber pressure-uniformity design}
\label{sec:cvd-allocation}

Finally, the adjoint method is applied to the inlet-flow allocation of a three-dimensional showerhead chamber, with the geometry held fixed. In such chambers, the key quantities are the pressure uniformity and the mean pressure over the wafer, which determine the within-wafer consistency of the deposition process and the reaction rate, respectively. Under a fixed total gas supply, the inlet flow rates are optimized to improve the pressure uniformity over the target wafer region while maintaining the mean pressure level.

The showerhead chamber in Fig.~\ref{fig:cvd-setup} supplies argon at a fixed total rate of 400~sccm through 124 holes in eight concentric rings, whose radii range from $10$ to $90$~mm and whose hole counts range from $4$ to $32$. The inlet gas and the solid walls are at 473~K, and the outlet is an annular pressure outlet at the chamber bottom, held at 1~Pa. Pressure uniformity is evaluated over the central region of the wafer within half its diameter, referred to as the target disk. With a reference length $l_{\mathrm{ref}}$ of 20~mm and the outlet state as the reference condition, the Knudsen number is approximately $0.55$: the chamber flow lies well inside the transition regime, where the Navier--Stokes equations cease to be valid and a kinetic description is required. The mesh contains 467,516 physical cells, and the velocity discretization uses $20^3$ uniform points over $[-6v_m,6v_m]^3$.

\begin{figure}[!htbp]
\centering
\subfloat[Geometry and boundary types]{\includegraphics[width=.40\textwidth,height=.26\textheight,keepaspectratio]{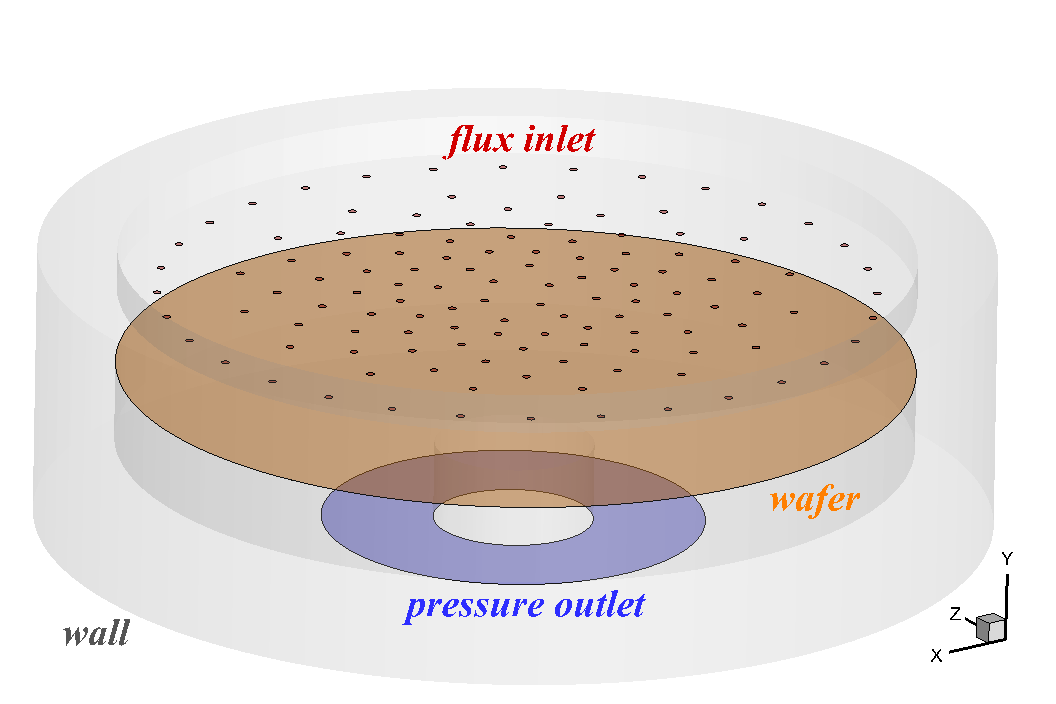}}\hspace{.02\textwidth}%
\subfloat[Symmetry-plane physical mesh]{\includegraphics[width=.40\textwidth,height=.26\textheight,keepaspectratio]{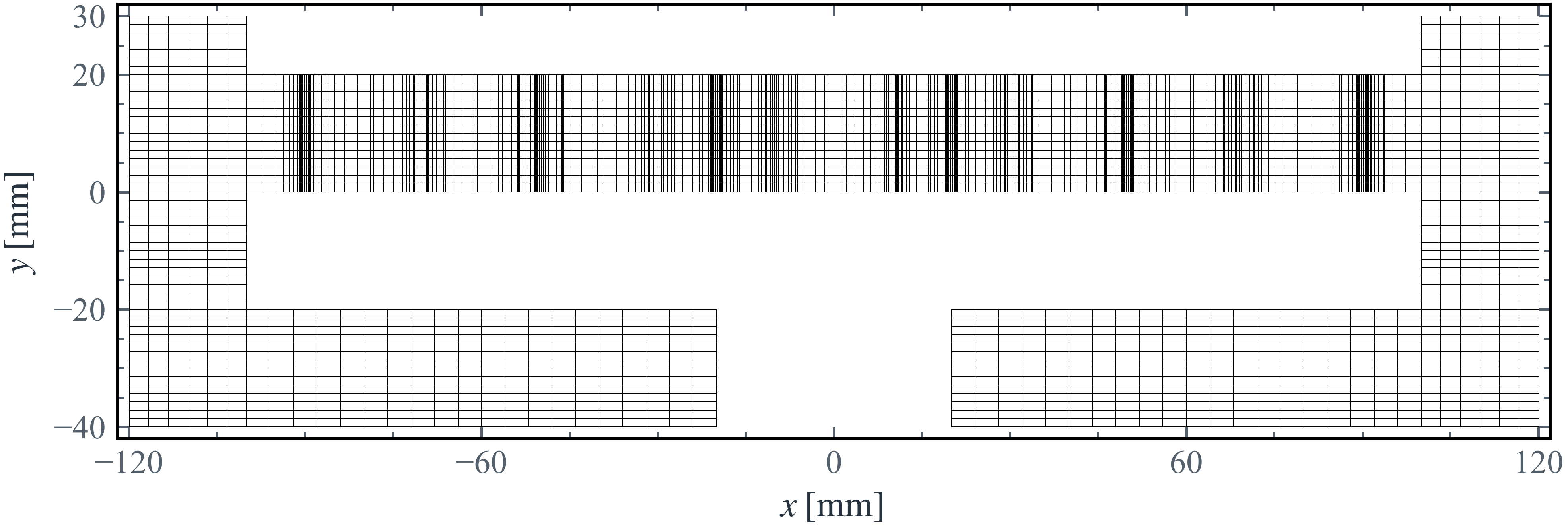}}
\caption{Chamber geometry, boundary conditions, and physical mesh.}
\label{fig:cvd-setup}
\end{figure}

Since the chamber and the outlet are axisymmetric, the inlet holes are grouped into the eight concentric rings, and all holes in a ring share the same flow rate. Let $Q_i$ be the flow rate in ring $i$ with $n_i$ holes. The map $Q_i=Q_{\rm tot}n_i e^{z_i}/\sum_{j=1}^{8}n_j e^{z_j}$, with $z_8=0$, fixes the total flow and leaves seven independent variables. The flow per hole is bounded between $0.125$ and $3$ times its initial value. The objective balances pressure uniformity and mean pressure,
\begin{equation}
 J=\frac{\mathrm{CV}^2}{\mathrm{CV}_{\rm ref}^2}+1-\frac{\overline p}{p_{\rm ref}},
 \qquad
 \mathrm{CV}=\frac{1}{\overline p}\left[\frac{1}{|A|}\int_A (p-\overline p)^2\,\mathrm{d}S\right]^{1/2},
 \qquad \overline p=\frac{1}{|A|}\int_A p\,\mathrm{d}S,
 \label{eq:cvd-objective}
\end{equation}
where $A$ denotes the target disk, evaluated using the intersections of the wafer faces with the disk, and $\mathrm{CV}$, the ratio of the area-weighted standard deviation to the mean pressure, serves as a quantitative measure of uniformity. The reference values $p_{\rm ref}\simeq4.38$~Pa and $\mathrm{CV}_{\rm ref}\simeq1.49\times10^{-2}$ are both obtained at the initial evaluation. The first term of the objective promotes the within-wafer deposition consistency, while the second term prevents the optimization from improving uniformity simply by lowering the overall pressure level.

The results (Fig.~\ref{fig:cvd-optimization} and Table~\ref{tab:cvd-selected}) demonstrate the rapid convergence of the adjoint-based optimization. Within only four evaluations, the CV drops from 1.49\% to 0.09\%, an improvement of more than one order of magnitude. After 30 evaluations, the CV reaches 0.06\%, about 25 times lower than the initial value, while the mean pressure slightly increases from 4.38 to 4.40~Pa, indicating that the uniformity gain is not obtained at the expense of the pressure level. The final evaluation gives the lowest objective in the run, $J\simeq-3.42\times10^{-3}$. The complete optimization takes 110~min on 576 cores of the CPU model ``\emph{Intel Xeon Platinum 9242}'', i.e., about 3.7~min per evaluation including one primal and one adjoint solve, with a memory cost of about 497~GiB. Since the cost of the adjoint gradient is independent of the number of design variables, this efficiency carries over directly to finer allocation schemes, such as per-hole optimization in chambers with non-axisymmetric outlets.

\begin{figure}[!htbp]
\centering
\includegraphics[width=.82\textwidth,height=.30\textheight,keepaspectratio]{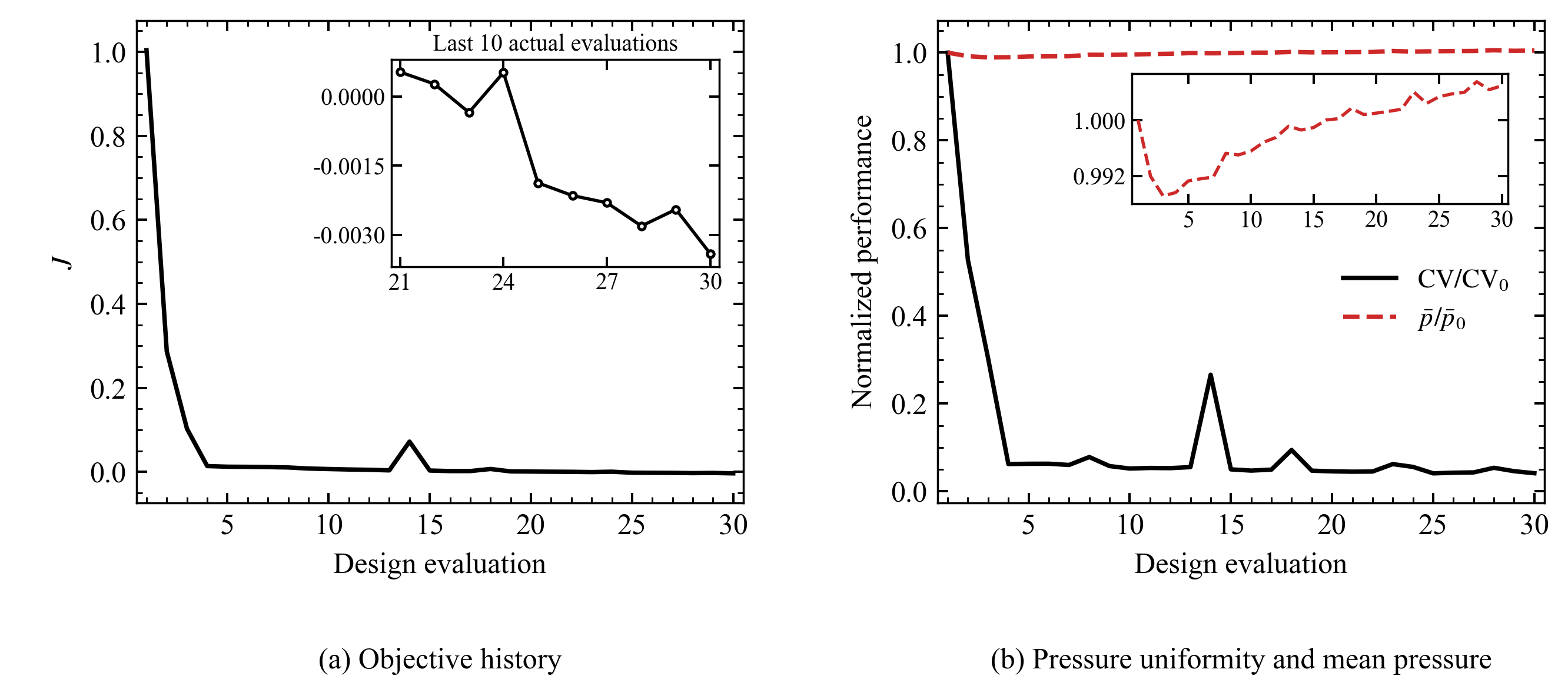}
\caption{Chamber optimization: (a) objective history, with the last ten evaluations in the inset; (b) CV and mean pressure normalized by their initial values.}
\label{fig:cvd-optimization}
\end{figure}

\begin{table}[!htbp]
\centering
\caption{Selected chamber-design evaluations. The variance is normalized by the squared mean pressure over the target disk.}
\label{tab:cvd-selected}
\begin{tabular}{rrrrr}
\hline
Evaluation & $J$ & $\overline p$ (Pa) & $\mathrm{CV}^2$ & CV (\%)\\
\hline
1 & $1.00$ & $4.38$ & $2.22\times10^{-4}$ & $1.49$\\
4 & $1.40\times10^{-2}$ & $4.33$ & $8.62\times10^{-7}$ & $0.09$\\
20 & $8.58\times10^{-4}$ & $4.38$ & $4.61\times10^{-7}$ & $0.07$\\
30 & $-3.42\times10^{-3}$ & $4.40$ & $3.79\times10^{-7}$ & $0.06$\\
\hline
\end{tabular}
\end{table}

\begin{figure}[!htbp]
\centering
\newsavebox{\cvdcomparisonbox}
\sbox{\cvdcomparisonbox}{\begin{minipage}{.9\textwidth}
\centering
\makebox[.38\linewidth]{Inlet allocation}\makebox[.60\linewidth]{Wafer pressure}\\[-.2em]
\subfloat[Evaluation 1]{\begin{minipage}{\linewidth}\centering
\includegraphics[width=.375\linewidth]{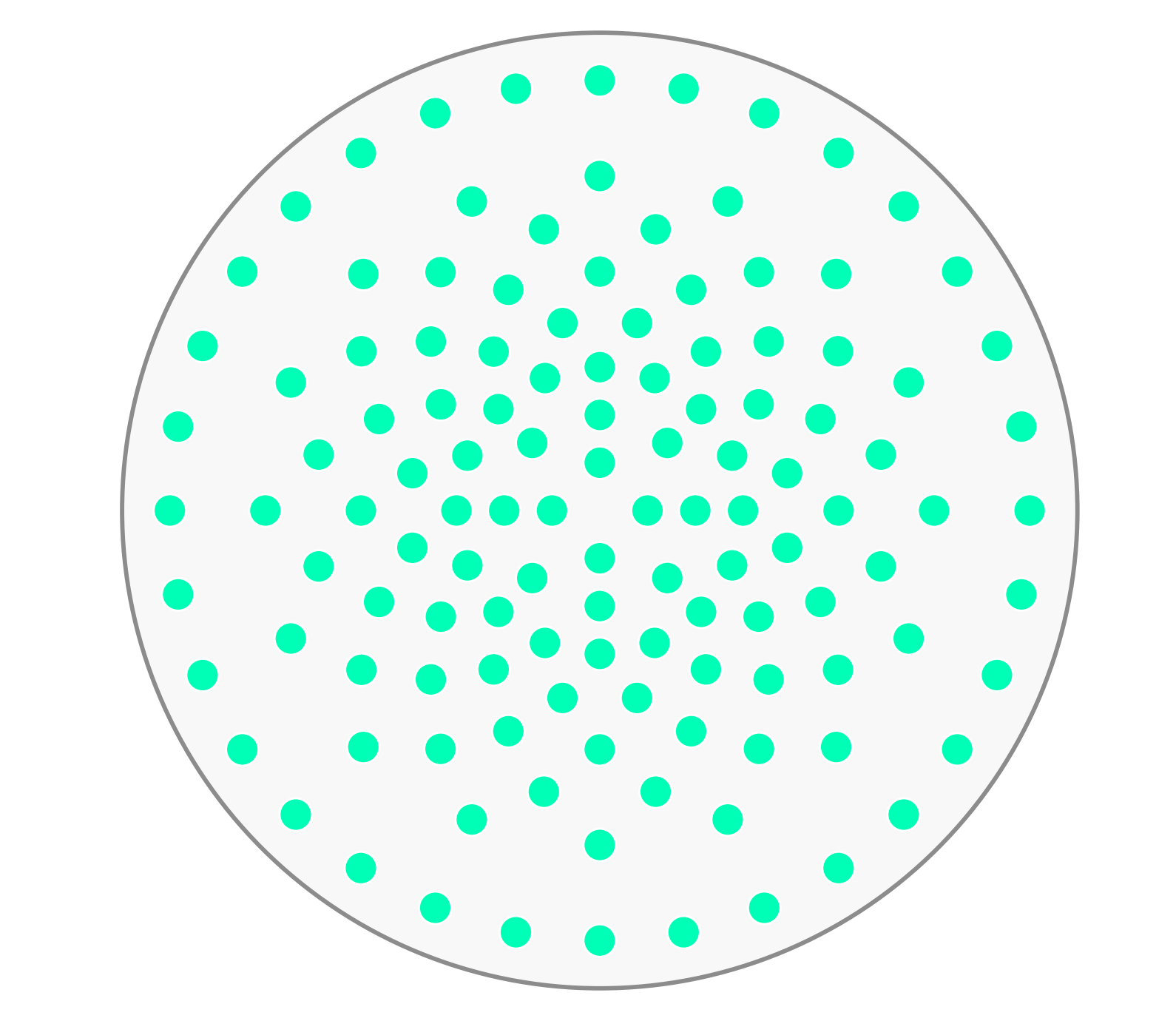}\hspace{.02\linewidth}%
\includegraphics[width=.595\linewidth]{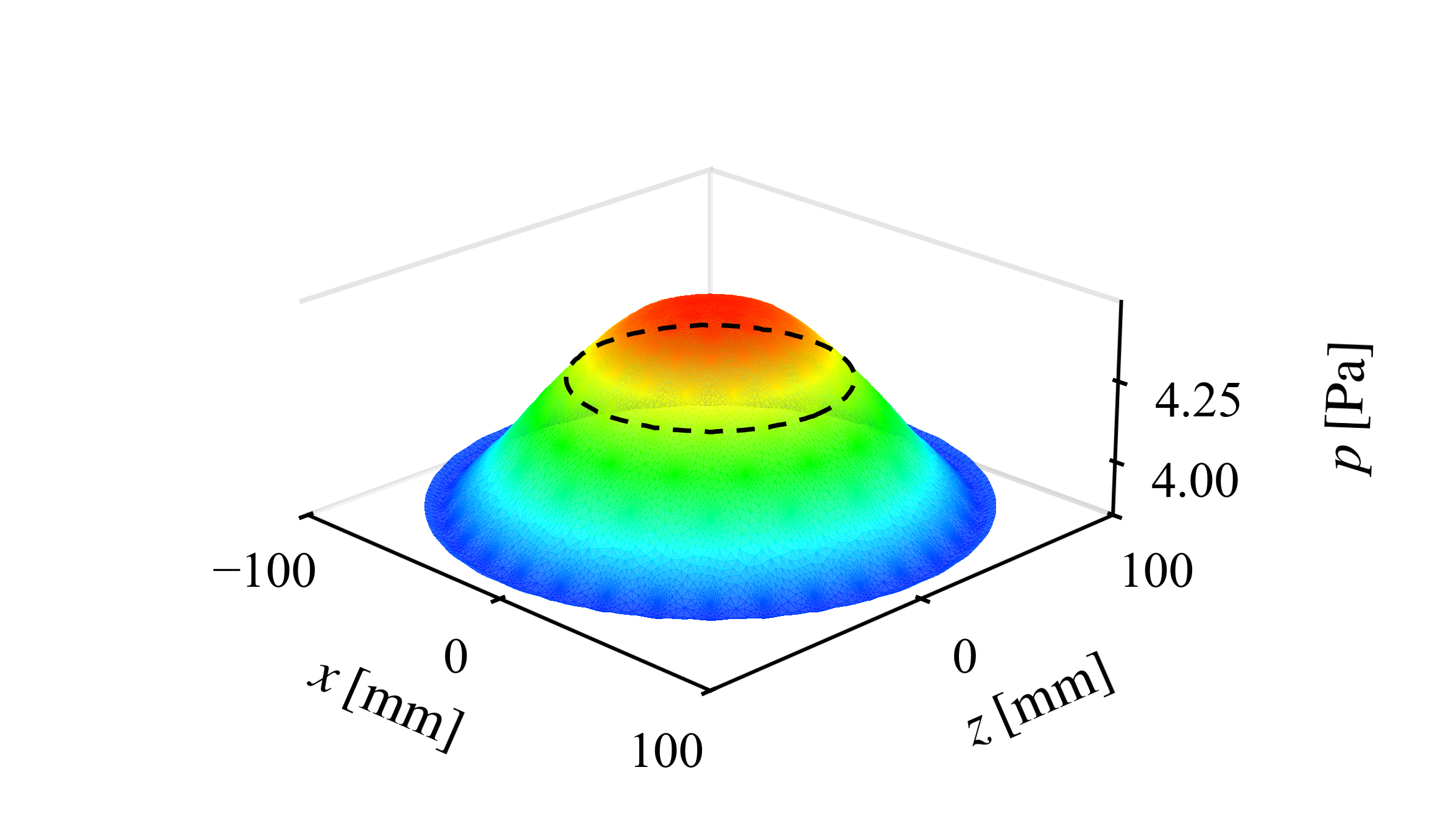}
\end{minipage}}\\[-.3em]
\subfloat[Evaluation 4]{\begin{minipage}{\linewidth}\centering
\includegraphics[width=.375\linewidth]{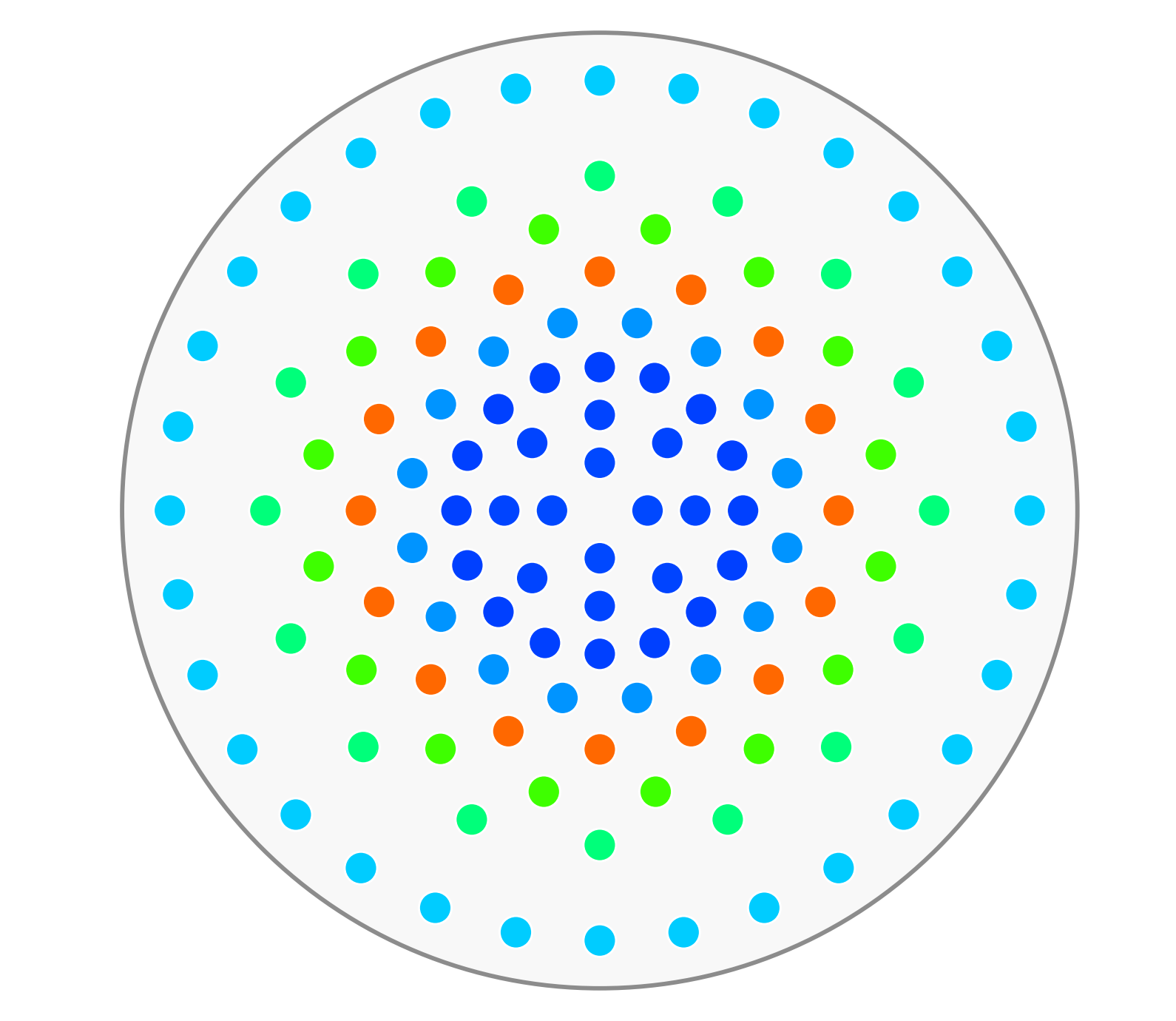}\hspace{.02\linewidth}%
\includegraphics[width=.595\linewidth]{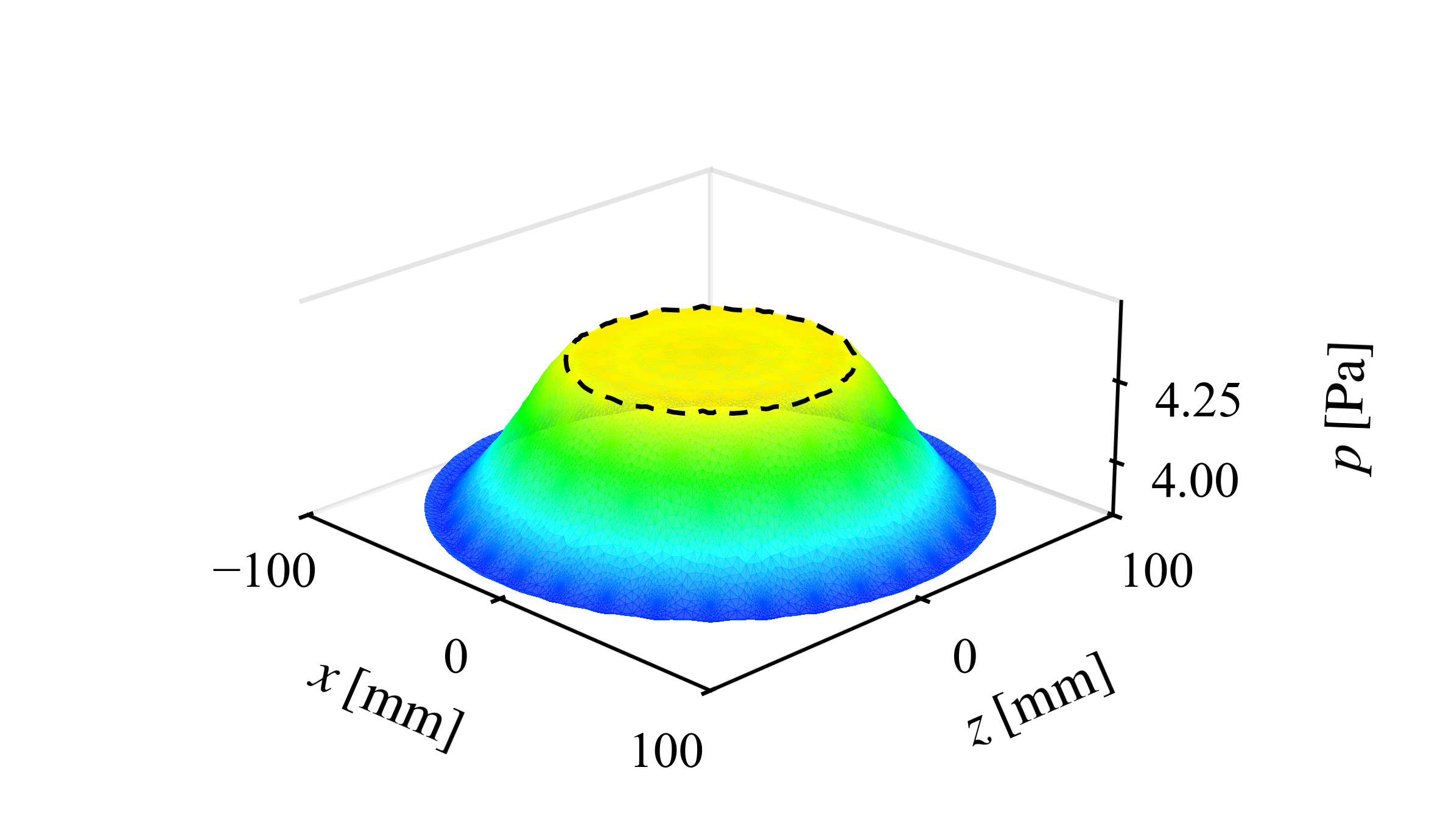}
\end{minipage}}\\[-.3em]
\subfloat[Evaluation 20]{\begin{minipage}{\linewidth}\centering
\includegraphics[width=.375\linewidth]{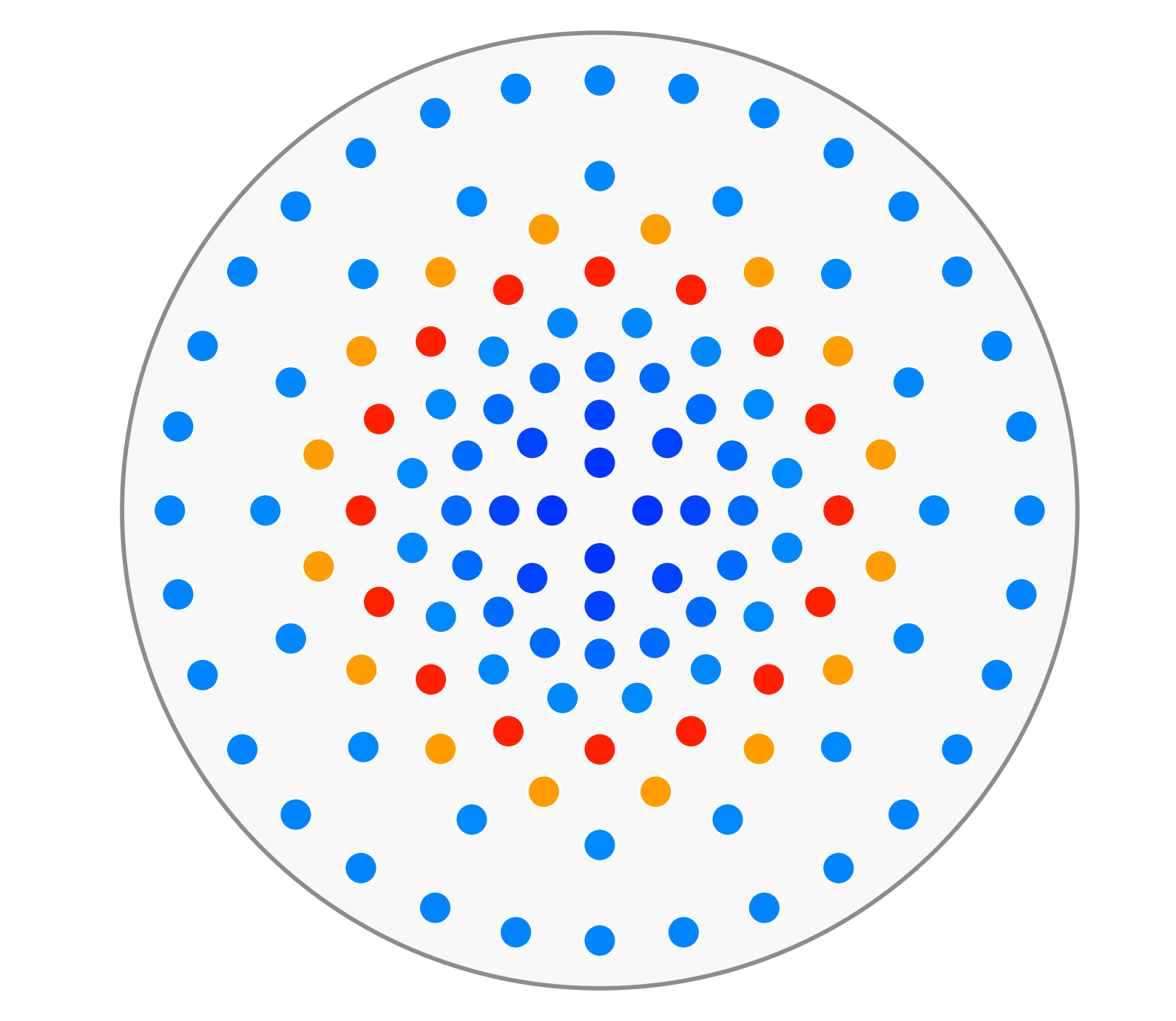}\hspace{.02\linewidth}%
\includegraphics[width=.595\linewidth]{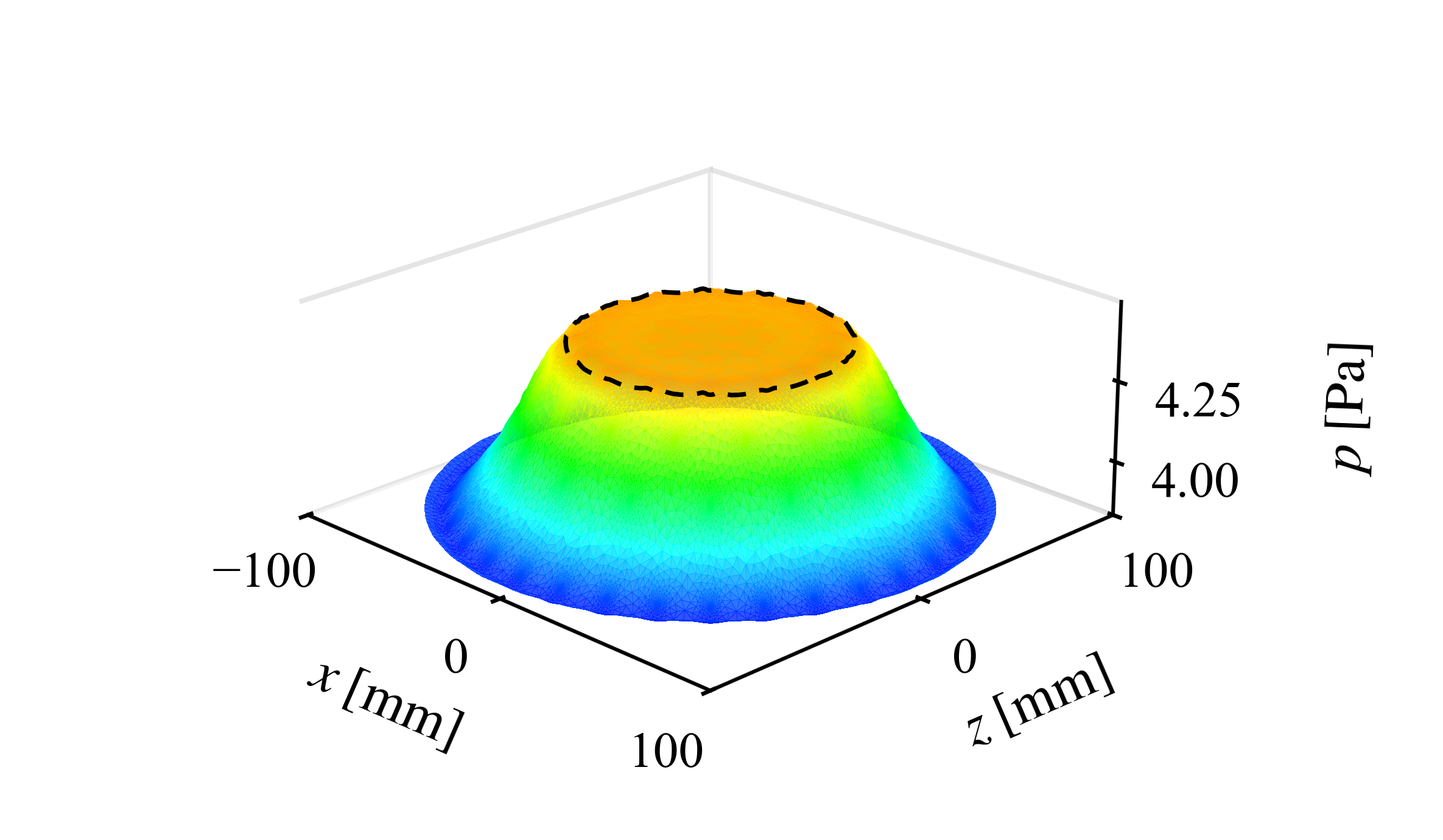}
\end{minipage}}\\[-.3em]
\subfloat[Evaluation 30]{\begin{minipage}{\linewidth}\centering
\includegraphics[width=.375\linewidth]{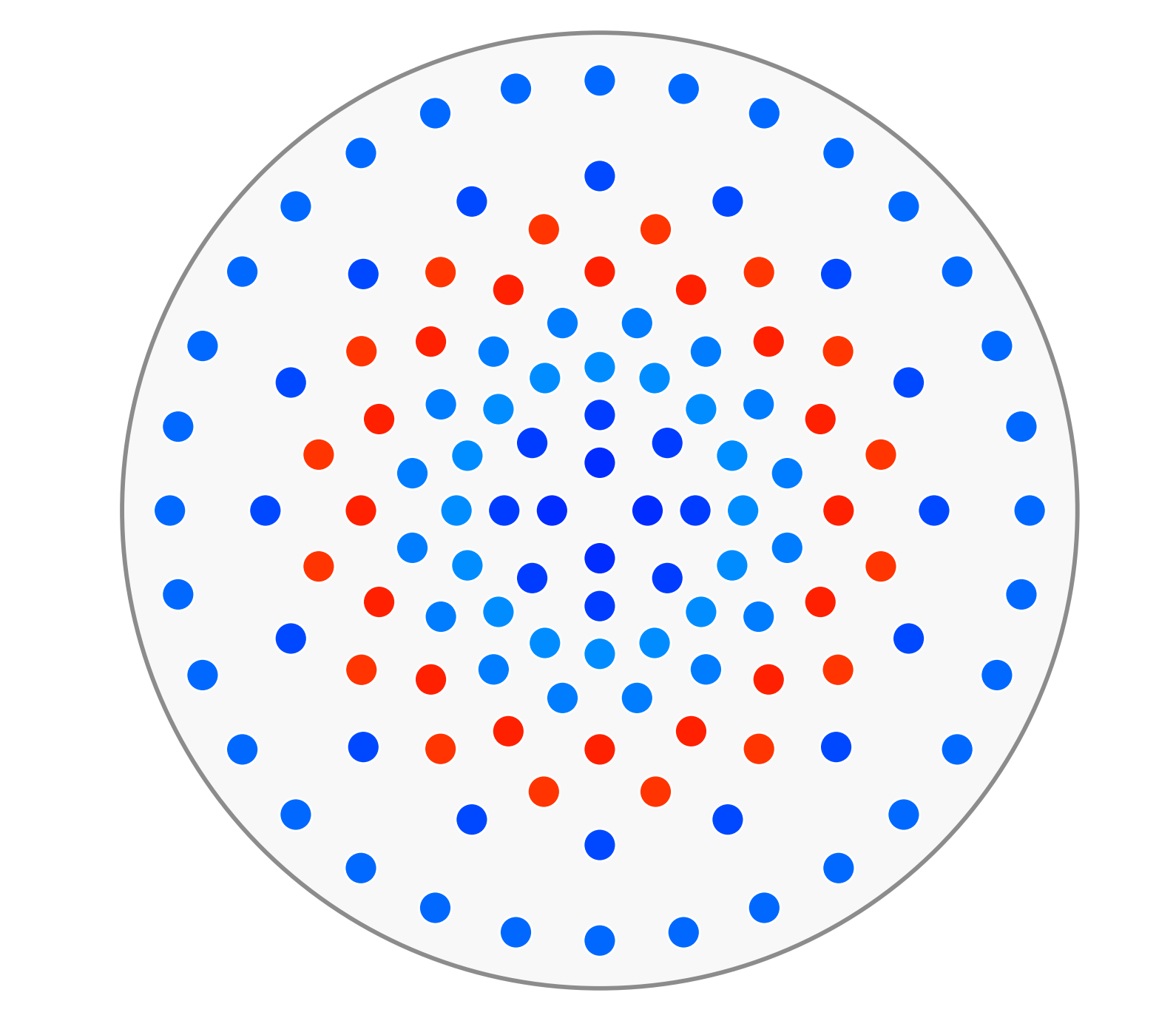}\hspace{.02\linewidth}%
\includegraphics[width=.595\linewidth]{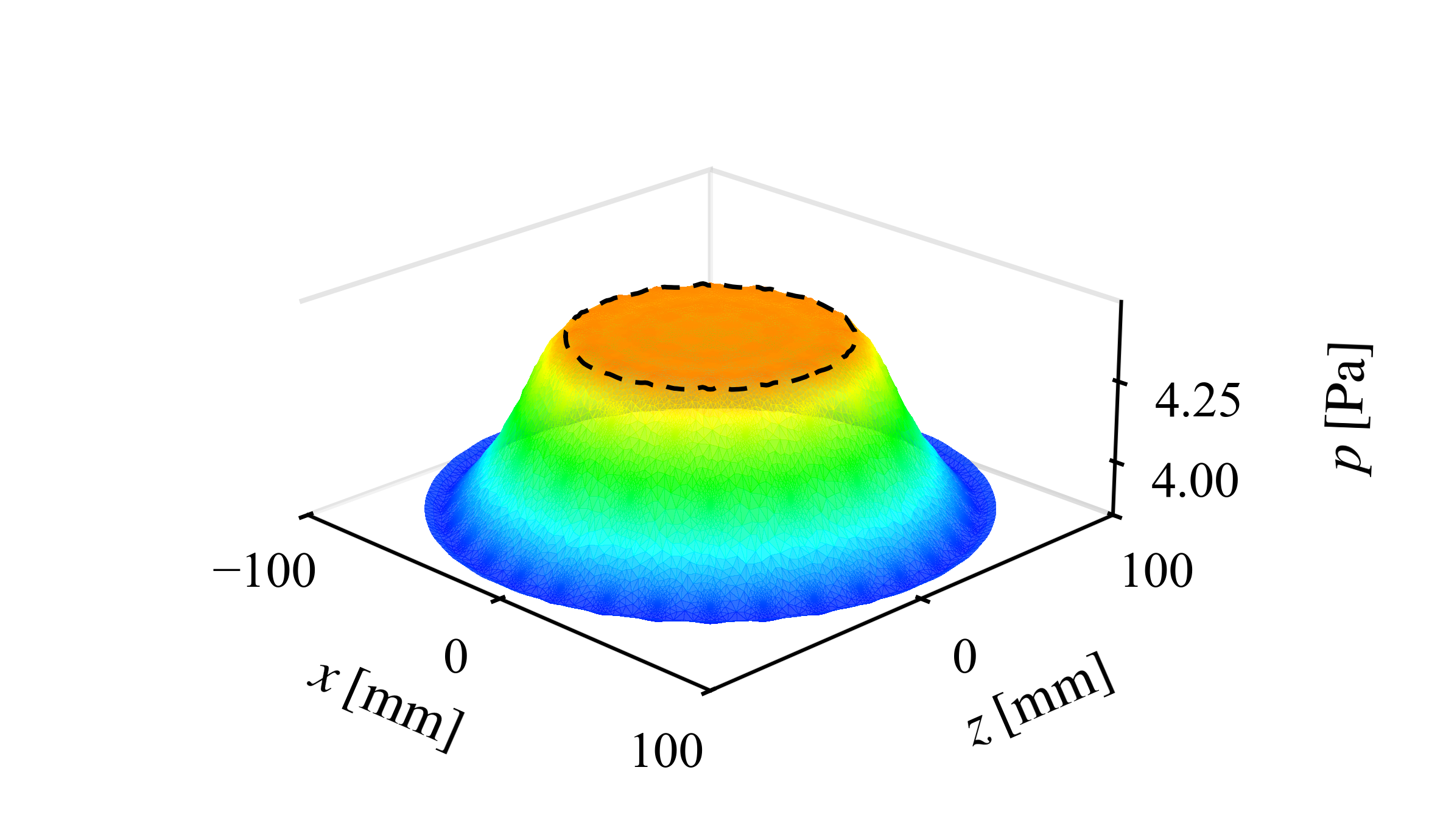}
\end{minipage}}\\[-.3em]
\includegraphics[width=\linewidth]{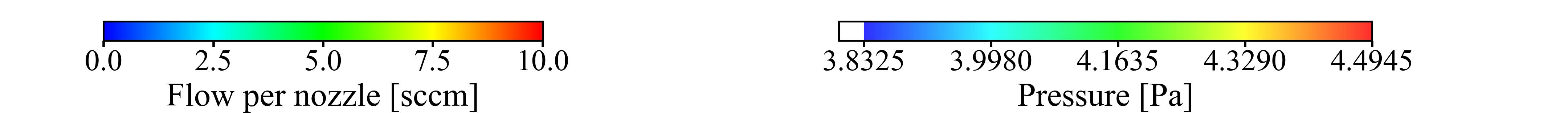}
\end{minipage}}
\ifdim\dimexpr\ht\cvdcomparisonbox+\dp\cvdcomparisonbox\relax>.8\textheight
  \resizebox*{!}{.8\textheight}{\usebox{\cvdcomparisonbox}}
\else
  \usebox{\cvdcomparisonbox}
\fi
\caption{Inlet flow per nozzle (left, sccm) and wafer pressure (right, Pa) at evaluations 1, 4, 20, and 30 (top to bottom). The dashed circle marks the target disk of radius 50~mm.}
\label{fig:cvd-pressure-allocation}
\end{figure}

\FloatBarrier

The evolution of the wafer pressure distribution is illustrated by the pressure maps in Fig.~\ref{fig:cvd-pressure-allocation} and the radial profiles in Fig.~\ref{fig:cvd-redistribution}(b). With the initial uniform allocation, the pressure peaks at the disk center and decreases toward its edge. Since the objective concerns only the target disk within a radius of 50~mm, the optimizer first reduces the flow of the central rings and increases the supply near the 50~mm ring, which directly removes the central peak and fills the under-pressurized edge region; a nearly flat plateau over the target disk is established already at evaluation 4 (Figs.~\ref{fig:cvd-redistribution}(a,b)), corresponding to the order-of-magnitude uniformity gain. The subsequent adjustment is slower: the optimization mainly raises the flow of the 60~mm ring to increase the gas supply over the target-disk region. The pressure plateau is thereby lifted as a whole with its shape essentially unchanged (evaluations 20 and 30 in Fig.~\ref{fig:cvd-redistribution}(b)), and the mean pressure recovers to a level no lower than the initial one. The symmetry-plane speed contours and streamlines of the initial and final allocations are compared in Fig.~\ref{fig:cvd-redistribution}(c,d), with weaker jets near the center and stronger jets near the edge of the target disk, consistently with the allocation change.
\begin{figure}[!htbp]
\centering
\includegraphics[width=.82\textwidth]{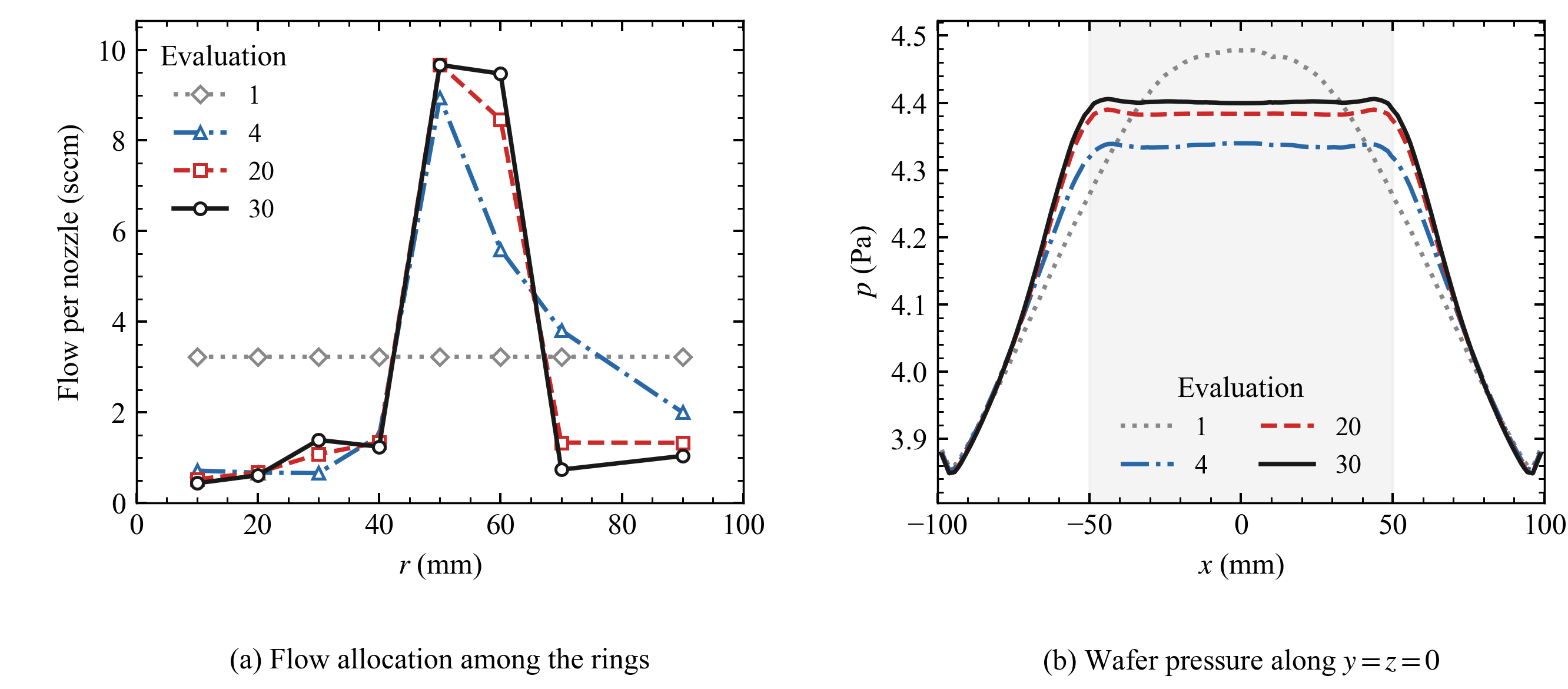}
\par\smallskip
\setcounter{subfigure}{2}
\subfloat[Initial allocation (evaluation 1)]{\includegraphics[width=.40\textwidth]{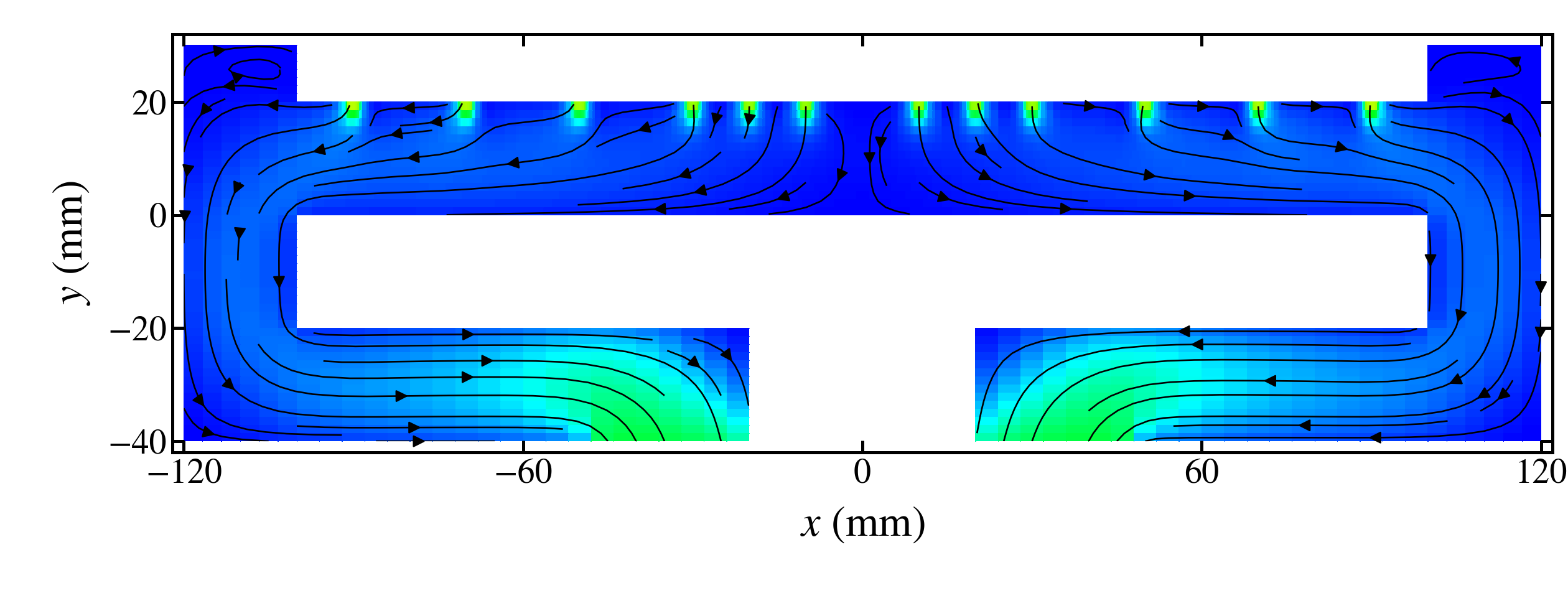}}\hspace{.02\textwidth}%
\subfloat[Optimized allocation (evaluation 30)]{\includegraphics[width=.40\textwidth]{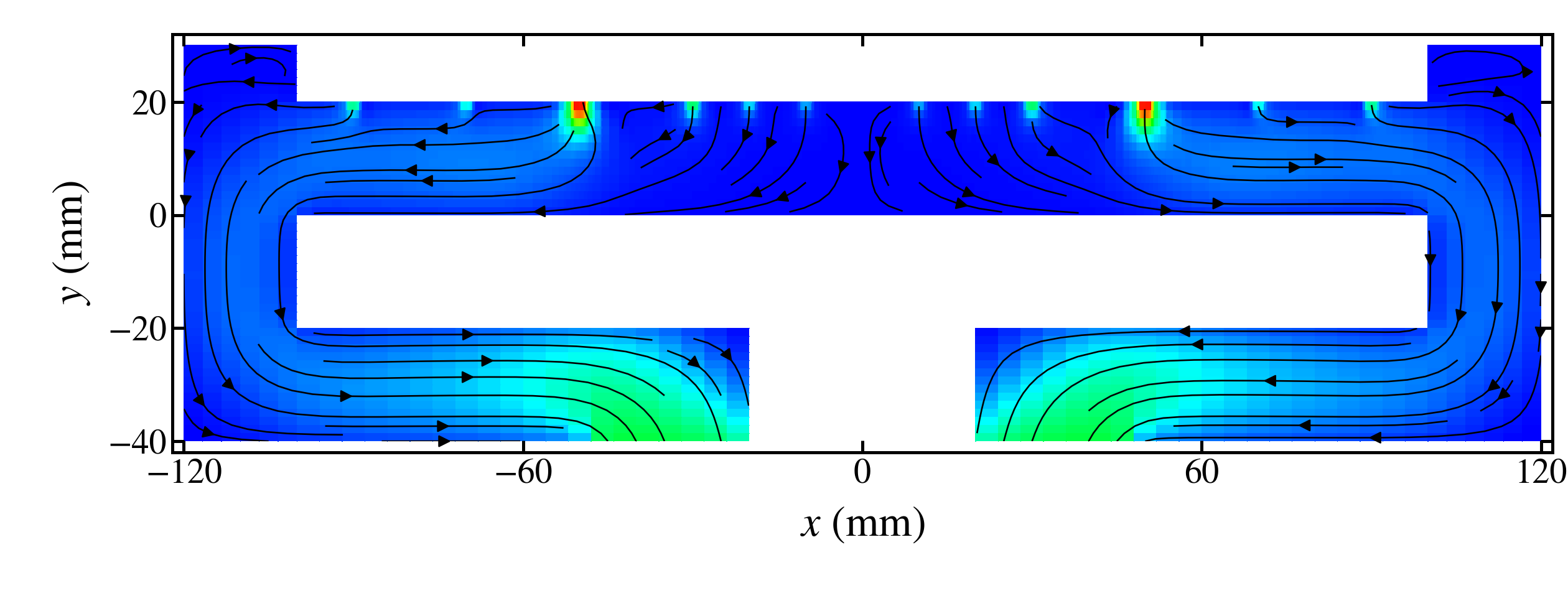}}
\par\smallskip
\includegraphics[width=.38\textwidth]{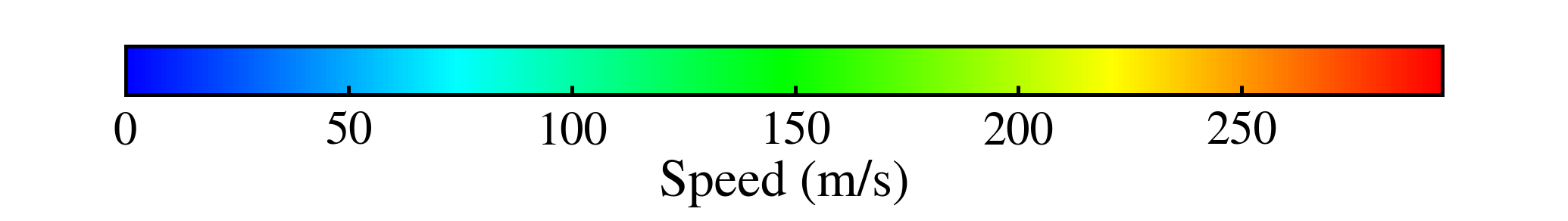}
\caption{Inlet flow per nozzle (a) and wafer pressure along $y=z=0$ (b) at evaluations 1, 4, 20, and 30. The shaded interval $|x|\le50$~mm in (b) marks the target disk. (c,d) Speed and projected streamlines on the symmetry plane $z=0$ for the initial and optimized allocations, respectively.}
\label{fig:cvd-redistribution}
\end{figure}
\FloatBarrier
\endgroup

 \section{Conclusions}\label{sec:conc}

In this paper, we have developed a discrete adjoint method for GSIS to enable efficient gradient-based optimization of gas flows across continuum and rarefied regimes. The construction starts from the fixed-point representation of the primal GSIS, in which the macroscopic synthetic equations are retained together with the kinetic equations. By incorporating both sets of equations into the discrete Lagrangian and imposing the stationarity conditions, we obtain a coupled mesoscopic--macroscopic adjoint system that directly inherits the macroscopic acceleration structure of the primal method. The resulting system is solved by alternating kinetic and macroscopic updates, in close analogy to the primal GSIS, while the inner subproblems can be treated using established discrete adjoint techniques for DVM and NS discretizations. Consequently, the complete adjoint scheme follows from systematic differentiation and transposition of the discrete primal residuals, without a separate Chapman--Enskog or asymptotic analysis of the adjoint equations. The formulation has been developed for the BGK model and extended to the Shakhov model while preserving the same iterative structure. Combined with design sensitivity evaluation based on the same discrete formulation, including a mesh adjoint for implicitly deformed body-fitted meshes, it provides design gradients for both shape optimization and fixed-geometry flow-control problems.

The accuracy and efficiency of the proposed method have been assessed through sensitivity verification, convergence tests, and two three-dimensional optimization problems. For the NACA0012 airfoil, the discrete shape sensitivities passed stringent verification against central finite differences, with relative $L_2$ differences of only $0.03\%$ and $0.02\%$ at $\mathrm{Kn}=0.01$ and $0.5$, respectively. The adjoint GSIS also achieved a convergence efficiency consistent with that of the primal GSIS in both near-continuum and rarefied flows. At $\mathrm{Kn}=0.01$, the primal and adjoint iteration counts were reduced relative to CIS by factors of approximately 22 and 23, respectively, while corresponding acceleration was also observed at $\mathrm{Kn}=0.5$. For the three-dimensional re-entry capsule, constrained shape optimization reduced the peak wall heat flux by 14.17\%, 10.68\%, and 6.54\% while achieving prescribed lift-to-drag gains of 10\%, 20\%, and 25\%, respectively, and satisfying the volume constraints. These results demonstrate the trade-off between aerodynamic performance and peak heating. In the showerhead-chamber design, inlet-flow allocation at fixed total supply reduced the coefficient of variation of the wafer pressure from 1.49\% to 0.06\%, while maintaining the mean pressure at approximately its initial level. Collectively, the numerical results confirm the accuracy and efficiency of the discrete adjoint formulation and demonstrate its applicability to both geometric and boundary-control design in large-scale three-dimensional problems.

Overall, this work establishes a systematic route for multiscale adjoint analysis in kinetic theory: a primal multiscale method is first expressed as a fixed-point system, its multiscale structure is then retained in the associated Lagrangian, and differentiation directly yields the corresponding multiscale adjoint system. This construction is concise and avoids the need to devise a separate acceleration strategy for the adjoint equations. It is particularly noteworthy that the proposed discrete-adjoint construction is not restricted to GSIS. By incorporating the macroscopic prediction equations into the fixed-point representation and associated Lagrangian, the method can be naturally extended to macroscopic-prediction-based UGKS and DUGKS~\cite{Zhu2016Implicit,yuan2021multi}. Future work will extend this framework to more complex kinetic models and multiphysics problems, including gas mixtures~\cite{zeng2024mixtures} and coupled gas--radiation transport~\cite{zeng2026radiation}.

\section*{Declaration of competing interest}
The authors declare that they have no known competing financial interests or personal relationships that could have appeared to influence the work reported in this paper.

\section*{Acknowledgments}
This work is supported by the National Natural Science Foundation of China ( Grant~No.~ 12402388) and the Research Grants Council of the Hong Kong Special Administrative Region, China, under the Early Career Scheme (Project~No.~HKUST~26211523). The authors acknowledge the computing resources from the Center for Computational Science and Engineering at the Southern University of Science and Technology.

\appendix

\section{Reconstruction operators and interface NS flux}\label{app:reconstruction}

 The operators $\CC_{ij}$ and   $\CC_{ib}$ map cell values to interior and boundary faces, respectively. Each distribution  $f_k$   and each primitive macroscopic variable is reconstructed independently, and the kinetic trace is upwinded by Eq.~\eqref{eq:fij}. For a scalar $\eta$, the trace from cell $i$ is
\begin{equation}
\eta_{ij}=\eta_i+\delta\eta_{ij},
\qquad
\delta\eta_{ij}=(\nabla\eta)_i\cdot(\vx_{ij}-\vx_i),
\label{eq:face-reconstruction}
\end{equation}
with the  cell   gradient evaluated by least squares,
\begin{equation}
(\nabla\eta)_i=\mathsf M_i^{-1}\sum_{j\in N(i)}w_{ij}\,\vec d_{ij}\,(\eta_j-\eta_i),
\qquad
\mathsf M_i=\sum_{j\in N(i)}w_{ij}\,\vec d_{ij}\vec d_{ij}^{\mathsf T},
\qquad
\vec d_{ij}=\vx_j-\vx_i,
\label{eq:least-squares}
\end{equation}
The stencil $N(i)$ contains the face-neighboring interior cells. Physical boundary faces and ghost cells do not contribute stencil members, and the outgoing trace at a boundary face is evaluated at the boundary-face center. Inverse-distance weights $w_{ij}=|\vec d_{ij}|^{-2}$ are used.

The   increment $\delta\eta_{ij}$ in Eq.~\eqref{eq:face-reconstruction}, applied to both the kinetic distributions and the primitive macroscopic variables, is limited by the smooth limiter of Venkatakrishnan~\cite{venkatakrishnan1995convergence}, whose smoothness allows the discrete adjoint to differentiate the actual limited reconstruction. The least-squares gradients also supply the velocity and temperature derivatives for the viscous flux. The two traces of the higher-order correction are averaged at the face as  in Eq.~\eqref{eq:Hij}. The discrete adjoint differentiates the actual stencils, limiters, and primitive-variable conversion, including their geometric dependence.
 
The interface NS flux in  Eq.~\eqref{eq:synthetic-macro-discrete}   is written as
$\bF_{\NS,ij}=\bF_{\mathrm{c},ij}+\bF_{\mathrm{v},ij}$, where the convective flux $\bF_{\mathrm{c},ij}$ uses the AUSM$^+$-up scheme~\cite{liou2006ausmup}, and the viscous flux $\bF_{\mathrm{v},ij}$ is computed by central differencing.
In addition, for the viscous flux, a correction coefficient is introduced on the viscosity $\mu$ and thermal conductivity $\kappa$ to suppress the robustness issue caused by the growth of the linear constitutive relation at large Knudsen numbers~\cite{yuan2021multi}: $\mu_{\mathrm{eff}}=\theta\mu$ and $\kappa_{\mathrm{eff}}=\theta\kappa$, with $\theta=1/\sqrt{1+(\mu/\mu_c)^2}$ and ${\mu _c} = \rho \sqrt {RT} {l_{{\rm{ref}}}}$. Note that this correction applies only to the linear constitutive relation of the NS flux, and the kinetic relaxation time $\tau=\mu/p$ is evaluated from the physical viscosity.

In the numerical implementation, to obtain a smooth initial field, the computation begins with a first-order NS solve using the Rusanov flux~\cite{rusanov1962calculation} for the convective terms, referred to as NS-pre\label{app:ns-pre}, followed by a few CIS steps based on the first-order DVM. This initialization supplies only the initial state for the subsequent CIS and GSIS iterations and is not involved in the adjoint analysis.

 \section{  Implicit inner  solvers }\label{app:inner-solvers}
 \subsection{Primal   increment equations }\label{app:primal-inner-solvers}
  Introducing the increment $\Delta f^{(m)}=f^{(m+1)}-f^{(m)}$ and a local pseudo-time step $\Delta t_{f,i,k}$, the backward-Euler discretization  of Eq.~\eqref{eq:gsis1-b}  in pseudo time gives the mesoscopic increment equation in delta form,
\begin{equation}
\left(\frac{V_i}{\Delta t_{f,i,k}}+a_{ii,k}\right)\Delta f_{i,k}^{(m)}
+\sum_{j\ne i}a_{ij,k}\Delta f_{j,k}^{(m)}
=-r_{i,k}^{(m)},
\label{eq:primal-kinetic-increment}
\end{equation}
where $r_{i,k}^{(m)}$ is the mesoscopic residual~\eqref{eq:rmeso-b} at the $m$-th inner iteration, and $a_{ij,k}=\delta_{ij}V_i/\tau_i+\partial(\DD_{i,k}+r_{\mathrm B,i,k}^{\mathrm{inc}})/\partial f_{j,k}$ contains the derivatives of the interior transport and the boundary correction at the same iteration. For the increment terms on the left-hand side, a compact-stencil, first-order upwind approximation to the Jacobian is used for $a_{ij,k}$. The resulting increment gives $f^{(m+1)}=f^{(m)}+\Delta f^{(m)}$. The increment equation~\eqref{eq:primal-kinetic-increment} is solved with lower--upper symmetric Gauss--Seidel (LU--SGS), with a scalar kinetic diagonal at each velocity node. For each velocity node, the cells are ordered along its upwind transport direction by Kahn's algorithm~\cite{kahn1962topological}. Under this velocity-dependent ordering the implicit operator is triangular, and a single forward sweep solves it exactly.

The macroscopic increment equation follows the same delta form, with $\Delta\bU^{(m)}=\bU^{(m+1)}-\bU^{(m)}$ and a local pseudo-time step $\Delta t_{U,i}$ ,
  \begin{equation}
\left(\frac{V_i}{\Delta t_{U,i}}\mathsf I+\mathsf J_{ii}\right)\Delta\bU_i^{(m)}
+\sum_{j\ne i}\mathsf J_{ij}\Delta\bU_j^{(m)}
=-\bR_i^{(m)},
\label{eq:primal-macro-increment}
\end{equation}
where $\mathsf I$ is the identity matrix, $\bR_i^{(m)}$ is the macroscopic residual~\eqref{eq:Rmacro-b} at the $m$-th inner iteration, and $\mathsf J_{ij}=\partial(\RR_{\NS,i}[\bU]+\RR_{\BM,i}[\bU])/\partial\bU_j$ is evaluated at the same iteration. The equations are solved component by component with scalar diagonals in the same LU--SGS sweeps, and the update is $\bU^{(m+1)}=\bU^{(m)}+\Delta\bU^{(m)}$ .

\subsection{  Adjoint increment equations }\label{app:adjoint-inner-solvers}
  In the adjoint equations, $a_{ij,k}$ and $\mathsf J_{ij}$ are the spatial coefficients of Eqs.~\eqref{eq:primal-kinetic-increment} and~\eqref{eq:primal-macro-increment} evaluated at the converged primal flow. The sources $b_{f,i,k}$ and $\bm b_{U,i}$ are the frozen right-hand sides of Eqs.~\eqref{eq:adj-meso-full} and~\eqref{eq:adj-macro-b}. As in the primal iteration, the collision moments and boundary terms in $b_{f,i,k}$ are evaluated from the latest adjoint variables and held fixed during the inner solve , and   $\bm b_{U,i}$ is updated from the latest mesoscopic adjoint. For the Shakhov model, the additional heat-flux moment enters the frozen kinetic source as described in Appendix~\ref{app:shakhov-adjoint}. Since the adjoint operator is the transpose of the primal linearization, the kinetic coefficients exchange the cell indices and the macroscopic blocks are transposed, while the collision term $V_i/\tau_i$ and the pseudo-time term remain unchanged.

Let $\Delta\phi^{(m)}=\phi^{(m+1)}-\phi^{(m)}$. With the same local pseudo-time step as in  Eq.~  \eqref{eq:primal-kinetic-increment}, the mesoscopic adjoint increment equation is
\begin{equation}
\left(\frac{V_i}{\Delta t_{f,i,k}}+a_{ii,k}\right)\Delta\phi_{i,k}^{(m)}
+\sum_{j\ne i}a_{ji,k}\Delta\phi_{j,k}^{(m)}
=b_{f,i,k}-\sum_j a_{ji,k}\phi_{j,k}^{(m)},
\label{eq:adjoint-kinetic-increment}
\end{equation}
with the update $\phi^{(m+1)}=\phi^{(m)}+\Delta\phi^{(m)}$.

 Let $\Delta\bwh^{(m)}=\bwh^{(m+1)}-\bwh^{(m)}$. With the same local pseudo-time step as in Eq.~\eqref{eq:primal-macro-increment}, the macroscopic adjoint increment equation is
\begin{equation}
\left(\frac{V_i}{\Delta t_{U,i}}\mathsf I+\mathsf J_{ii}^{\mathsf T}\right)\Delta\bwh_i^{(m)}
+\sum_{j\ne i}\mathsf J_{ji}^{\mathsf T}\Delta\bwh_j^{(m)}
=\bm b_{U,i}-\sum_j\mathsf J_{ji}^{\mathsf T}\bwh_j^{(m)},
\label{eq:adjoint-macro-increment}
\end{equation}
with the update $\bwh^{(m+1)}=\bwh^{(m)}+\Delta\bwh^{(m)}$.

The transposed products are evaluated matrix-free: for the mesoscopic equation, the contribution of each face is accumulated to its face neighbors velocity by velocity, with the collision term applied pointwise; for the macroscopic equation, the flux derivatives at the converged primal flow are stored and applied component by component. The kinetic storage scales as $O(N_{\mathrm{cell}}N_{\mathrm{vel}})$ as in the primal solve, where $N_{\mathrm{cell}}$ and $N_{\mathrm{vel}}$ are the numbers of cells and velocity nodes.

Both adjoint increment equations are solved in the same manner as their primal counterparts. The primal cell ordering is reused: the transposed operator reverses the upwind dependencies, and a single sweep in the reverse ordering solves Eq.~\eqref{eq:adjoint-kinetic-increment} exactly.

 \section{Derivation details of the discrete adjoint system}\label{app:adjoint-derivations}
This appendix gives the principal differentiation steps leading to the adjoint equations in Section~\ref{sec:method}. The mesh, velocity quadrature, and all independent variables other than the one being differentiated are held fixed. Following Section~\ref{sec:interior-adjoint}, the objective contribution and explicit boundary terms are set aside in \ref{app:interior-derivatives} and~\ref{app:interior-stationarity} and included in \ref{app:boundary-stationarity}.

\subsection{Derivatives of the interior residuals}\label{app:interior-derivatives}
Because mesoscopic reconstruction acts separately at each velocity node, the transport derivative is
\begin{equation}
\frac{\partial\DD_{i,k}[f]}{\partial f_{\ih,\kh}}
=\delta_{k\kh}\frac{\partial\DD_{i,\kh}[f]}{\partial f_{\ih,\kh}}.
\label{eq:dDdf}
\end{equation}
Using Eq.~\eqref{eq:dUdf}, the collision and NS residual derivatives are
\begin{align}
\frac{\partial g_k(\bU_i)}{\partial f_{\ih,\kh}}
&=\delta_{i\ih}\frac{\partial g_k(\bU_i)}{\partial\bU_i}\cdot\bpsi_{\kh}\dXi_{\kh},\label{eq:dgdf}\\
\frac{\partial\tau_i}{\partial f_{\ih,\kh}}
&=\delta_{i\ih}\frac{\partial\tau_i}{\partial\bU_i}\cdot\bpsi_{\kh}\dXi_{\kh},\label{eq:dtaudf}\\
\frac{\partial\RR_{\NS,i}[\bU]}{\partial f_{\ih,\kh}}
&=\frac{\partial\RR_{\NS,i}[\bU]}{\partial\bU_{\ih}}\cdot\bpsi_{\kh}\dXi_{\kh}.\label{eq:dRNSdf}
\end{align}
The derivative in Eq.~\eqref{eq:dRNSdf} includes every occurrence of $\bU_{\ih}$ in the face fluxes and their reconstruction stencils.

For the higher-order correction, the cell-center NS flux first gives
\begin{equation}
\frac{\partial\vec{\bF}_{\NS,j}[\bW]}{\partial\bW_{\ih}}
=\delta_{j\ih}\left.\frac{\partial\vec{\bF}_{\NS}}{\partial\bW}\right|_j
+\left.\frac{\partial\vec{\bF}_{\NS}}{\partial\bm G}\right|_j
:\frac{\partial\GG_j[\bW]}{\partial\bW_{\ih}}.
\label{eq:dFNSdW}
\end{equation} 
The argument $\bm G$ collects the velocity and temperature gradients in Eq.~\eqref{eq:FNS-cell}. The vertical bar denotes evaluation at $(\bW_j,\GG_j[\bW])$. The two flux partial derivatives respectively hold   $\bm G$  and the local state fixed. The derivative of $\GG_j[\bW]$ includes the primitive-variable conversion and the gradient stencil. Differentiating Eq.~\eqref{eq:Hcell} then yields
\begin{equation}
\frac{\partial\vec{\bH}_j}{\partial f_{\ih,\kh}}
=\left[\delta_{j\ih}\vv_{\kh}\bpsi_{\kh}
-\frac{\partial\vec{\bF}_{\NS,j}[\bW]}{\partial\bW_{\ih}}\cdot\bpsi_{\kh}\right]\dXi_{\kh}.
\label{eq:dHdf}
\end{equation}
Since the higher-order face reconstruction depends on the flow only through $\vec{\bH}$, the remaining chain rule is
\begin{equation}
\frac{\partial\RR_{H,i}[\bH]}{\partial f_{\ih,\kh}}
=\sum_j\frac{\partial\RR_{H,i}[\bH]}{\partial\vec{\bH}_j}
:\frac{\partial\vec{\bH}_j}{\partial f_{\ih,\kh}}=\left[\frac{\partial\RR_{H,i}[\bH]}{\partial\vec{\bH}_{\ih}}:(\vv_{\kh}\bpsi_{\kh})
-\bK_{i\ih}\cdot\bpsi_{\kh}\right]\dXi_{\kh},
\label{eq:dRHdf}
\end{equation} 
where $\bK$ is defined in Eq.~\eqref{eq:Kdef}. The combined stencil therefore includes the dependencies from both the higher-order face reconstruction and the NS gradient reconstruction.

For differentiation with respect to the independent macroscopic correction, $\partial\bU_i/\partial\bw_{\ih}=\delta_{i\ih}\mathsf I$, with $\mathsf I$ the identity in conservative-variable space. Thus
 \begin{equation}
\frac{\partial g_k(\bU_i)}{\partial\bw_{\ih}}
=\delta_{i\ih}\frac{\partial g_k(\bU_i)}{\partial\bU_i},
\qquad
\frac{\partial\tau_i}{\partial\bw_{\ih}}
=\delta_{i\ih}\frac{\partial\tau_i}{\partial\bU_i},\qquad
\frac{\partial\RR_{\NS,i}[\bU]}{\partial\bw_{\ih}}
=\frac{\partial\RR_{\NS,i}[\bU]}{\partial\bU_{\ih}}.
\label{eq:dw}
\end{equation} 
Although $\vec{\bH}_j$ depends on multiple cells through $\GG_j[\bW]$, none of these kinetic moments depends on $\bw$. Consequently,
\begin{equation}
\frac{\partial\vec{\bH}_j}{\partial\bw_{\ih}}=\bm0,
\qquad
\frac{\partial\RR_{H,i}[\bH]}{\partial\bw_{\ih}}=\bm0.
\label{eq:dHdw}
\end{equation}

\subsection{Expansion of the interior stationarity conditions}\label{app:interior-stationarity}
For the interior terms of Eq.~\eqref{eq:lagrangian-full}, differentiation with respect to $f_{\ih,\kh}$ produces contributions from transport, the collision target, direct relaxation, the relaxation time, and the two macroscopic flux components. Dividing by the common velocity weight and using Eq.~\eqref{eq:CWhat} gives
\begin{equation}
\begin{aligned}
\frac{1}{\dXi_{\kh}}\frac{\partial L}{\partial f_{\ih,\kh}}
={}&\sum_i\phi_{i,\kh}\frac{\partial\DD_{i,\kh}[f]}{\partial f_{\ih,\kh}}
+\frac{V_{\ih}}{\tau_{\ih}}\phi_{\ih,\kh}
-\frac{V_{\ih}}{\tau_{\ih}}\bWh_{\ih}\cdot\bpsi_{\kh}
+V_{\ih}\bCh_{\ih}\cdot\bpsi_{\kh}\\
&+\sum_i\bwh_i\cdot\frac{\partial\RR_{\NS,i}[\bU]}{\partial\bU_{\ih}}\cdot\bpsi_{\kh}
+\sum_i\bwh_i\cdot\left[
\frac{\partial\RR_{H,i}[\bH]}{\partial\vec{\bH}_{\ih}}:(\vv_{\kh}\bpsi_{\kh})
-\bK_{i\ih}\cdot\bpsi_{\kh}\right].
\end{aligned}
\label{eq:dLdf}
\end{equation} 
In particular, the positive sign of the $\bCh$ term follows by differentiating the full collision residual $-V_i(g_k-f_{i,k})/\tau_i$ with respect to $\tau_i$. Using Eq.~\eqref{eq:dHdw}, stationarity in $\bw_{\ih}$ gives
\begin{equation}
\frac{\partial L}{\partial\bw_{\ih}}
=\sum_i\bwh_i\cdot\frac{\partial\RR_{\NS,i}[\bU]}{\partial\bU_{\ih}}
-\frac{V_{\ih}}{\tau_{\ih}}\bWh_{\ih}+V_{\ih}\bCh_{\ih}.
\label{eq:dLdw}
\end{equation}
Applying the stationarity conditions~\eqref{eq:stationarity} to Eqs.~\eqref{eq:dLdf} and~\eqref{eq:dLdw} and rearranging recovers Eqs.~\eqref{eq:adj-meso} and~\eqref{eq:adj-macro}.

\subsection{Boundary stationarity conditions}\label{app:boundary-stationarity}
The independent boundary distributions satisfy
\begin{equation}
\frac{\partial f_{ib,k}}{\partial f_{\ibh,\kh}}
=\delta_{i\ih}\delta_{b\bh}\delta_{k\kh}.
\label{eq:dfib}
\end{equation}
The Jacobian of their constraints is
\begin{equation}
\frac{\partial r_{ib,k}}{\partial f_{\ibh,\kh}}
=\begin{cases}
\delta_{i\ih}\delta_{b\bh}\delta_{k\kh},&k\in M(ib,+),\\
\delta_{i\ih}\delta_{b\bh}\delta_{k\kh},&k\in M(ib,-),\ \kh\in M(\ibh,-),\\
-\delta_{i\ih}\delta_{b\bh}\dfrac{\partial g_{\mathrm r,\ibh,k}}{\partial f_{\ibh,\kh}},
&k\in M(ib,-),\ \kh\in M(\ibh,+).
\end{cases}
\label{eq:drib-dfib}
\end{equation}
The boundary multiplier term can accordingly be split over $M(ib,+)$ and $M(ib,-)$. The two flow-residual terms of the Lagrangian contribute
\begin{align}
\frac{\partial}{\partial f_{\ibh,\kh}}
\sum_i\sum_k\phi_{i,k}r_{i,k}\dXi_k
&=A_{\ibh}(\vv_{\kh}\cdot\vn_{\ibh})\phi_{\ih,\kh}\dXi_{\kh},\label{eq:dphi-dfib}\\
\frac{\partial}{\partial f_{\ibh,\kh}}\sum_i\bwh_i\cdot\bR_i
&=A_{\ibh}(\vv_{\kh}\cdot\vn_{\ibh})(\bwh_{\ih}\cdot\bpsi_{\kh})\dXi_{\kh}.\label{eq:dw-dfib}
\end{align}
The second line comes from the kinetic boundary-flux moment in Eq.~\eqref{eq:RB}. There is no extra term from $\GG_j$ or $\vec{\bH}_j$, which are independent of $f_{ib,k}$. Using Eq.~\eqref{eq:Bhat}, the boundary multiplier contribution is
\begin{equation}
\frac{\partial}{\partial f_{\ibh,\kh}}
\sum_{ib}\sum_k\varphi_{ib,k}r_{ib,k}A_{ib}\dXi_k
=\begin{cases}
A_{\ibh}\varphi_{\ibh,\kh}\dXi_{\kh},&\kh\in M(\ibh,-),\\
A_{\ibh}\big(\varphi_{\ibh,\kh}\dXi_{\kh}-\hat B_{\ibh,\kh}\big),&\kh\in M(\ibh,+).
\end{cases}
\label{eq:dvarphi-dfib}
\end{equation}
Adding Eq.~\eqref{eq:dJ}, imposing Eq.~\eqref{eq:stationarity}, and dividing by $A_{\ibh}\dXi_{\kh}$ gives the boundary relations~\eqref{eq:phi-final-neg} and~\eqref{eq:phi-final-pos}. The factor $\dXi_{\kh}$ inside the diffuse-reflection derivative in Eq.~\eqref{eq:reflection-derivative} is retained in $\hat B$ before this division.

For the interior distribution, only the outgoing branch of the boundary constraint has a direct reconstruction derivative:
\begin{equation}
\frac{\partial r_{ib,k}}{\partial f_{\ih,\kh}}
=-\delta_{k\kh}\Heavi(\vv_k\cdot\vn_{ib})
\frac{\partial\CC_{ib}[f_{\kh}]}{\partial f_{\ih,\kh}}.
\label{eq:drib-df}
\end{equation}
Its contribution to the interior stationarity condition is
\begin{equation}
\frac{\partial}{\partial f_{\ih,\kh}}
\sum_{ib}\sum_k\varphi_{ib,k}r_{ib,k}A_{ib}\dXi_k
=-\sum_{ib}A_{ib}\varphi_{ib,\kh}\dXi_{\kh}\Heavi(\vv_{\kh}\cdot\vn_{ib})
\frac{\partial\CC_{ib}[f_{\kh}]}{\partial f_{\ih,\kh}}.
\label{eq:adj-bdry-extra}
\end{equation}
After dividing by $\dXi_{\kh}$ and substituting Eq.~\eqref{eq:phi-final-pos}, the direct $\phi$ term has a positive sign and remains on the left of Eq.~\eqref{eq:adj-meso-full}. Moving the remaining boundary terms to the right gives its last line, with $\hat B/\dXi_{\kh}$ positive and the direct macroscopic-adjoint and objective terms negative. Finally, differentiation of $\RR_{\BM,i}[\bU]-\RR_{\BM,i}[\bW]$ supplies the difference of the two boundary-flux Jacobians in the mesoscopic equation. Only the first of these depends on the independent $\bw$, yielding the boundary contribution on the left of Eq.~\eqref{eq:adj-macro-b}. This completes the full stationarity system without freezing any residual dependency during differentiation.

\section{Extension to the Shakhov model}\label{app:shakhov-extension}
The Shakhov model modifies the collision target through a heat-flux correction~\cite{shakhov1968generalization,yuan2021multi}. The discrete adjoint construction remains based on the same independent variables and residual constraints. This appendix identifies the additional distribution dependence and the resulting adjoint source term.

\subsection{Shakhov target distribution and heat-flux moment}\label{app:shakhov-model}
To distinguish quantities extracted from the predicted macroscopic state from those extracted directly from the kinetic moments, write
\begin{equation}
\begin{aligned}
\bU_i&=(\tilde\rho_i,\tilde\rho_i\tilde{\vec u}_i,\tilde\rho_i\tilde E_i)^{\mathsf T},\\
\tilde T_i&=\frac{\gamma-1}{R}\left(\tilde E_i-\frac{|\tilde{\vec u}_i|^2}{2}\right),
\qquad \tilde p_i=\tilde\rho_i R\tilde T_i.
\end{aligned}
\label{eq:U-primitive}
\end{equation}
The tildes identify components of $\bU_i$, not iteration levels or adjoint variables. Untilded quantities continue to be defined by $\bW_i$. With $\tilde{\vec c}_{i,k}=\vv_k-\tilde{\vec u}_i$, the Shakhov target is
\begin{equation}
g_{\mathrm S,k}(\bU_i,\vec q_i)
=g_{\mathrm M,k}(\tilde\rho_i,\tilde{\vec u}_i,\tilde T_i)
\left[1+(1-\mathrm{Pr})\frac{\vec q_i\cdot\tilde{\vec c}_{i,k}}{5\tilde p_i R\tilde T_i}
\left(\frac{|\tilde{\vec c}_{i,k}|^2}{R\tilde T_i}-5\right)\right].
\label{eq:shakhov-target}
\end{equation}
Here the Prandtl number is $\mathrm{Pr}=c_p\mu/\kappa$, where $\kappa$ is the heat conductivity. It controls heat-flux relaxation relative to momentum transport.   The Shakhov calculations use $\mathrm{Pr}=2/3$ .

The interior mesoscopic residual becomes
\begin{equation}
r_{i,k}^{\mathrm S}=\DD_{i,k}[f]
-V_i\frac{g_{\mathrm S,k}(\bU_i,\vec q_i)-f_{i,k}}{\tau_i},
\label{eq:S-rmeso}
\end{equation}
with the boundary contribution added exactly as in Eq.~\eqref{eq:rmeso-b}. The heat flux in Eq.~\eqref{eq:shakhov-target} is evaluated using the kinetic mean velocity, not $\tilde{\vec u}_i$. Define
\begin{equation}
\vec c_{i,k}=\vv_k-\vec u_i,
\label{eq:S-c}
\end{equation}
and the following kinetic moments:
\begin{equation}
\begin{aligned}
\vec q_i&=\frac{1}{2}\sum_k\vec c_{i,k}|\vec c_{i,k}|^2 f_{i,k}\dXi_k,\\
E_{\mathrm{th},i}&=\frac{1}{2\rho_i}\sum_k|\vec c_{i,k}|^2 f_{i,k}\dXi_k,
\qquad
\mathsf P_i=\sum_k\vec c_{i,k}\vec c_{i,k}f_{i,k}\dXi_k.
\end{aligned}
\label{eq:S-q-defs}
\end{equation}
Here $E_{\mathrm{th},i}$ is the specific thermal energy and $\mathsf P_i$ the kinetic pressure tensor. Both $\rho_i$ and $\vec u_i$ are determined from $\bW_i$; consequently, $\vec q_i$ depends on $f$ but not on the independent correction $\bw$. No separate heat-flux state or additional multiplier is introduced.

\subsection{Derivatives of the Shakhov target distribution}\label{app:shakhov-derivatives}
The velocity entering Eq.~\eqref{eq:S-c} varies with the distribution according to
\begin{equation}
\frac{\partial\vec u_i}{\partial f_{\ih,\kh}}
=\delta_{i\ih}\frac{\vec c_{i,\kh}}{\rho_i}\dXi_{\kh}.
\label{eq:S-dudf}
\end{equation}
Differentiating the heat-flux moment must therefore retain the change in this mean velocity. Using Eq.~\eqref{eq:S-q-defs}, the exact discrete derivative is
\begin{equation}
\frac{\partial\vec q_i}{\partial f_{\ih,\kh}}
=\delta_{i\ih}\vec\zeta_{i,\kh}\dXi_{\kh},
\qquad
\vec\zeta_{i,\kh}
=\frac{1}{2}\vec c_{i,\kh}|\vec c_{i,\kh}|^2
-E_{\mathrm{th},i}\vec c_{i,\kh}
-\frac{1}{\rho_i}\mathsf P_i\cdot\vec c_{i,\kh}.
\label{eq:S-dqdf}
\end{equation}
The last two terms arise from the dependence of $\vec u_i$ on $f$. Together with Eq.~\eqref{eq:dUdf}, this gives
\begin{equation}
\frac{\partial g_{\mathrm S,k}(\bU_i,\vec q_i)}{\partial f_{\ih,\kh}}
=\delta_{i\ih}\left[
\frac{\partial g_{\mathrm S,k}}{\partial\bU_i}\cdot\bpsi_{\kh}
+\frac{\partial g_{\mathrm S,k}}{\partial\vec q_i}\cdot\vec\zeta_{i,\kh}
\right]\dXi_{\kh}.
\label{eq:S-dgdf}
\end{equation}
The partial derivative with respect to $\bU_i$ holds $\vec q_i$ fixed, and that with respect to $\vec q_i$ holds $\bU_i$ fixed. Since the kinetic heat flux is independent of $\bw$,
\begin{equation}
\frac{\partial g_{\mathrm S,k}(\bU_i,\vec q_i)}{\partial\bw_{\ih}}
=\delta_{i\ih}\frac{\partial g_{\mathrm S,k}}{\partial\bU_i}.
\label{eq:S-dgdw}
\end{equation}

\subsection{Modified adjoint source terms}\label{app:shakhov-adjoint}
Replace the collision moments in Eq.~\eqref{eq:CWhat} by their Shakhov counterparts and introduce the heat-flux adjoint moment:

\begin{equation}
\begin{aligned}
\bWh_i&=\sum_k\phi_{i,k}\frac{\partial g_{\mathrm S,k}(\bU_i,\vec q_i)}{\partial\bU_i}\dXi_k,
\quad
\hat{\vec Q}_i=\sum_k\phi_{i,k}\frac{\partial g_{\mathrm S,k}(\bU_i,\vec q_i)}{\partial\vec q_i}\dXi_k,\\
\bCh_i&=\left(\sum_k\phi_{i,k}\frac{g_{\mathrm S,k}(\bU_i,\vec q_i)-f_{i,k}}{\tau_i^2}\dXi_k\right)
\frac{\partial\tau_i}{\partial\bU_i}.
\end{aligned}
\label{eq:S-WQChat}
\end{equation}

With these moments, the stationarity conditions~\eqref{eq:stationarity} give the modified interior mesoscopic adjoint equation,
\begin{equation}
\begin{aligned}
&\sum_i\phi_{i,\kh}\frac{\partial\DD_{i,\kh}[f]}{\partial f_{\ih,\kh}}
+\frac{V_{\ih}}{\tau_{\ih}}\phi_{\ih,\kh}\\
&\quad=\frac{V_{\ih}}{\tau_{\ih}}\left(\bWh_{\ih}\cdot\bpsi_{\kh}
+\hat{\vec Q}_{\ih}\cdot\vec\zeta_{\ih,\kh}\right)
-V_{\ih}\bCh_{\ih}\cdot\bpsi_{\kh}
-\sum_i\bwh_i\cdot\frac{\partial\RR_{\NS,i}[\bU]}{\partial\bU_{\ih}}\cdot\bpsi_{\kh}\\
&\qquad-\sum_i\bwh_i\cdot\left[
\frac{\partial\RR_{H,i}[\bH]}{\partial\vec{\bH}_{\ih}}:(\vv_{\kh}\bpsi_{\kh})
-\bK_{i\ih}\cdot\bpsi_{\kh}\right].
\end{aligned}
\label{eq:S-adj-meso}
\end{equation} 
The additional coupling is the source $V_{\ih}\hat{\vec Q}_{\ih}\cdot\vec\zeta_{\ih,\kh}/\tau_{\ih}$. The gradient-reconstruction contribution remains fully represented by $\bK$. The macroscopic adjoint retains the form of Eq.~\eqref{eq:adj-macro}, or Eq.~\eqref{eq:adj-macro-b} with boundaries, using the moments in Eq.~\eqref{eq:S-WQChat}; no heat-flux term is generated by differentiation in $\bw$.

The boundary adjoint relations and boundary contributions in Eq.~\eqref{eq:adj-meso-full} are unchanged. After their substitution, the direct $\phi$ contribution is still updated on the left, while the remaining boundary terms are frozen on the right. At the start of each mesoscopic adjoint inner solve, the additional moment $\hat{\vec Q}$ is evaluated from the latest available $\phi$ and frozen together with $\bWh$ and $\bCh$. Thus the three-step iteration of Section~\ref{sec:adjoint-iteration} and the design-sensitivity procedure of Section~\ref{sec:sens} apply without structural modification.

%----------------------------------------------------------
% 4. References
%----------------------------------------------------------

\bibliographystyle{elsarticle-num}
\bibliography{2026_dadjoint}

\end{document}